\documentclass[twocolumn,apj,twocolappendix,numberedappendix]{openjournal}
\usepackage{natbib}
\usepackage{graphicx,amsmath,amssymb,amstext}
\usepackage{amsbsy,amsfonts,amsthm,color,xcolor}

\usepackage{siunitx}
\usepackage[colorlinks,linkcolor=blue,citecolor=blue,urlcolor=blue ]{hyperref}
\usepackage[utf8]{inputenc}
\usepackage{float}
\usepackage[caption=false]{subfig}
\usepackage{upgreek}

\usepackage{orcidlink}

\newcommand{\eagle}{\mbox{EAGLE}}
\newcommand{\euclid}{\mbox{\it Euclid}}
\newcommand{\hst}{\mbox{\it HST}}
\newcommand{\jwst}{\mbox{\it JWST}}
\newcommand{\flares}{\mbox{FLARES}}
\newcommand{\rst}{\mbox{\it Roman Space Telescope}}

\newcommand{\synth}{\texttt{synthesizer}}
\newcommand{\hii}{H\textsc{ii}}
\newcommand{\cloudy}{\texttt{cloudy}}
\newcommand{\ippe}{$\xi_{\rm ion}$}
\newcommand{\Ha}{H$\alpha$}
\newcommand{\Hb}{H$\beta$}
\newcommand{\Msun}{{\rm M_{\odot}}}

\newcommand{\um}{\SI{}{\micro\meter}}
\newcommand{\eg}[0]{$\textnormal{e.g. }$}
\newcommand{\ie}[0]{$\textnormal{i.e. }$}

\defcitealias{lovell2020}{\flares\,I}
\newcommand\blfootnote[1]{%
  \begingroup
  \renewcommand\thefootnote{}\footnote{#1}%
  \addtocounter{footnote}{-1}%
  \endgroup
}

\begin{document}

\title{Interpreting nebular emission lines in the high-redshift Universe\vspace{-3em}}
\author{Aswin P. Vijayan$^{1,\star}$\orcidlink{0000-0002-1905-4194}} % Orcid: 0000-0002-1905-4194

\author{Robert M. Yates$^{2}$\orcidlink{0000-0001-9320-4958}}

\author{Christopher C. Lovell$^{3}$\orcidlink{0000-0001-7964-5933}}
\author{William J. Roper$^{1}$\orcidlink{0000-0002-3257-8806}}
\author{Stephen M. Wilkins$^{1}$\orcidlink{0000-0003-3903-6935}}

\author{Hiddo S. B. Algera$^{4}$\orcidlink{0000-0002-4205-9567}}
\author{Shihong Liao$^{5}$\orcidlink{0000-0001-7075-6098}}
\author{Paurush Punyasheel$^2$\orcidlink{0009-0006-0037-1014}}
\author{Lucie E. Rowland$^{6}$}
\author{Louise T. C. Seeyave$^{1}$\orcidlink{0000-0002-7020-3079}}

\blfootnote{$^{\star}$\mbox{Corresponding author, email: \href{mailto:A.Payyoor-Vijayan@sussex.ac.uk}{A.Payyoor-Vijayan@sussex.ac.uk}}}

% List of institutions
\affiliation{$^1$Astronomy Centre, University of Sussex, Falmer, Brighton BN1 9QH, UK}

\affiliation{$^2$Centre for Astrophysics Research, University of Hertfordshire, Hatfield, AL10 9AB, UK}

\affiliation{$^3$Institute of Cosmology and Gravitation, University of Portsmouth, Burnaby Road, Portsmouth, PO1 3FX, UK}

\affiliation{$^4$Institute of Astronomy and Astrophysics, Academia Sinica, 11F of Astronomy-Mathematics Building, No.1, Sec. 4, Roosevelt Rd, Taipei 106216, Taiwan, R.O.C.}

\affiliation{$^5$Key Laboratory for Computational Astrophysics, National Astronomical Observatories, Chinese Academy of Sciences, Beijing 100101, China}

\affiliation{$^6$Leiden Observatory, Leiden University, P.O. Box 9513, 2300 RA Leiden, The Netherlands}

\begin{abstract}
One of the most remarkable outcomes from \textit{JWST} has been the exquisite UV-optical spectroscopic data for galaxies in the high-redshift Universe ($z \geq 5$), enabling the use of various nebular emission lines to infer conditions of the interstellar medium. In this work, we assess the reliability of commonly used diagnostics for estimating the star formation rate (SFR), the ionising photon production efficiency ($\xi_{\rm ion}$), and the gas-phase oxygen abundance, focusing on dust corrections based on A$_{\rm V}$ (V-band attenuation) and the Balmer decrement. Using forward-modelled galaxy spectra from idealised toy models and the \flares\ cosmological hydrodynamical simulations, we examine how variations in stellar populations and star-dust geometry affect these diagnostics. 
We find that the clumpy nature of \flares\ galaxies lead to strong internal variation in age, metallicity and dust attenuation, biasing the inferred quantities. 
In \flares\ the SFRD at the bright-end of the SFR function can be underestimated by as much as $30\%$ compared to the true values. 
While the intrinsic $\xi_{\rm ion}$ in \flares\ is nearly constant with stellar mass, estimates derived from \Ha\ or \Hb\ can be underestimated by more than 0.5 dex at high stellar masses ($>10^{9.5}$ M$_{\odot}$), introducing an artificial declining trend.
Similarly, the dust-corrected mass-metallicity relation inferred from line ratios is significantly flatter than the intrinsic mass-weighted relation.
These systematic offsets arise from the coupling between heterogeneous stellar populations and non-uniform star-dust geometry and depend on the diagnostic and the dust-correction method employed. 
No single dust-correction approach yields unbiased estimates of all quantities simultaneously, highlighting the need for forward modelling and comparisons in observed space for robust high-redshift inference.

\end{abstract}

\maketitle

%%%%%%%%%%%%%%%%%%%%%%%%%%%%%%%%%%%%%%%%%%%%%%%%%%

%%%%%%%%%%%%%%%%% BODY OF PAPER %%%%%%%%%%%%%%%%%%

\section{Introduction}\label{sec:intro}

Understanding the formation and evolution of early galaxies is a key priority for many current and upcoming observatories, including ALMA, ELT, \euclid, \hst, and \jwst\ \cite[]{stark_2016_review,robertson2022_review}.
Complementing these observational efforts, substantial progress has also been made on the theoretical front, particularly through the development of sophisticated simulations of galaxy formation and evolution \cite[][]{Dave_somerville_2015review,Vogelsberger2020,crain_voort_review}. 
In the current extragalactic astronomy paradigm, this understanding 
% manifests 
is achieved by deriving the
physical properties of galaxies such as the stellar mass, star formation rates, and gas-phase metallicities, and comparing these across observations and simulations. These comparisons provide critical insight into how galaxies build up their mass and chemically enrich over cosmic time.

Recent years have seen crucial strides in this direction achieved through the advent of \jwst. \jwst\ has enabled the study of the physical properties of galaxies in the high-redshift Universe through its powerful photometric and spectroscopic capabilities in the observed near- and mid-infrared. The spectroscopic capabilities have been crucial in obtaining nebular emission and absorption lines, providing insight into the physical conditions of the interstellar medium (ISM), including ionisation state \cite[]{Cameron2023,Reddy2023}, electron densities \cite[]{Isobe2023ne,Li2025_ne}, gas-phase abundances \cite[]{Arellano2022,Curti2023,Schady2024,Chakraborty2025}, and dust content \cite[]{Sandles2024,Burgarella2025}, as well as the intrinsic stellar population properties \cite[]{Rowland2025_rebels} of these early galaxies.

These physical quantities are primarily inferred from multiple emission lines originating from the nebular regions of  galaxies. By analysing these line ratios via scaling relations or empirical/theoretical calibrations \cite[]{Kewley2019}, the community has inferred quantities such as the star formation rate, ionising photon production efficiency, dust attenuation, and metallicity of these galaxies. These calibrations rely on photoionisation models, Stellar Population Synthesis (SPS) models, assumptions on the conditions of the ISM or secondary fits to well-established relations. Many of these relations have also been filtered down into analytical forms, such as polynomial fits, enabling quick estimation of galaxy properties from spectroscopic data \cite[\eg][]{Nakajima2023_metallicity,Sanders2024_calibration}. However, these relationships are only as robust as the validity of the underlying assumptions and models used to derive them. 

% hence are bound to give a `numerical value'. 
Theoretical models are essential tools to provide insight into how these observables relate to the underlying physical properties. 
In this work, we explore whether the commonly-used empirical or theoretical calibrations break-down for high-redshift galaxies containing stellar populations with varying ages, metallicity, and star-dust geometry\footnote{instead of uniform dust screen models that assume a single value of dust attenuation for a galaxy, also see \cite{salim_narayanan_2020,FLARES-XII}}.
% the reliability of these measurements for galaxies with variations in the  underlying physical properties, such as their ages, metallicities, and star-dust geometry \cite[instead of uniform dust screen models that assume a single value of dust attenuation for a galaxy, also see][]{salim_narayanan_2020,FLARES-XII}. 
% We will also explore how, by using commonly used calibrations or methodology, one can come to very different conclusions on the physical properties of galaxies in the early Universe. 
We will explore this by combining simple toy models and forward modelled galaxies from the First Light And Reionisation Epoch Simulations \cite[\flares,][]{FLARESI,FLARESII}, a suite of hydrodynamical resimulations based on the \eagle\ \cite[]{schaye_eagle_2015,crain_eagle_2015} model. 
Using these theoretical models, we will demonstrate that a clearer understanding of the reliability of these emission line diagnostics
% of these measures 
are required to make robust conclusions about the physics of galaxy formation, and its evolutionary trends in the high-redshift Universe.

The paper is structured as follows. In Section~\ref{sec:modeling} we introduce our modelling of the spectral energy distribution (SED) of the toy model galaxies and those from the \flares\ \cite[]{FLARESI,FLARESII} suite of hydrodynamical simulations. In Section~\ref{sec:geometry} we explore why the star-dust geometry within galaxies plays an important role in the observed galaxy properties, such as those derived using emission lines. In Section~\ref{sec:results} we present our results, focusing on the reliability of emission lines to estimate the star formation rate (Section~\ref{sec:results.sfrd}), the ionising photon production rate (Section~\ref{sec:results.ippe}) and the mass-metallicity relation (Section~\ref{sec:results.mzr}) using toy model galaxies and \flares\ galaxies. We present some of the caveats to consider when interpreting our results in Section~\ref{sec:caveats}. We finally summarise and present our conclusions in Section~\ref{sec:conclusions}.

Throughout this work we assume a Planck year 1 cosmology
\cite[$\Omega_{\rm m} = 0.307$, $\Omega_{\rm \Lambda} = 0.693$, ${\rm h} = 0.6777$,][]{planck_collaboration_2014} and solar metallicity, Z$_{\odot} = 0.014$.

\section{Modelling and Simulations}\label{sec:modeling}
\subsection{Synthetic spectral energy distributions}\label{sec:modelling.forward}
An important aspect in this study is the generation of spectra from toy models and \flares\ galaxies. In this subsection we describe how we forward model our galaxies. 

Young massive stars, mainly the O and B stars, produce significant amount of energetic photons which ionise their immediate surroundings. These so called \hii\ regions are the source of nebular line and continuum emission that provide crucial information about the nature of these stars, and the physical conditions of the emitting gas. Many prominent UV-optical-NIR (\eg [{\rm O\textsc{ii}}]$\lambda 3727,29$\AA, \Hb\ ($4861.33$\AA), Pa$\alpha$ ($1.87510$\um)) lines are used to probe this \cite[see][for a review]{Kewley2019}. 

We generate SEDs of galaxies using the Binary Population And Spectral Synthesis \cite[\textsc{BPASS,}][]{BPASS2.2.1} v2.2.1 Stellar Population Synthesis (SPS) models, assuming a \cite{ChabrierIMF} initial mass function (IMF) with an upper mass limit of 300 M$_{\odot}$. To generate the nebular continuum and line emission we employ version 23.01 of the \cloudy\ \cite[][]{Cloudy17.02,cloudy2023_23.01} photoionisation code.

In this work we follow previous modelling \cite[][see also \citet{Feltre2016,Gutkin2016,Byler2017,Newman2025} for similar approaches]{Wilkins2020_nebular}, and use an average reference ionisation parameter, U$_{\rm ref}=0.01$, which is defined at an age of 1 Myr and metallicity, Z=0.01. We assume a spherical geometry for the \hii\ region, and thus the size of the Str\"omgren sphere scales as the cube-root of the ionising photon luminosity, Q (obtained from the SSP). We then parametrise the relationship between the ionising photon luminosity and the input ionisation parameter to \cloudy\ using,
\begin{equation}\label{eq:Uref}
    U = U_{\rm ref} \bigg(\frac{Q}{Q_{\rm ref}}\bigg)^{1/3}.
\end{equation}
We assume the metallicity of the gas is the same as the illuminating stars, and fix the hydrogen number density of the \hii\ region, n$_{\rm H}=10^{2.5}$ cm$^{-3}$. The abundance ratio of the different elements at a particular metallicity is assumed to scale with the Milky Way stellar abundance from \cite{GalacticConcordance2017}. We also assume that all the lyman-continuum photons are processed within the \hii\ regions \ie the escape fraction is zero. We set the stopping criteria for the \cloudy\ calculation to be when the ionised fraction in the gas drops to $1\%$.

\subsection{Toy galaxies}\label{sec:modelling.toygal}
\begin{figure*}
    \centering
    \includegraphics[width=0.28\textwidth]{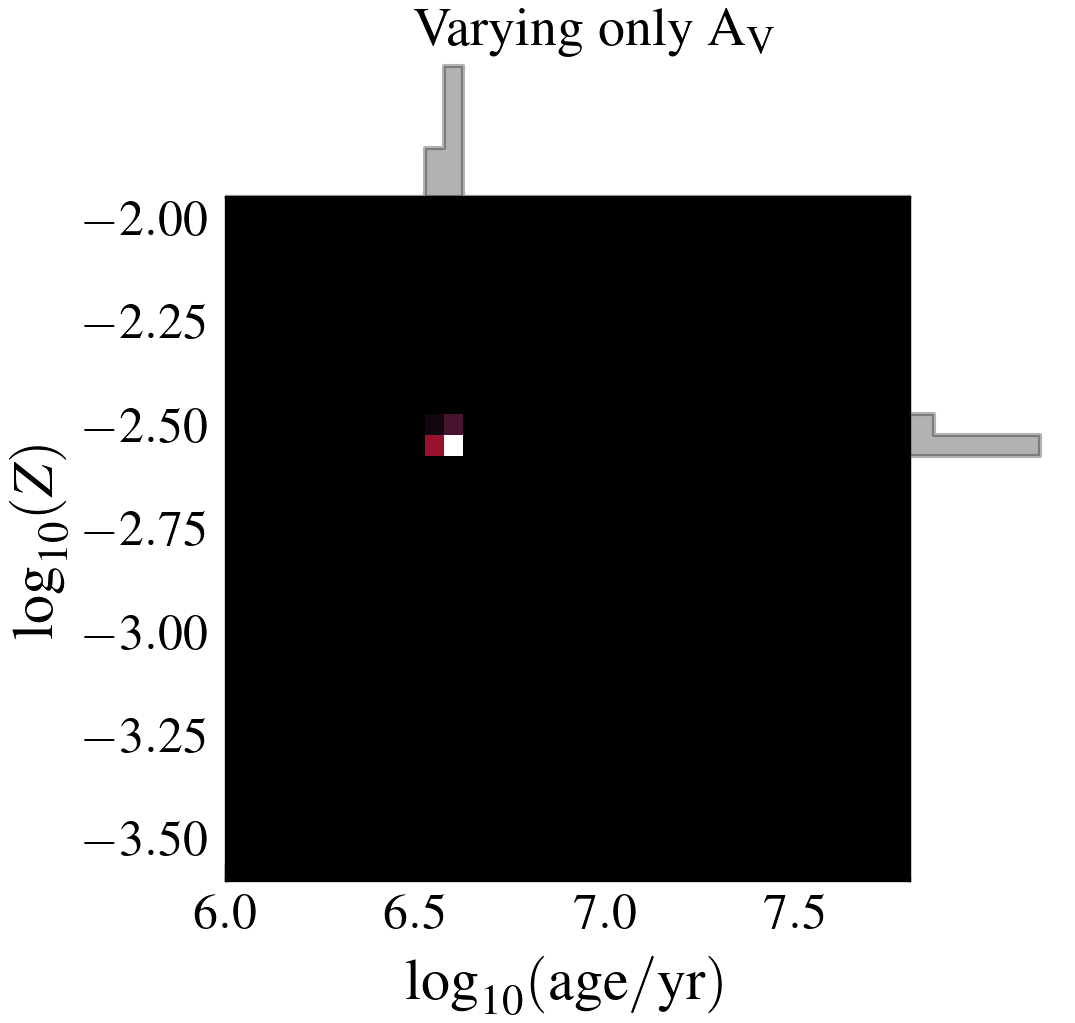}
    \hspace*{-0.5cm}\includegraphics[width=0.26\textwidth]{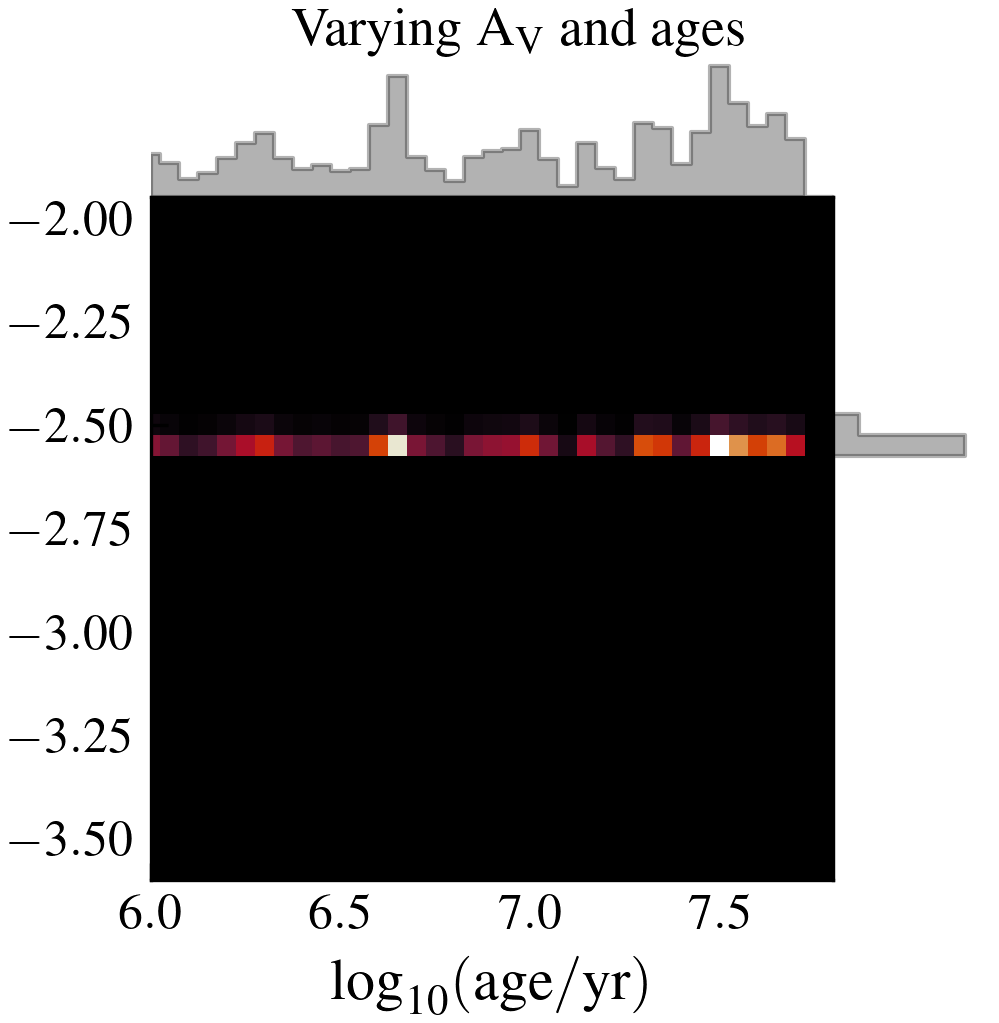}
    \hspace*{-0.5cm}\includegraphics[width=0.26\textwidth]{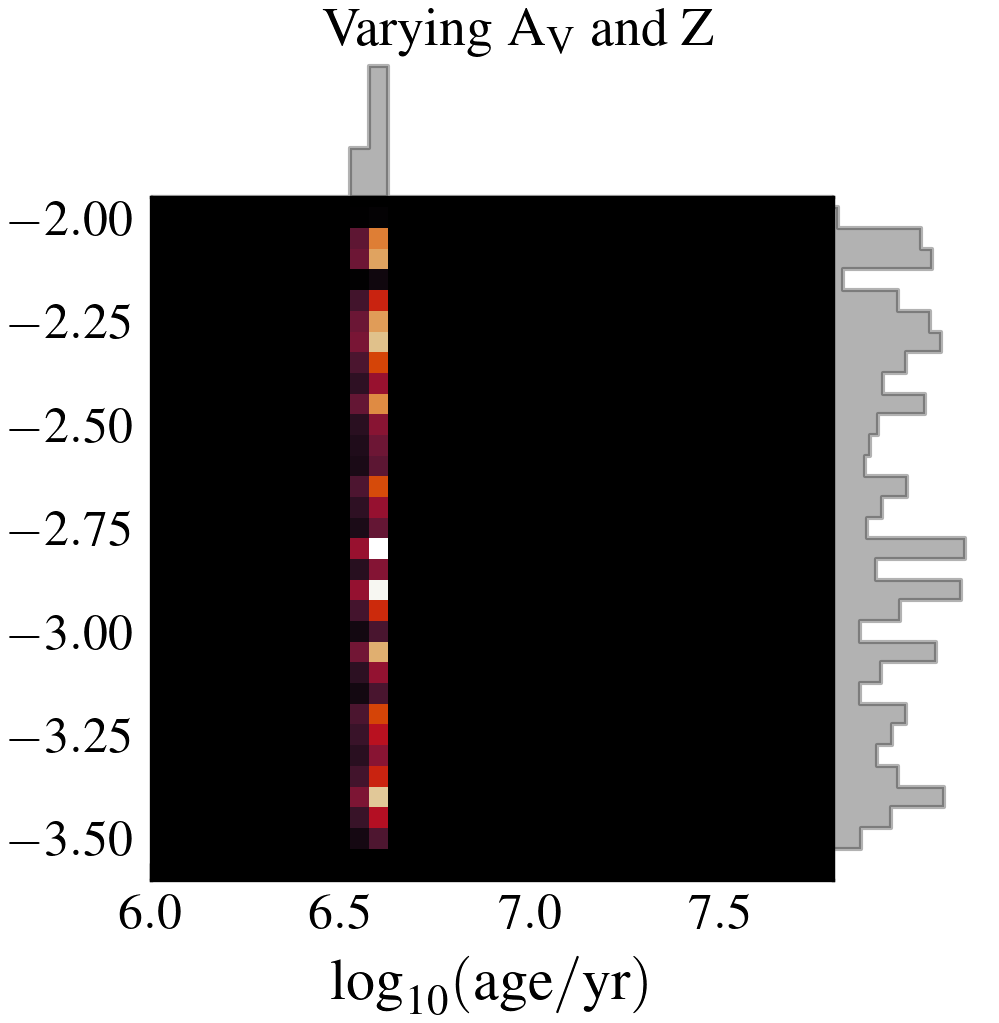}
    \hspace*{-0.5cm}\includegraphics[width=0.26\textwidth]{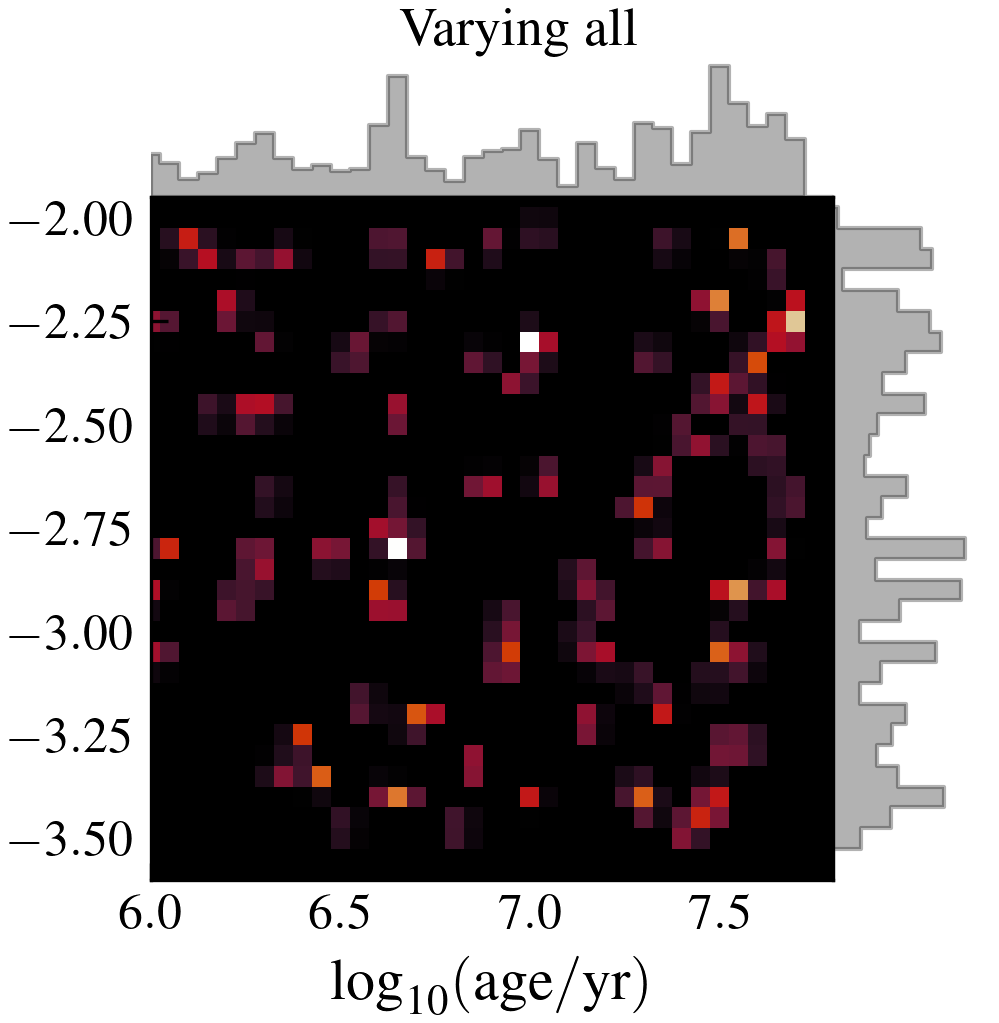}
    \caption{The star formation and metal enrichment history of the 4 toy model galaxies.}
    \label{fig:toy_sfzh} 
\end{figure*}
In order to better understand the impact of galaxies containing multiple stellar populations with different ages and metallicities, 
% having a range of physical properties of the underlying stellar population in a galaxy, 
as well as the effect of star-dust geometry, we model toy galaxies with varying dust attenuation across their multiple stellar populations,
% its stellar population, 
similar to what was presented in \cite{FLARES-XII}. In this work we will use \synth \footnote{\href{https://github.com/synthesizer-project/synthesizer}{https://github.com/synthesizer-project/synthesizer}} \cite[]{synth2_2025,synth1_2025}, an open-source forward modelling package, to explore this variation.

We construct a set of toy galaxies by assuming they are at $z=6$, each composed of 100 star-forming clumps, characterised by their mass, age and metallicity (Z), assuming each clump can be described by a simple stellar population (SSP). The composite SED of the resulting galaxy is obtained by summing the individual SEDs of these star-forming clumps. For this study, we construct four toy galaxies, whose underlying properties are given as follows.

For all these galaxies we pick the individual clump masses randomly from a uniform distribution between $[2-5]\times10^{6}$ M$_{\odot}$. All the four different toy galaxies are assigned the same masses, and for our chosen random number seed, this sums to $\simeq 10^{8.5}$ M$_{\odot}$ for each galaxy. 
To explore the dependence of the observed properties on the underlying variations in the physical properties of the star-forming clumps, we systematically vary the age and metallicity of the clumps across the four different galaxy models:
\begin{enumerate}
    \item Clumps have the same age and metallicity.
    \item Clumps vary in age, but metallicity is fixed.
    \item Clumps vary in metallicity, while age is fixed.
    \item Clumps vary in both age and metallicity.
\end{enumerate}
When varying the ages, we randomly sample from a uniform distribution between $1-50$ Myr (in logspace). For the metallicity, we sample from a uniform distribution in log-space, with $-3.5\le {\rm log}_{10}(Z) \le -2$. 
In the toy galaxies with fixed ages or metallicities, we set these values to match the mass-weighted age or metallicity of young star particles ($\le 10$ Myr) from the varying sample. 
This ensures that all the toy galaxies have comparable average physical properties (stellar mass, age and metallicity). For the chosen random number seed the mass-weighted age is $\sim 4$ Myr and the mass-weighted metallicity, $Z\sim0.003$ \cite[or $12 + {\rm log(O/H)} \sim 8.0$, assuming solar metallicity, $Z_{\odot}=0.014$,][]{Asplund2009}. 
We assume that the stars form in an instantaneous burst with this mass, age and metallicity.
Figure~\ref{fig:toy_sfzh} shows the star formation and metal enrichment history of the 4 model galaxies.

To explore the effect of a varying star-dust geometry, we set different levels of dust attenuation for these star forming clumps within each galaxy. We set the optical depth in the $V$-band for each clump by randomly drawing from a normal distribution with a mean ($\mu_0$) of 0.3 and standard deviation ($\sigma_0$) of 0.4. We quantify the spread in the dust attenuation across the galaxy by varying the standard deviation from $0-6\sigma_0$ as follows:
\begin{equation}\label{eq:tau_vi}
    \tau_{\rm V,i} = \mu_0 + \mathcal{N}(0, n\sigma_0),
\end{equation}
where $\tau_{\rm V,i}$ is the optical depth in the V-band of the i$^{\rm th}$ clump and $n\in[0,6]$. We set any negative $\tau_{\rm V,i}$ to 0.01. We choose such a spread in the dust attenuation to mimic the variation seen in the \flares\ galaxies \cite[for an in-depth discussion, see Section 3.2 of][]{FLARES-XII}. 
In the context of specifying the spread in the dust attenuation, from now on (mainly in the figures), we will quote only `n', where $\sigma= n\sigma_0$. The scenario with $n=0$ can be thought of as being analogous to a uniform screen of dust in front of the stars.

\begin{figure}
    \centering
    \includegraphics[width=\columnwidth]{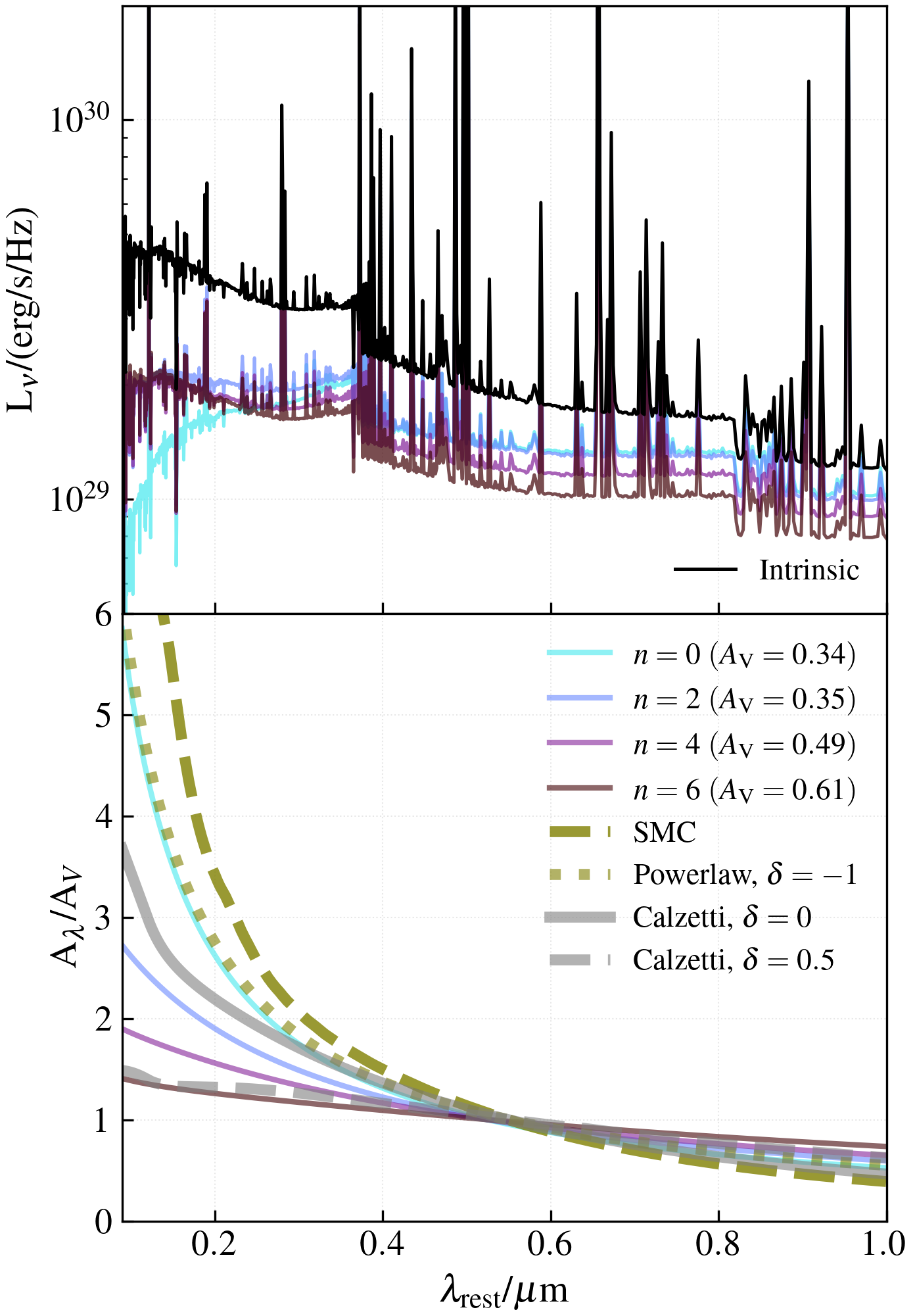}
    \caption{\textbf{Top}: The total intrinsic (black) and the dust-attenuated (for different spread in the dust attenuation, in coloured) SEDs of the `Varying only A$_{\rm V}$' model galaxy, truncated to $0.1-1$ \um.  \textbf{Bottom}: Attenuation curves of the resultant galaxy when varying the spread in the dust attenuation across the different star-forming clumps. We also show the power-law extinction curve (slope $=-1$) that was used for each stellar clumps, as well as the SMC and Calzetti (for slope$=0,0.5$) attenuation curve.}
    \label{fig:toy_att_curves}
\end{figure}

To relate the V-band optical depth to other wavelengths, we adopt an extinction curve that follows a power-law form with a slope of $-1$. This relationship can be expressed as follows:
\begin{equation}
    \tau_{\lambda} = \tau_{\rm V}\, (\lambda\, / \, 5500\, {\rm\AA})^{-1}.
\end{equation}
This extinction curve is flatter in the UV than the SMC extinction curve \cite[]{SMC1992}, yet steeper than the Calzetti attenuation curve \cite[][]{Calzetti2000} (also seen in Figure~\ref{fig:toy_att_curves}).
Given the low-metallicity environments of high-redshift galaxies, extinction curves which are steep in the UV, similar to the SMC extinction curve, serves as a reasonable physical assumption.
With this we generate 4 forward modelled spectra, with varying star-dust geometries, of the 4 model galaxies.
From now on, we will refer to the four model galaxies defined above as `Varying only A$_{\rm V}$', `Varying A$_{\rm V}$ and ages', `Varying A$_{\rm V}$ and Z' and `Varying all', when describing them.

We obtain the `intrinsic' SED of the model galaxy by summing up the SEDs of the individual clumps without applying dust attenuation, while the `observed' SED is the sum of the SEDs accounting for dust. It goes without saying that the intrinsic SED of each model galaxy remains the same when varying the spread in dust attenuation.
Figure~\ref{fig:toy_att_curves} shows the SED (top panel) of the `Varying only A$_{\rm V}$' model galaxy. We can clearly see the increasing effect of dust attenuation on the galaxy, which has reduced the magnitude of the observed SED from the intrinsic (dust-free) model. All the other model galaxies exhibit the same effect, with the intrinsic and observed SED slightly different from the one shown. The bottom panel shows the effect of the variation in the star-dust geometry on the resultant attenuation curve. Higher values of $\sigma$ \cite[also giving higher A$_{\rm V}$, also see][]{Boquien2022} make the resulting attenuation curve flatter than the input curve. This is similar to previous works \cite[\eg][]{Witt2000,Inoue2005}, which have shown that, in a clumpy medium, one can get a Calzetti-like curve from a SMC type dust law or grain composition which is intrinsically significantly steeper. It is also quite clear that the largest variation in the shapes of the different attenuation curves are at the shorter wavelength (in the UV and near-optical).

\subsection{FLARES}
In order to study the impact of varying properties of the underlying stellar population as well as the star-dust geometry for galaxies with realistic star formation and metal enrichment histories, we use the FLARES suite of simulations. 
\flares\ \cite[]{FLARESI,FLARESII} is a suite of zoom hydrodynamical simulations of 40 regions of radius 14 cMpc/h, representing a wide range of densities, taken from a (3.2 cGpc)$^3$ dark matter only simulation box (hereafter, the parent box). The regions were picked so that a large fraction of them (16/40) were biased towards the most overdense regions in the parent box, with other regions sampled from lower overdensities, mean densities and extremely underdense regions. This approach allows us to weight these different regions, so as to produce distribution functions that are representative of the overall parent box. The bias towards extreme overdensities allows us to explore the more massive and brighter galaxies that will be identified in current or future surveys from telescopes such as \jwst, \euclid, or \rst. 

The re-simulations of these regions follow the same physics model as the \eagle\ simulations \cite[see][]{schaye_eagle_2015,crain_eagle_2015}. It differs from the fiducial \eagle\ model with respect to the AGN physics. \flares\ uses the AGNdT9 version, where the AGN feedback is more energetic but less frequent compared to the fiducial implementation, giving better hot gas properties in galaxy clusters. For more details see \cite{FLARESI}. 
The simulations assume a Planck year 1 cosmology \cite[]{planck_collaboration_2014}.

\subsubsection{Galaxy selection}
Galaxies in \flares\ are identified using the \textsc{Subfind} \citep{Springel2001subfind,Dolag2009subfind} algorithm, which runs on bound groups identified by a Friends-of-Friends \cite[\textsc{fof},][]{Davis1985FOF} algorithm. In this work, similar to other \eagle\ and \flares\ works, the quoted stellar mass, star formation rate (SFR) or luminosities are calculated within 30 kpc 3D apertures centred on the most bound particle of the self-bound substructures identified by \textsc{Subfind}. 

We use the mass-weighted metallicity computed using the mass of the stellar particle within the galaxy. We limit our analysis to galaxies with a stellar mass $> 10^8$ M$_{\odot}$ at all redshifts, as this approximately corresponds to 100 stellar particles and can be considered to be resolved in the simulation.

\subsubsection{Galaxy spectral energy distribution}\label{sec:modelling.flares.sed}
The full details of the SED modelling in \flares\ is presented in \cite{FLARESII}. The nebular emission modelling is the same as that described in Section~\ref{sec:modelling.forward}, whereby we treat each stellar particle in the simulation as an SSP. Here, we will describe very briefly how we model dust attenuation in \flares.

% We treat each stellar particle in the simulation as a Simple Stellar Population (SSP) and assign them an SED based on their age and metallicity. We model their emission using the BPASS v2.2.1 SPS model \cite[]{BPASS2.2.1}, with a \cite{ChabrierIMF} IMF. We assign a pure stellar spectrum based on the SPS model. We also associate young stars that produce lyman-continuum photons with \hii\ regions, using the \textsc{cloudy} \cite[v17.02,][]{Cloudy17.02} photoionisation modeling code. This model also assumes that all the lyman-continuum photons are processed within the \hii\ regions \ie the escape fraction is zero.

\flares\ does not model the formation and destruction of dust. Hence we model the dust attenuation by converting the line-of-sight (along the z-axis, \ie the viewing angle here) metal column density for each star particle into a dust column density. This conversion uses the dust-to-metal (DTM) ratio obtained using the fitting function presented in \cite{Vijayan2019}, obtained from the age and metallicity of the galaxy. We use the following equation to calculate the V-band optical depth due to the intervening diffuse dust in the ISM,
\begin{gather}
	\tau_{\textrm{ISM,V}}(x,y) = \mathrm{DTM}\,\kappa_{\textrm{ISM}}\,\Sigma\,(x,y),
\end{gather}
where $\tau_{\textrm{ISM,V}}(x,y)$ and $\Sigma\,(x,y)$ are the V-band optical-depth and integrated metal column density respectively along the line-of-sight at position (x,y) of the star particle.
We also associate extra dust attenuation from stellar birth clouds, which dissipate over a timescale of 10 Myr \cite[]{CF00}. We scale their dust content with the star particle metallicity, described as follows,
\begin{gather}
\tau_{\textrm{BC,V}}(x,y) = 
	\begin{cases}
		\kappa_{\textrm{BC}} (\textrm{Z}_{\star}/0.01)\, & \text{t} \leq 10\, \textrm{Myr}\\
		0\, & \text{t} > 10\, \textrm{Myr}\:,
	\end{cases}
\end{gather}
where $\tau_{\textrm{BC,V}}(x,y)$ is the V-band optical-depth due to the stellar birth-cloud and \textrm{Z}$_{\star}$ is the smoothed metallicity of the young stellar particle. In the above equations,
$\kappa_{\textrm{ISM}}=0.0795$ and $\kappa_{\textrm{BC}}=1.0$, and are obtained by calibrating the $z=5$ UV luminosity function (UVLF) from \cite{Bouwens2015}, $z=5$ UV continuum-slope from \cite{Bouwens2012,Bouwens2014} and the [O\textsc{iii}]$\lambda4959,5007$ + H$\beta$ equivalent width distribution at $z=8$ from \cite{deBarros19_OIIIHbeta}.

We relate the optical depth in the V-band to other wavelengths using a simple power law extinction curve as:
\begin{equation}\label{eq:tau_lambda}
	\tau_{\lambda} = (\tau_{\textrm{ISM}} + \tau_{\textrm{BC}}) \times\,(\lambda/5500\,\textrm{\AA})^{-1}\:.
\end{equation}
We refer the interested readers to Section 2.3, 2.4 and Appendix A of \cite{FLARESII} for an in-depth discussion of the photometry generation in \flares.  This line-of-sight dust attenuation model differs from using a uniform screen of dust across a galaxy (usually assumed in SED fitting), with spatially distinct stellar populations experiencing differing amounts of dust attenuation based on their environment.

The galaxy observables have also been found to successfully match several observables in the high-redshift Universe such as the \Ha\ LF \cite[]{CoveloPaz2025_halpha}, [O\textsc{iii}]$\lambda$5007\AA\ LF \cite[]{FLARES-XI,Meyer2024Oiii}, redshift evolution of the UV continuum slope \cite[]{Tacchella2022,Cullen2024uvslope}, and the JWST filter colour - redshift evolution \cite[]{FLARES-VI}. Hence we can with some confidence make use of the model and extend the predictions to new galaxy observables.

\section{Impact of Star-dust geometry}\label{sec:geometry}
In a previous work \cite[]{FLARES-XII}, we explored how the star-dust geometry and the spread in the underlying properties of the stellar population within galaxies plays a crucial role in the observed galaxy properties.
In that work, we found that there is a large spread in the fraction of the UV luminosity that is obscured across a galaxy in \flares\ at all redshifts (figure 4 in that work). This obscured fraction was shown to be dependent on the SFR of the galaxy, with galaxies having a higher SFR displaying a larger spread in UV obscuration and vice-versa. This led to the attenuation curve being flatter compared to the input extinction curve, exactly what we saw in Section~\ref{sec:modelling.toygal}. This is also one of the main motivations for the extreme spread in the dust optical depths chosen for the toy model galaxies.

We know galaxies are a composite mixture of stellar populations with different ages and metallicities, which determine their ISM properties and intrinsic emission line strengths. The observed (i.e. attenuated) line strength depends on the dust along the line-of-sight to these different stellar populations. The total
line strength is then the sum of these individual values. The resultant line luminosities are the dust weighted sum of the intrinsic luminosities \cite[also see][]{XihanJi2023dust} emanating from distinct regions with different physical properties. 
% Thus, when calculating line ratios, which are widely used for estimating galaxy properties, the numerator and denominator represent the sum of these individual lines, not an average of the individual ratios. 
% The variation in line ratios arises because the sum of the observed ratios does not equal the sum of the intrinsic ratios multiplied by a single effective attenuation. 
This can be more clearly understood by discretising the galaxy into multiple nebular regions, each contributing to the total line emission. In this case, the observed line ratio can be expressed as:
\begin{equation}\label{eq:line_ratio}
    \centering
    \frac{\sum_{k}\mathrm{F}_{i,k}\,\times\,\mathrm{T}_{i,k}}{\sum_{k}\mathrm{F}_{j,k}\,\times\,\mathrm{T}_{j,k}} \neq \frac{\sum_{k}\mathrm{F}_{i,k}}{\sum_{k}\mathrm{F}_{j,k}} \times \frac{\mathrm{T}_{\mathrm{eff},i}}{\mathrm{T}_{\mathrm{eff},j}},
\end{equation}
where F$_{i,k}$ and F$_{j,k}$ are the intrinsic line fluxes of lines `i' and `j' from the k$^{\rm th}$ nebular region, with T$_{i,k}$ and T$_{j,k}$ the attenuation experienced by lines `i' and `j', respectively in that region.
T$_{\mathrm{eff, i}}$ and T$_{\mathrm{eff, j}}$, represent the effective attenuation across the galaxy for lines `i' and `j', respectively.

For pairs of lines that are closer in wavelength, the ratio $\mathrm{T}_{\mathrm{eff},i}/\mathrm{T}_{\mathrm{eff},j}$ can be roughly approximated as 1. In all observational studies, in such a scenario the effect of dust attenuation is neglected.
When the lines are not close in wavelength, one needs to measure the dust attenuation through other methods such as the Balmer decrement or SED fitting. 
This can then be combined with an attenuation law, used to infer the dust-corrected value.
However, significant deviations from this assumption of effective attenuation will occur, when there is a varying amount of dust along the line-of-sight towards stellar populations, each with their own differing intrinsic physical properties (age or metallicity). 
It should be noted that the effect of dust becomes negligible for lines that are in the far-IR and trace \hii\ region properties \cite[\eg][]{Nagao2011,Fernandez2021}.

In \cite{FLARES-XII}, we already covered how the varying star-dust geometry can affect the Balmer decrement (resulting in a higher dust attenuation correction than expected from Calzetti like curves) and the BPT diagram (moving the points away from the intrinsic value), using toy model galaxies (similar to those in Section~\ref{sec:modelling.toygal}) and \flares\ galaxies. 
In this work we will explore some of the wider implications, focusing on how this affects the estimation of galaxy SFRs, ionising photon production efficiencies and metallicities.

In Appendix~\ref{sec:app.vary_mu}, we demonstrate, through increasing the value of $\mu_{0}$ in equation~\ref{eq:tau_vi}, that the main driver of the unreliability in the dust-correction and therefore uncertainty in the inferred quantities is the variation in the star–dust geometry.

\section{Dust Correction}\label{sec:dustcorr}
In this section we will provide a description of how we perform dust correction in two different ways. We will also later show how the star-dust geometry and the assumed attenuation curve used for the correction affect the different lines we are considering.
For the purpose of this discussion we will use the \Ha\ (6562.80\AA) line as the reference for the dust correction. This can be extended to any luminosity/flux of lines or filters.

We will perform the dust correction of the observed H$\alpha$ in two ways:
\begin{enumerate}
    \item Correction using A$_{\rm V}$ (A$_{\rm V}$ method): We measure the A$_{\rm V}$ and apply the Calzetti attenuation curve to compute the attenuation at H$\alpha$. The nebular attenuation is given by
    \begin{equation}\label{eq:dustcorr}
        A_{\rm H\alpha} = E(B-V) k_{\rm H\alpha},
    \end{equation}
    where the colour excess E(B-V) is related to A$_{\rm V}$ through
    \begin{equation}
        E(B-V) = A_{\rm V} / k_{\rm V}.
    \end{equation}
    Here k$_{\rm H\alpha}$ and k$_{\rm V}$ denotes the value of the Calzetti attenuation curve at the wavelength of H$\alpha$ and V-band respectively. This can be extended to any wavelength on the assumption of an attenuation curve. We would like to remind the reader that flatter attenuation curve than the Calzetti curve will give higher E(B-V) for the same A$_{\rm V}$ and vice versa.
    % Many works in the literature assume that the nebular attenuation is equal to the stellar attenuation when applying this method. We follow the same in this work.

    \item Correction using the Balmer decrement (Balmer decrement method): In this case we measure the Balmer decrement of H$\alpha$ and H$\beta$ (4861.33\AA), and use the Calzetti attenuation curve to perform the dust correction. We use the same equation~\ref{eq:dustcorr} as above, but now,
    \begin{equation}
        E(B-V) = \frac{2.5}{k_{\rm H\beta} - k_{\rm H\alpha}} {\rm log}_{10} \bigg[\frac{({\rm H\alpha}/{\rm H\beta})_{\rm obs}}{({\rm H\alpha}/{\rm H\beta})_{\rm int}} \bigg],
    \end{equation}
    where k$_{\rm H\alpha}$ and k$_{\rm H\beta}$ are the the Calzetti attenuation curve values at the wavelength of H$\alpha$ and H$\beta$ respectively. We use the dust-free or intrinsic ratio between H$\alpha$ and H$\beta$, (H$\alpha$/H$\beta$)$_{\rm int}=2.79$ \cite[for electron density $10^{2} \, \mathrm{cm}^{-2}$ and temperature $15,000\,$K, conditions typically expected for high-redshift galaxies,][]{Sanders2023} and (H$\alpha$/H$\beta$)$_{\rm obs}$ is the observed ratio. 
\end{enumerate}
We will also use the two methods to correct for dust and recover the required properties in the following section ($\S$\ref{sec:results}).
The two dust correction methods enable a comparative analysis of how different assumptions about dust attenuation affect the results of analysis that are commonly presented in the literature.

It should be noted that previous studies \cite[e.g.][]{Reddy2023_decrement} have shown that the Balmer decrement can miss star formation that is optically thick in the Balmer lines. This can result in the underestimation of $\sim25\%$ or more \cite[in dusty star-forming galaxies,][]{Chen2020} of the total star formation in galaxies. 

The dust-corrected H$\alpha$ luminosity can be obtained from the observed luminosity using
\begin{equation}\label{eq:Ha_dustcorr}
    {\rm L_{H\alpha}}_{\rm dust\, corr} = {\rm L_{H\alpha}}_{\rm obs} \times 10^{A_{\rm H\alpha}/2.5}.
\end{equation}
This formulation can be applied analogously to other filters or emission-line fluxes and luminosities.

Another factor to note is the relationship of \Ha\ or \Hb\ luminosity with metallicity. L$_{\rm H\alpha}$ and L$_{\rm H\beta}$ varies by $\sim 0.1$ dex with metallicity across a metallicity range of $10^{-5} - 10^{-3}$, with this quickly changing to $\sim 0.4$ dex across $10^{-3} - 10^{-2}$ \cite[]{Wilkins2020_nebular}. 

When implementing these dust corrections, the choice of attenuation curve plays a critical role in determining the magnitude of the correction. 
The Calzetti attenuation curve is flatter than the SMC extinction curve (which resembles a power law with slope $-1$, as assumed here) at the shorter wavelengths. 
Consequently, for a given inferred A$_{\rm V}$ or Balmer decrement, the Calzetti curve produces a smaller dust correction than the SMC curve.
Therefore, if the true dust attenuation curve is steeper, then for the same inferred dust content (from A$_{\rm V}$ or the Balmer decrement), the resulting correction will be smaller, increasingly underestimating the attenuation toward shorter wavelengths.

Another often overlooked subtlety is that, as shown in Figure~\ref{fig:toy_att_curves}, while the curves flatten off at shorter wavelengths due to the spread in the dust optical depths, they simultaneously become progressively steeper at longer wavelengths ($>5500$\AA) compared to the Calzetti or SMC curves.

\section{Results}\label{sec:results}
In this section we will explore the effect of star-dust geometry on inferring the physical properties of galaxies from observables, focusing on line emission. In this section we do not attempt to fit any of the galaxy photometry or spectrum using any SED fitting tools, and derive the A$_{\rm V}$ (V-band attenuation) or stellar masses similar to observational studies.
Instead, we take them directly from the toy model galaxies or from the \flares\ catalogue.
This will provide a more accurate value of these quantities compared to those derived from SED fitting \cite[]{Narayanan2024,Harvey2025}.
\subsection{Star formation rate function and the unobscured fraction}\label{sec:results.sfrd}
\begin{figure*}
    \centering
    \includegraphics[width=0.75\textwidth]{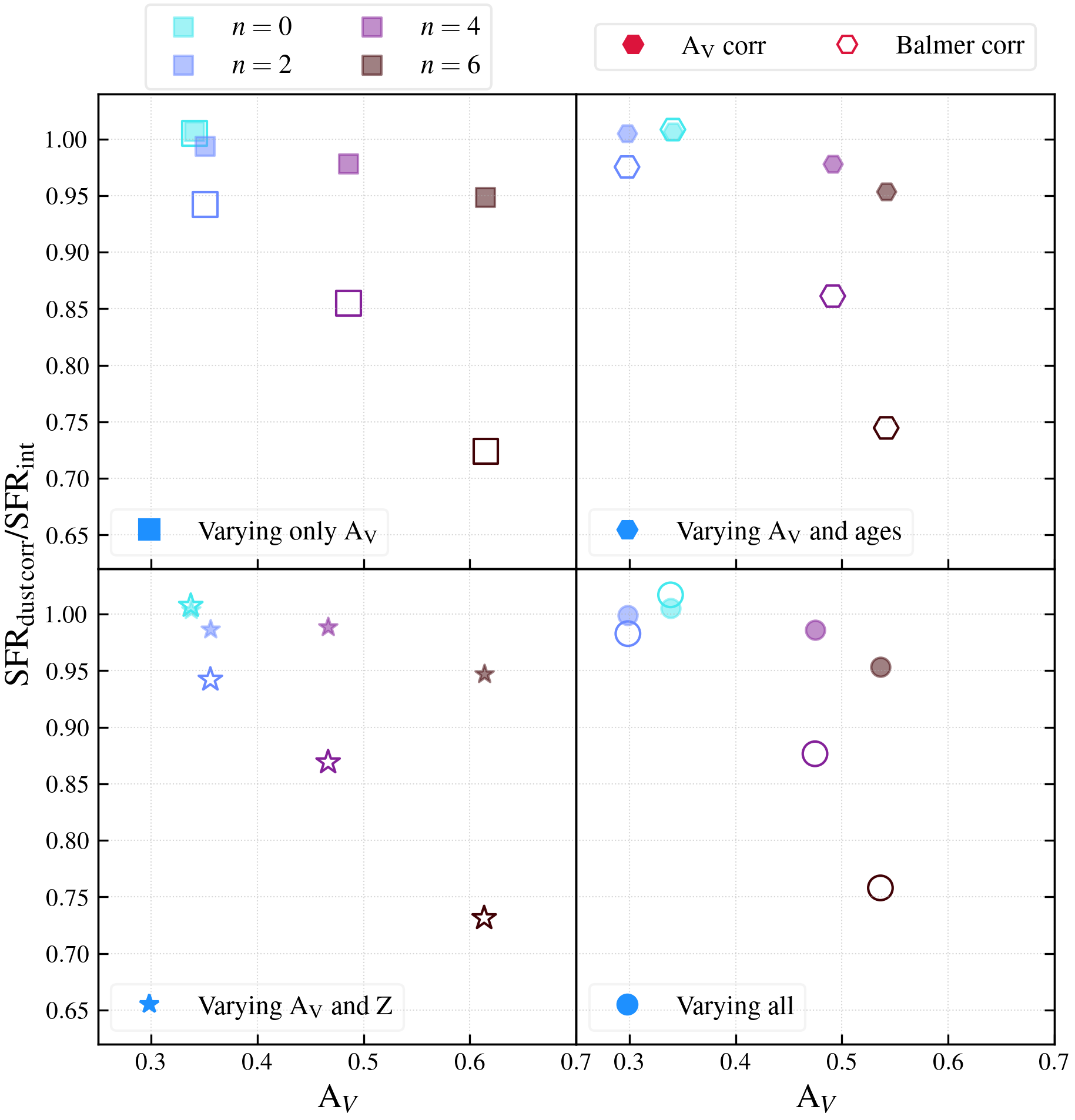}
    \caption{Fraction of the recovered SFR after correcting for dust obscuration for the different toy galaxies in the model. The different markers denote the different models, with the different colours denoting the spread in the obscuration along the line-of-sight to the different stellar clumps within the galaxy. The filled and open markers denote the SFR fraction recovered using the A$_{\rm V}$ method and the Balmer decrement method respectively. It is easier to follow the plot by concentrating on the same colour and shape (along the same A$_{\rm V}$); one can see the pattern that the A$_{\rm V}$ method recovers a higher fraction of the SFR, while the Balmer decrement method recovers a lower fraction.}
    \label{fig:toy_sfr_frac}
\end{figure*}
Star formation rate is a crucial diagnostic of how actively galaxies convert their gas into stars, shedding light on its current evolutionary state. 
The SFR is closely linked to other properties of galaxies such as its stellar mass (for star forming galaxies). At fixed stellar mass, galaxies typically exhibit higher SFR with increasing redshift \cite[]{Speagle2014}, termed the evolution of the main-sequence relation.
The star formation rate distribution function (SFRF) quantifies the number of galaxies per unit volume per unit SFR interval. Integration of the SFRF above a defined lower-limit is used to get the star formation rate density (SFRD). The redshift evolution of the SFRD is also employed in the calibration of some cosmological simulations \cite[\eg][]{Pillepich2018a}. 

In many observational studies utilising optical spectra, the SFR of a galaxy is estimated by converting its observed H${\alpha}$ (6562.80\AA) luminosity, after correcting for dust, into a SFR \cite[]{Kennicut1998,Wilkins2019}.  This conversion usually assumes a star formation history and an IMF. In this section we will adopt the same observational methodology, and explore its impact on the recovered SFRF and the recovered SFRD.

As shown in \S\ref{sec:dustcorr}, the dust-corrected H$\alpha$ luminosity can be obtained from the observed luminosity using equation~\ref{eq:Ha_dustcorr}.
This can be converted to a SFR by using the relation from \cite{Kennicut1998}, converted to a \cite{ChabrierIMF} IMF by multiplying with $1.8$,
\begin{equation}\label{eq:sfr_kennicut}
    {\rm SFR_{H\alpha}} ({\rm M_{\odot}/yr}) = 10^{-41.36}\, {\rm L_{H\alpha}}_{\rm dust\, corr}\, ({\rm erg/s}).
\end{equation}
We also calculate the dust-corrected SFR using the dust-corrected H${\rm \alpha}$ luminosity.

Figure~\ref{fig:toy_sfr_frac} shows the fraction of recovered SFR (SFR$_{\rm dust\, corr}$/SFR$_{\rm int}$) using both methods (the unobscured fraction generally decreases with higher A$_{\rm V}$). When the variation in the dust attenuation across the galaxy is minimal (i.e. $\sigma\sim0$), both methods effectively recover the obscured star formation ($\sim 1$). However, as the spread in the dust obscuration as well as A$_{\rm V}$ increases, both methods do not perform as well. Notably, the Balmer decrement method does worse, recovering as low as $\sim 70\%$ of the total SFR in our toy galaxies. In contrast, the A$_{\rm V}$ method consistently recovers over $\sim 95\%$ of the SFR across all toy galaxies. This is because in our analysis we take the A$_{\rm V}$ directly from the toy galaxies. Observations require SED fitting to recover the  A$_{\rm V}$, which can be unreliable due to degeneracies with other parameters such as stellar ages \cite[]{Calzetti2013seg}. 

\subsubsection{FLARES}
\begin{figure*}
    \centering
    \includegraphics[width=\linewidth]{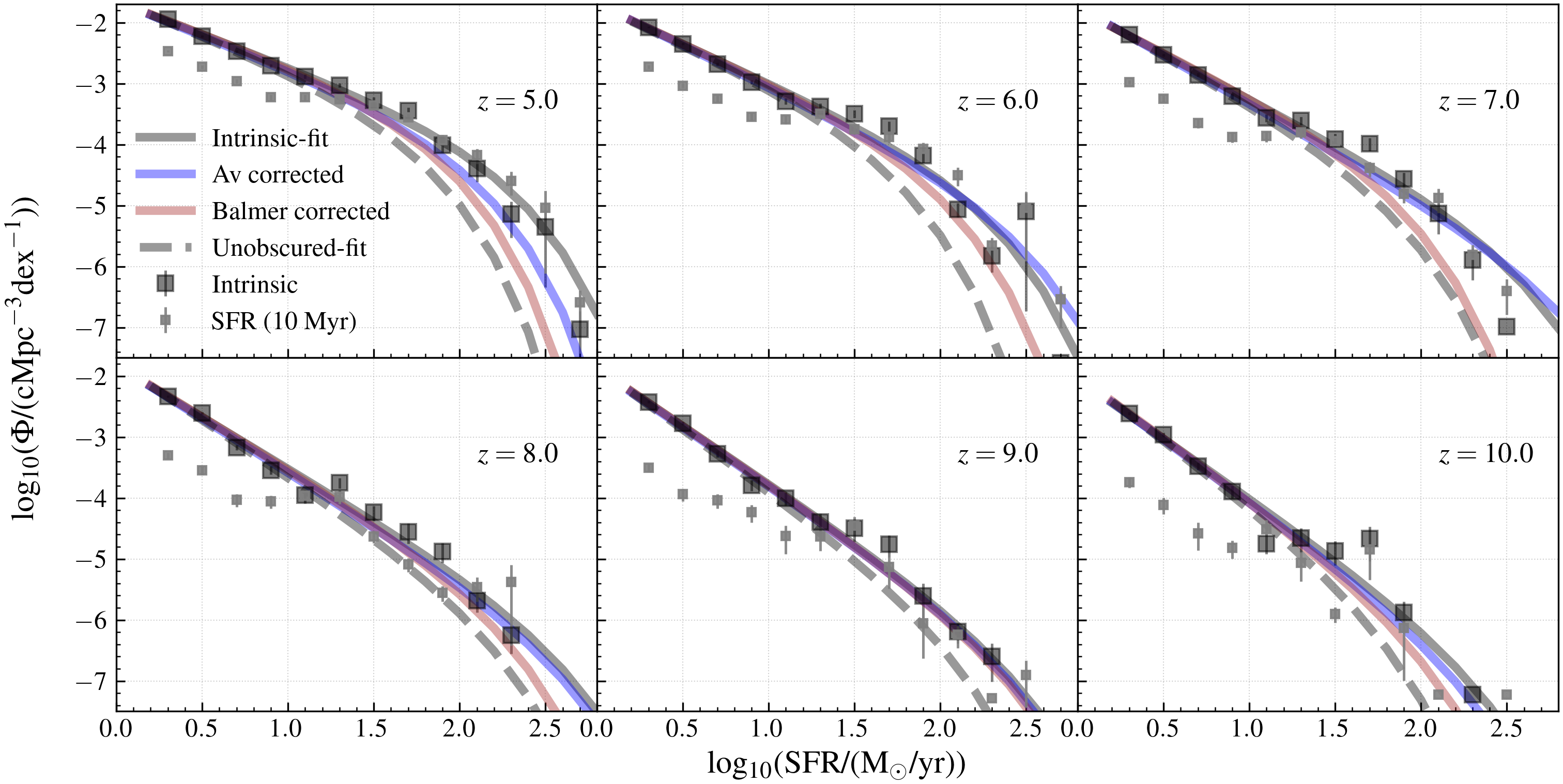}
    \caption{The different panel shows the star formation rate function for the galaxies in \flares\ using different dust corrections for $z\in[5,10]$. We also plot the observed and intrinsic SFRF. The black scatter points with the errorbars are the densities taken directly from the simulation, shown for comparing to the Schechter fit. We also plot the SFR derived by directly summing up the mass of stars formed in the last 10 Myr, denoted as `SFR (10 Myr)'. We do not plot the corresponding Schechter fit for this.}
    \label{fig:flares_sfr_func}
\end{figure*}
\begin{figure*}
    \centering
    \includegraphics[width=0.8\textwidth]{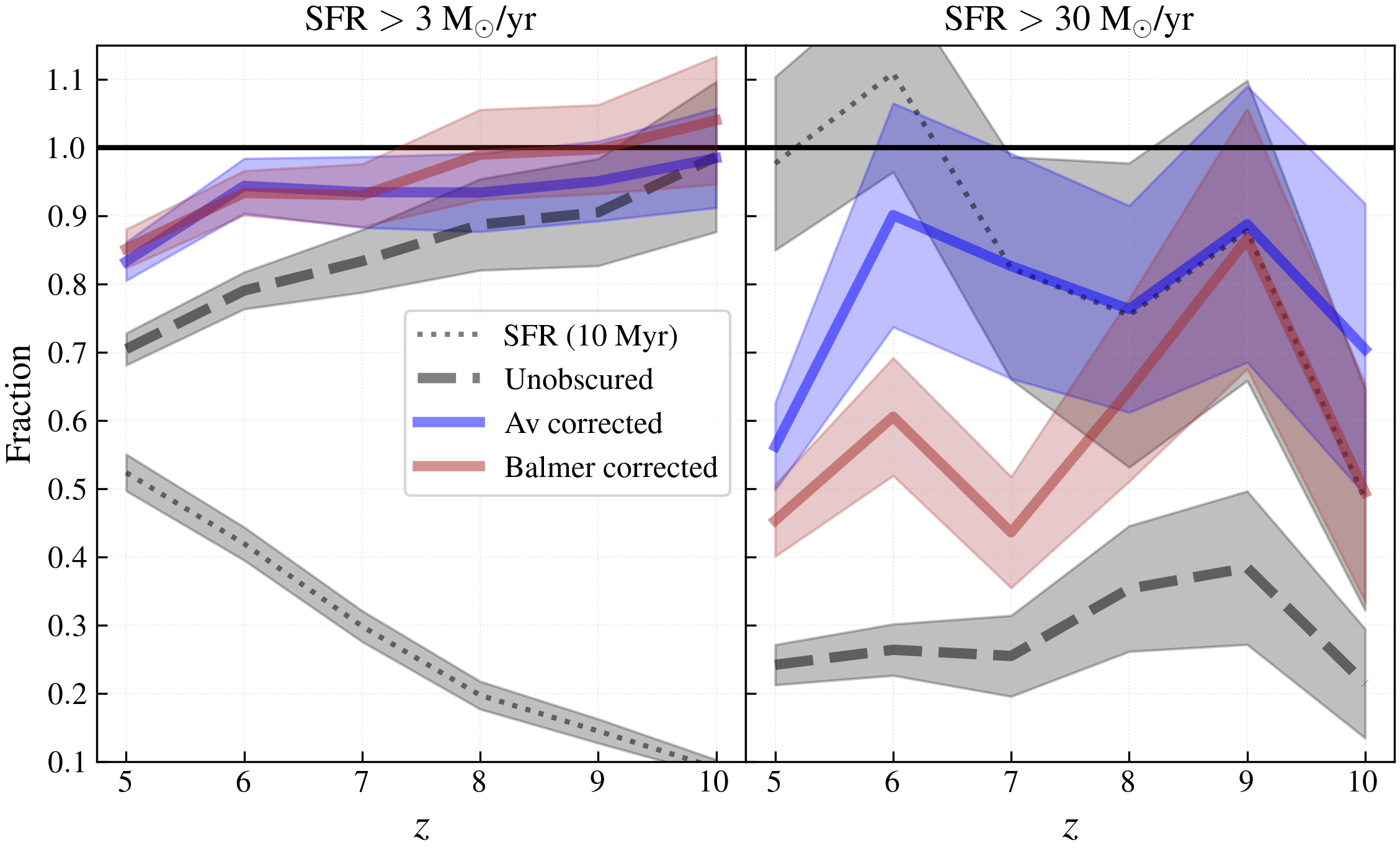}
    \caption{The recovered SFRD fraction for \flares\ galaxies using different dust corrections for $z\in [5,10]$. We also plot the unobscured fraction and SFR (10 Myr) for comparison. 
    The shaded region indicate the $1-\sigma$ scatter associated with the fit obtained by sampling the covariance matrix of the fit.}
    \label{fig:flares_sfrd_frac}
\end{figure*}
We now apply both methods to the \flares\ galaxies. Figure~\ref{fig:flares_sfr_func} shows the SFRF Schechter function \cite[]{Schechter1976} fit of the \flares\ galaxies for $z\in[5,10]$ \cite[also see Figure 9 in][for the \Ha\ LF under different Calzetti attenuation curve slope for \flares]{FLARES-XII}. We provide our method for fitting the Schechter function and the corresponding fit parameters in Appendix~\ref{sec:app.schecter}. 
We denote the SFRF fit obtained from the intrinsic (dust-free), unobscured (no dust correction), Balmer decrement corrected and A$_{\rm V}$ corrected as `Intrinsic-fit', `Unobscured-fit', `Balmer corrected' and `A$_{\rm V}$ corrected' respectively. We also show the binned SFRF from the intrinsic \Ha\ (`Intrinsic') as well as that obtained directly from the simulation by summing up the mass of stars formed in the last 10 Myr (`SFR (10 Myr)').

It should be noted that the binned data from \flares\ at $z=10$ exhibit significant scatter around the knee, meaning that the fit is less robust in this case. Also there is a sharp drop-off at the high-SFR end for the unobscured and Balmer corrected SFRF, such that the Schechter fit is steeper at the low-SFR end, sometimes higher than the fit to the intrinsic function ($z=6,7,8,10$). Thus it is also important to take account of such biases (steepening of the Schechter function), when using them to derived the SFRD evolution.

At low SFRs ($\le 10$ M$_{\odot}$/yr), all methods yield nearly identical results, indicating negligible dust attenuation, seen by the intrinsic and unobscured SFRF coinciding. 
At higher SFRs, where star-dust geometry and dust-attenuation play a more significant role, the methods start to diverge, similar to what we saw before.
The Balmer decrement method does worse than the A$_{\rm V}$ method. The difference between the two methods generally becomes smaller with increasing redshift, reflecting the reduction in the amount of dust obscuration. 

If we compare the `Intrinsic' and the `SFR (10 Myr)' value, we see that the latter shows a dip at the low SFR end. This is due to the galaxies at the low-SFR end are low-mass and low-metallicity galaxies, where the ionisation production rate is much higher than that expected from the calibration we use (equation~\ref{eq:sfr_kennicut}). 
A simple stellar population with metallicities of $10^{-4}$ ($12+{\rm log}_{10}({\rm O/H})\sim6.5$) and $0.02$ ($12+{\rm log}_{10}({\rm O/H})\sim8.8$) can differ approximately by an order of magnitude in the ionising photon production rate for stellar ages below 10 Myr \cite[see Figure A1 in][]{Wilkins2020_nebular}. Due to this dip in the SFRF at the low-end, the SFRD estimated would be significantly lower compared to that obtained from \Ha.

To compute the SFRD, we integrate the SFRF obtained using the two methods. Instead of quoting the absolute SFRD at each redshift from both the methods, we show the recovered fraction of the SFRD in Figure~\ref{fig:flares_sfrd_frac}. The figure also shows the effect of two different lower integration limits of $3$ and $30$ M$_{\odot}$/yr. 
For the lower limit, the recovered fraction is similar between the two methods across redshifts and approaches unity with increasing redshift. 
This is because there are more galaxies at low-SFR (where the effect of dust is negligible) at all redshifts compared to the high-SFR galaxies, and the low-SFR galaxies dominate the number density.
However, for the higher integration limit, the recovered fraction decreases, particularly for the Balmer decrement method, recovering $\lesssim 70\%$ of the SFRD at all redshifts (`A$_{\rm V}$ method' does better since we take A$_{\rm V}$ directly from the SED). This illustrates that we could be underestimating the SFRD for the extremely star forming galaxies in the early Universe, when using these methods.

Another interesting thing, but also expected from our previous discussion, is that the SFRD recovered using `SFR (10 Myr)' is always less than that obtained directly from \Ha. At fixed SFR, the metallicity of the \flares\ galaxies decreases with increasing  redshift.
This causes the discrepancy to increase with increasing redshift as seen clearly in Figure~\ref{fig:flares_sfrd_frac}.
\subsection{Ionising photon production efficiencies}\label{sec:results.ippe}
\begin{figure}
    \centering
    \includegraphics[width=\linewidth]{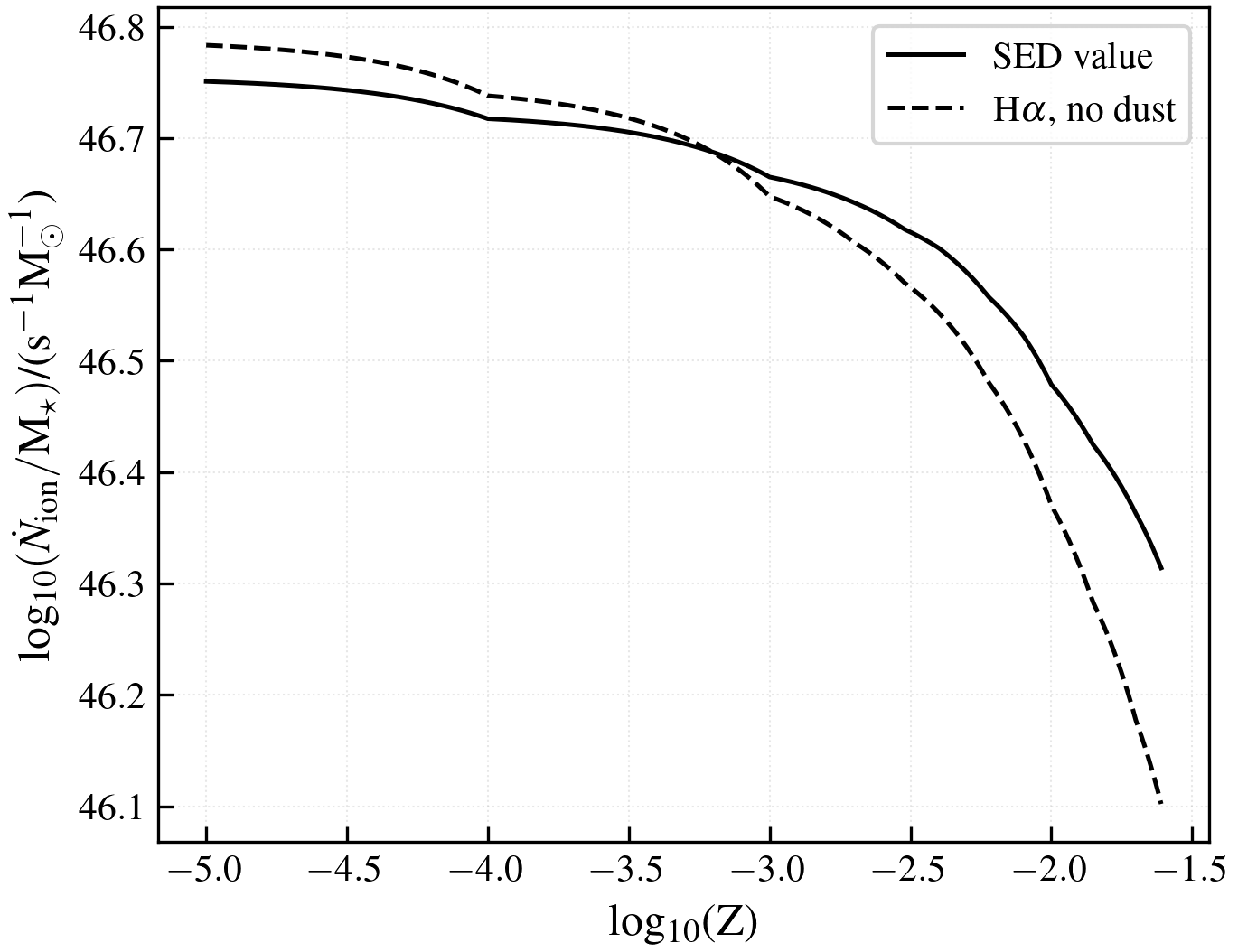}
    \caption{The ionising photon production rate normalised by the formed stellar mass ($\dot{N}_{\rm ion}/$M$_{\star}$) for a constant star formation of 10 Myr for different metallicities.
    The solid and the dashed line denote the value obtained directly from the SED and the \Ha\ luminosity, respectively.
    }
    \label{fig:nion_CSP}
\end{figure}
\begin{figure*}
    \centering
    \includegraphics[width=0.75\textwidth]{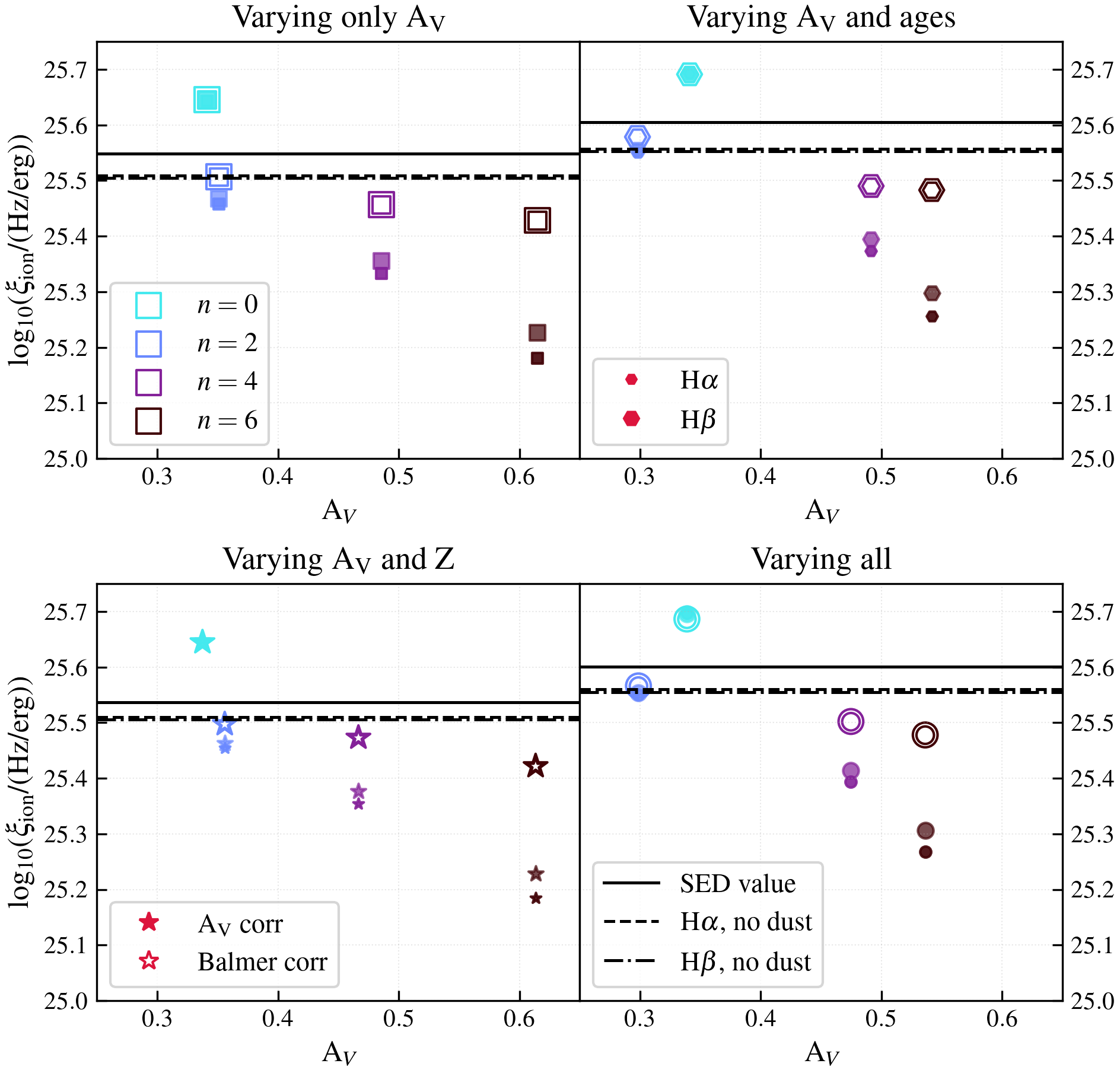}
    \caption{The different panels show the ionising photon production efficiency ($\xi_{\rm ion}$) for the toy galaxies and value recovered under different dust corrections. The different markers denote the 4 different toy model galaxies, with the colours denoting the spread in the dust attenuation across the star clusters within these galaxies. We denote the `SED value', `\Ha, no dust' and `\Hb, no dust' with solid, dashed and dashdot black lines respectively. The filled and open markers denote the dust correction performed using the A$_{\rm V}$ method and the Balmer decrement method respectively. The bigger markersize denote $\xi_{\rm ion}$ recovered using \Hb, while the smaller markersize denote the value recovered using \Ha. In the case of Balmer decrement derived $\xi_{\rm ion}$, the \Ha\ and \Hb\ derived values lie on top of each other.}
    \label{fig:toy_ion_ppe}
\end{figure*}
\begin{figure*}
    \centering
    \includegraphics[width=0.95\textwidth]{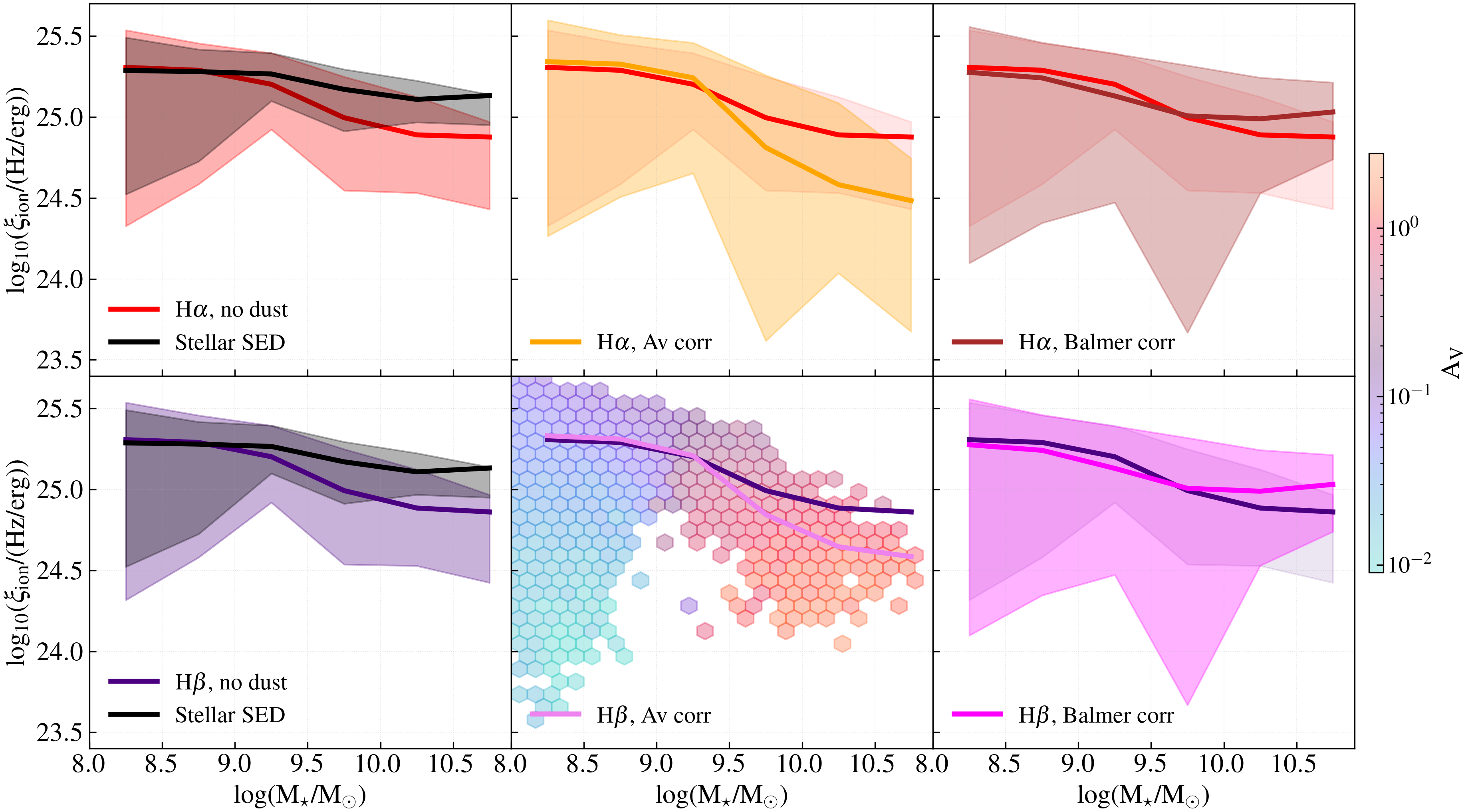}
    \caption{The different panels show the median ionising photon production efficiency calculated using different methods for the galaxies in \flares\ at $z=6$. The solid lines are the median values, with the shaded region showing the 16-84 percentile spread.
    In the panel showing the ionising photon production efficiency calculated from \Hb\ luminosity dust corrected using the `A$_{\rm V}$ method', we plot underlying data in hexbins, coloured by the median A$_{\rm V}$ of the bin.}
    \label{fig:flares_ion_ppe_z6}
\end{figure*}
\begin{figure*}
    \centering
    \includegraphics[width=0.95\textwidth]{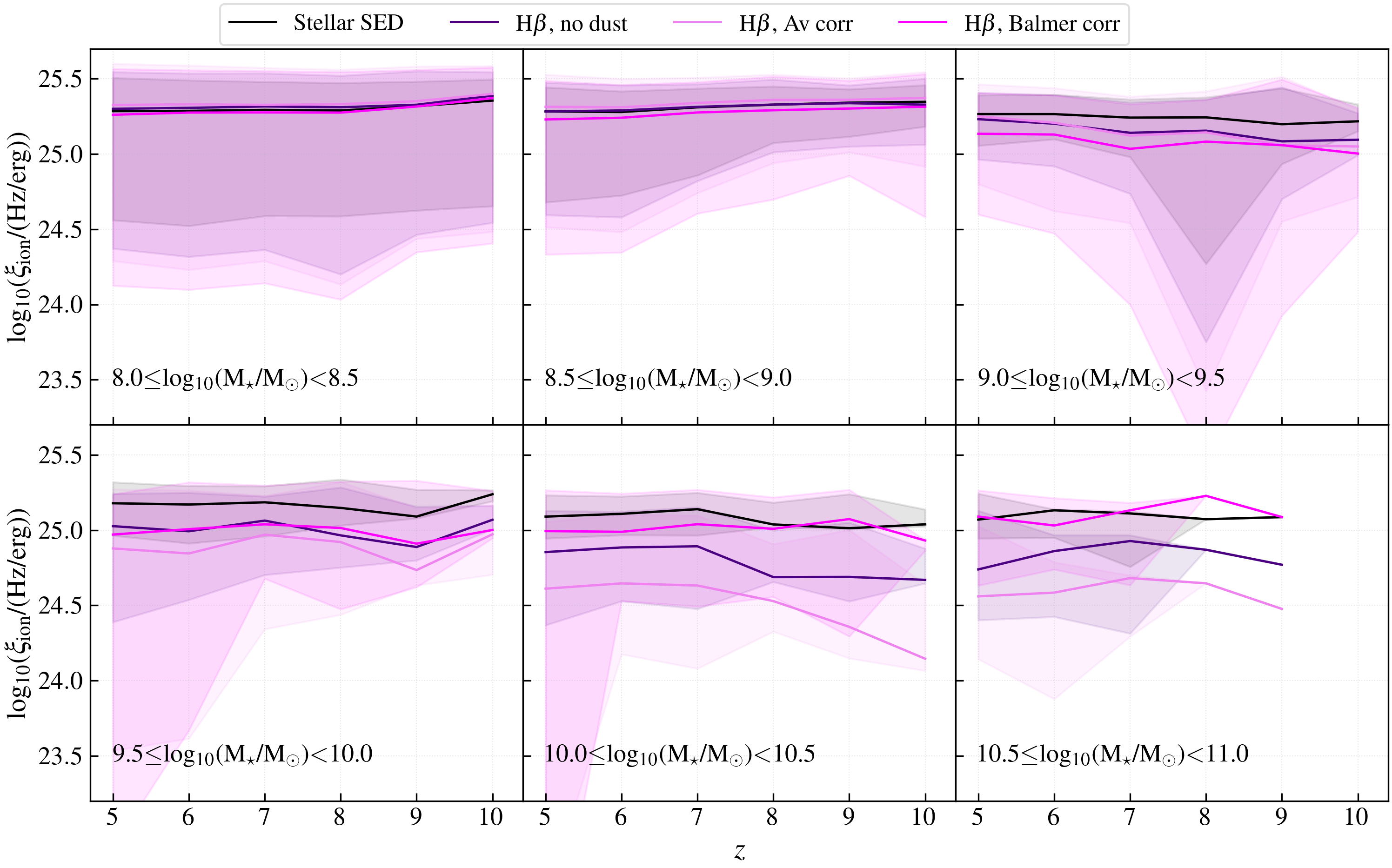}
    \caption{The different panels show the redshift evolution of the ionising photon production efficiency for the \flares\ galaxies for different stellar mass bins. We only plot in this figure the value obtained using \Hb\ luminosity, since the \Ha\ based value gives very similar results, and \Hb\ is accessible across all redshifts solely through NIRSpec on \jwst. The solid lines are the median values, with the shaded region showing the 16-84 percentile spread.  There are no galaxies with stellar mass $>10^{10.5}$ M$_{\odot}$ in \flares\ at $z=10$.}
    \label{fig:flares_ion_ppe_z_evo}
\end{figure*}
The ionising photon production efficiency (\ippe, in Hz/erg) is an important parameter that is used to probe how efficiently galaxies, across different stellar masses or UV luminosities, contributed to the reionisation of the Universe. This quantity is commonly expressed as
\begin{equation}
    \xi_{\rm ion} = \frac{\dot{N}_{\rm ion,\, intr}}{{\rm L}_{\rm FUV}},
\end{equation}
where $\dot{N}_{\rm ion,\, intr}$ is the intrinsic ionising photon production rate (s$^{-1}$) and L$_{\rm FUV}$ is the UV luminosity (erg/s/Hz) measured at 1500\AA. 

In theoretical models or simulations, $\xi_{\rm ion}$ can be calculated directly by integrating the stellar SED (\ie spectra free from nebular reprocessing) below the Lyman limit (912\AA) and dividing by the integrated UV luminosity of the galaxy (hereafter denoted as the `SED value'). This was already explored for \flares\ in \cite{FLARES_XIII}, including  how it evolves with redshift and correlates with the physical and observed properties of the host galaxies for $z\in[5,10]$. It can also be derived by using the \Ha\ or \Hb\ line luminosity of the galaxy, as it is typically done for observed galaxies. This is usually derived under the assumption that no ionising photons escape (\ie they are fully reprocessed in the \hii\ region) and case-B recombination applies. Under these assumptions, there is a one-to-one relationship between the ionising photon production rate and the \Ha\ or the \Hb\ luminosity via:
\begin{equation}
    {\rm L}_{\rm H\alpha}\, ({\rm erg/s}) = 1.36 \times 10^{-12} \dot{N}_{\rm ion,\, intr}\, ({\rm s^{-1}}),
\end{equation}
and 
\begin{equation}
    {\rm L}_{\rm H\beta}\, ({\rm erg/s}) = 4.87 \times 10^{-13} \dot{N}_{\rm ion,\, intr}\, ({\rm s^{-1}}),
\end{equation}
derived in \cite{Leitherer1995}. This can then be used to derive $\dot{N}_{\rm ion,\, intr}$:
\begin{equation}\label{eq:xion_ha}
    \xi_{\rm ion} = \frac{{\rm L}_{\rm H\alpha}\,}{1.36 \times 10^{-12} \times {\rm L}_{\rm FUV}}.
\end{equation}
or
\begin{equation}\label{eq:xion_hb}
    \xi_{\rm ion} = \frac{{\rm L}_{\rm H\beta}\,}{4.87 \times 10^{-13} \times {\rm L}_{\rm FUV}}.
\end{equation}

As in the previous subsection, we can perform a similar analysis to correct the different observables for dust; in this case the observed L$_{\rm FUV}$, L$_{\rm H\alpha}$ and L$_{\rm H\beta}$.

We will briefly describe here the effect of age, metallicity and dust on \Ha\ (or \Hb) luminosity and L$_{\rm FUV}$, so that it will be easier to follow the discussion.
In Figure~\ref{fig:nion_CSP} we show the variation of the intrinsic ionising photon production rate obtained using the SED and \Ha\ luminosity (equation~\ref{eq:xion_ha}) respectively as a function of the metallicity. This is the value obtained a composite stellar population with constant star formation rate for 10 Myr normalised by the stellar mass formed.
From the figure it is quite clear that at low metallicities there is little variation in the relationship between \Ha\ luminosities with the number of ionising photons, while a higher variation happens at higher metallicities ($\sim 0.6$ dex). 
This is the same for FUV luminosities, with a lower amount of change at higher metallicities of $\sim 0.2$ dex.
This implies that for the same range $\xi_{\rm ion}$ can vary by $\sim 0.4$ dex.
This is assuming all the Lyman continuum photons are reprocessed by the gas.
However, the change in the ionising photon production rate calculated from the SED over the same range is smaller, $\sim 0.4$ dex. 
Thus one can expect a difference of about $0.2$ dex between the SED value and the \Ha\ value.
We refer the reader to \S~2 of \cite{FLARES_XIII} for an in-depth discussion about the different effects of star formation history and metallicities on the ionising production efficiency and the ionising photon production rate.

In Figure~\ref{fig:toy_ion_ppe} we plot the ionising photon production efficiency for the toy model galaxies. We denote the values derived from the intrinsic (dust-free) \Ha\ and \Hb\ as `\Ha, no dust' and `\Hb, no dust', respectively, alongside the value obtained directly by integrating the galaxy SED, denoted as `SED value'. They are significantly closer to one another ($\lesssim 0.05$ dex), with the `SED value' lying above the other two coinciding with what is shown in Figure~\ref{fig:nion_CSP} for $\sim0.05$ dex difference around Z $\sim10^{-3}$.
We also show values corrected for dust attenuation using the `A$_{\rm V}$ method' and `Balmer decrement method', employing different marker styles, sizes and colours.
Similar to the previous subsection, when there is negligible dust attenuation, the estimates coincide well with the `stellar SED' and `no dust' estimates.
As expected for the case of little scatter in the optical depths, the derived $\xi_{\rm ion}$ values are higher than expected from all the `intrinsic' values, due to the FUV being under-corrected for dust, because the actual attenuation curve is steeper than Calzetti.

As A$_{\rm V}$ increases, $\xi_{\rm ion}$ can be lower by $\sim 0.5$ dex. 
Notably, in this instance, the Balmer decrement corrected values are significantly closer to the intrinsic value than those obtained using the `A$_{\rm V}$ method', opposite to the trend that was observed previously in Section~\ref{sec:results.sfrd}.
This difference arises due to the fact that L$_{\rm FUV}$ correction is minor when using the Balmer decrement, while the `A$_{\rm V}$ method' over corrects the value. 
This is because the dust correction inferred from the Balmer decrement is lower when using a steeper dust attenuation curve, when the `true' curve is flatter \cite[see Figure 8 in][]{FLARES-XII}. 
Since A$_{\rm V}$ is directly taken from the SED, the UV is over-corrected when using a steeper curve compared to the `true' attenuation curve.
This causes the overall ratio to be higher when using the Balmer decrement compared to the `A$_{\rm V}$ method'. The `A$_{\rm V}$ method' also shows a stronger response to dust, with the values diverging for higher values of A$_{\rm V}$.

\subsubsection{\flares}
Figure~\ref{fig:flares_ion_ppe_z6} shows the ionising photon production efficiency ($\xi_{\rm ion}$) as a function of stellar mass for  \flares\ galaxies at $z=6$.
The top and bottom panel show $\xi_{\rm ion}$ estimated from \Ha\ and \Hb\ luminosities, respectively.
The three columns compare the intrinsic values (no dust attenuation) with values obtained directly from the SED, and with those derived after dust correction using the A$_{\rm V}$ method and Balmer method.
As the $\xi_{\rm ion}$ values inferred from \Ha\ and \Hb\ are nearly identical, we show the distribution of values in hexbins coloured by their median A$_{\rm V}$ only for the \Hb\ luminosity panel dust corrected using the A$_{\rm V}$  method.

The same general trends seen in our toy model galaxies are also apparent here. $\xi_{\rm ion}$ estimated using the dust-free \Ha\ and \Hb\ luminosity closely match the value computed directly from the stellar SED. 
At the high mass end (M$_{\star}>10^9\Msun$), stellar SED value is higher than the \Ha\ (or \Hb) value by $\sim 0.2$ dex as expected from deviation due to higher metallicity at these stellar masses.
Similar to the results of \cite{FLARES_XIII}, we see a slight decrease in the ionising photon production efficiency with increasing stellar mass, due to older stellar ages and increasing metallicity.
Also in general, with increasing stellar mass, the dust attenuation increases, and the dust corrected $\xi_{\rm ion}$ values lie systematically below the  intrinsic values.
Furthermore, the Balmer-corrected values are in general higher compared to that from the A$_{\rm V}$ corrected ones. It is quite intriguing that there is a similar dip in the $\xi_{\rm ion}$ at high stellar masses ($>10^9$ M$_{\odot}$) to that seen in the observed results of \cite{Llerena2024} (their Figure 4), while the \flares\ stellar SED values are flatter compared to their results. This implies that the underlying variation in the star-dust geometry could be playing a role, thus mimicking certain observed trends.

To better understand the trends, we examine the second column of the second row, where the data are also shown as hexbin plots coloured by the median A$_{\rm V}$ in each bin, providing additional insight into the distribution of galaxies around the median. 
At low stellar masses ($<10^9$ M$_{\odot}$), the scatter towards very low $\xi_{\rm ion}$ values arises from quiescent galaxies with minimal dust attenuation.
In this regime, galaxies above the median have moderate dust attenuation with little spread, implying a dust attenuation curve steeper than Calzetti at short wavelengths, which under predicts the dust correction for the FUV luminosity.
With increasing stellar mass, there is a clear increase in A$_{\rm V}$. 
As shown in \cite{FLARES-XII}, galaxies exhibit a dichotomy between systems with little spread and those with extreme spread in dust obscuration (Figure~4).
At high stellar masses, galaxies above the median of the SED relation are lower-A$_{\rm V}$ systems with less spread in dust attenuation, again leading to the underprediction of the FUV.
In contrast, ones below the median have over-corrected FUV luminosities and  under-corrected \Hb\ luminosities (due the resultant curves being steeper at longer wavelengths), resulting in lower $\xi_{\rm ion}$ values.

Figure~\ref{fig:flares_ion_ppe_z_evo} shows the redshift evolution of the $\xi_{\rm ion}$ of the \flares\ galaxies for different stellar mass bins. We do not plot the \Ha\ luminosity derived values, since they show only negligible differences from those obtained from \Hb, and since \Hb\ is accessibly throughout the redshift range through NIRSpec on \jwst\ (\Ha\ falls out of the wavelength coverage for $z\ge7$). In general, there is very little evolution in the median ionising photon production efficiency values for the different stellar mass ranges across redshifts. This is not true only for the $10^{10} - 10^{10.5}$M$_{\odot}$ stellar mass bin in the case of the A$_{\rm V}$ corrected values. There is a strong preference towards lower $\xi_{\rm ion}$ values at $z>8$.
%why is that? Is it due to the FUV being over-corrected?
Similar to Figure~\ref{fig:flares_ion_ppe_z6}, we see that all the different methods coincide at the lowest stellar mass bin, where the effect of dust is the least. 
The values start to diverge for higher stellar masses, where the effect of dust becomes prominent. At the highest stellar masses ($\ge 10^{10}$ M$_{\odot}$), the `A$_{\rm V}$ method' can give values which can be different up to a dex.

\subsection{Mass-metallicity relation}\label{sec:results.mzr}
Metallicity is a critical diagnostic in understanding the evolutionary history of galaxies. It encodes key information about a galaxy's star formation history, gas outflows, inflows of pristine gas, mergers and feedback processes \cite[]{Kewley2019,MM2019review}. The chemical composition of the ISM, particularly the relative abundance of different metals, provides insight into the type of stars and the initial mass function within galaxies and their evolutionary stage. 

In observations, gas-phase metallicity is commonly quantified using the oxygen-to-hydrogen abundance ratio (by number), specifically `$12+{\rm log}_{10}{\rm (O/H)}$'.
Numerous recent studies \cite[\eg][]{Heintz2023,Nakajima2023_metallicity,Curti2024_metallicity,Sanders2024_calibration} have examined the evolution of the mass-metallicity relation (MZR) with redshift. 
Both observations and cosmological simulations suggest that, at fixed stellar mass, the metallicity  increases with decreasing redshift \cite[\eg][]{Torrey2019,Yates2021a,Curti2024_metallicity,Nakajima2023_metallicity}. This trend is intuitive, since chemical enrichment is inherently time-dependent, requiring successive generation of stars to enrich the ISM. Notably, these studies suggest only mild evolution in the mass-metallicity relation above $z\ge3$.

Gas-phase metallicity of a galaxy is usually inferred from emission or absorption lines in their spectrum \cite[\eg][]{Schady2024}. 
In this study, we will focus on emission lines originating from the \hii\ regions of galaxies. 
Another thing to note about metallicity in cosmological simulations is uncertainty associated with the metal yields from stellar populations. Even for a fixed IMF, yields of different metals can be uncertain by $0.3-0.5$ dex \cite[][]{Wiersma2009,Kobayashi2020}. Hence, its important to understand that comparisons across observations and simulations should be made with the relative shape of the mass-metallicity relation, rather than the absolute value.

There are two main methods for estimating metallicity. 
The first one is called the `direct method', which relies
on detecting auroral lines such as [O\textsc{iii}]$\lambda 4363$ to measure the electron temperature. 
This is usually the preferred method, however, these lines are usually weak and hence often not detected in individual spectra. 
Thus `strong line' ratios (such as R23 $= ([{\rm O\textsc{iii}}]\lambda 4959,5007 + [{\rm O\textsc{ii}}]\lambda 3727,29)/{\rm H\beta}$) calibrated to `direct method' based metallicities or photoionisation models are widely adopted to derive metallicities. 
%Additionally, theoretical methods that use photoionisation models (\eg through \textsc{cloudy} or \textsc{Mappings}) use fitting functions of multiple line ratios to derive the metallicities based on the model spectrum.

In this section, we evaluate the robustness of some empirically-calibrated strong-line methods.
The aim of this section is not to introduce a new empirical calibration for estimating galaxy metallicities \cite[\eg][]{Hirschmann2023,Cornejo2025}. Rather, our focus will be to understand the impact of varying the properties of the underlying stellar population and the star-dust geometry on the resulting metallicity estimates.
As such, we adopt the strong-line calibrations presented in \cite{Laseter2024} and \cite{Curti2024_metallicity} \cite[see Table B.1 in][]{Curti2024_metallicity}, which are based on direct method metallicities calibrated to a sample of low-redshift galaxies, adapted to better probe the low-metallicity end. Similar to \cite{Curti2024_metallicity}, we use a combination of the strong-line metallicity calibrations of R2 (log$_{10}([{\rm O\textsc{ii}}]\lambda3727,29 / {\rm H\beta}$)), R3 (log$_{10}([{\rm O\textsc{iii}}]\lambda 5007 / {\rm H\beta}$)), O32 (log$_{10}([{\rm O\textsc{iii}}]\lambda5007 / [{\rm O\textsc{ii}}]\lambda3727,29$)), Ne3O2 (log$_{10}([{\rm Ne\textsc{iii}}]\lambda3869 / [{\rm O\textsc{ii}}]\lambda 3727,29$)) and $\hat{R}$ ($0.47\times {\rm R2} + 0.88 \times {\rm R3}$). To obtain a `best' metallicity estimate, we perform a grid search on the metallicity range of $12+{\rm log}_{10}({\rm O/H})=7.0-9.69$ to minimise the log-likelihood, defined as
\begin{equation}
    {\rm log}(\mathcal{L}) \propto \sum_{i}\frac{R_{{\rm sim},i} - R_{{\rm cal},i}}{\sigma^2_{{\rm sim},i} + \sigma^2_{{\rm cal}, i}},
\end{equation}
where the sum is performed over the above defined diagnostics. Here $R_{\rm sim}$ are the line ratios of the simulated galaxies and $R_{\rm cal}$ are the expected line ratios for the metallicity across the search grid. 
$\sigma_{\rm cal}$ is the dispersion associated with individual strong-line calibrations and $\sigma_{\rm sim}$ is the uncertainty on the simulated line ratios, which we fix to 0.01, which is also the spacing on the grid search.
This method using a combination of multiple strong-line calibrations attempts to avoid the potential biases that could be introduced by using single calibrations. 

It is important to note that direct comparisons of absolute metallicity values across different set of calibrations are generally discouraged, as each calibration is normalised by the particular sample it uses. Therefore, we employ the same diagnostics \cite[R2, R3, O32, Ne3O2 calibration set from][]{Curti2024_metallicity} consistently across both the toy galaxy models and the \flares\ galaxies to ensure reliable relative comparisons.

\begin{figure*}
    \centering
    \includegraphics[width=0.75\textwidth]{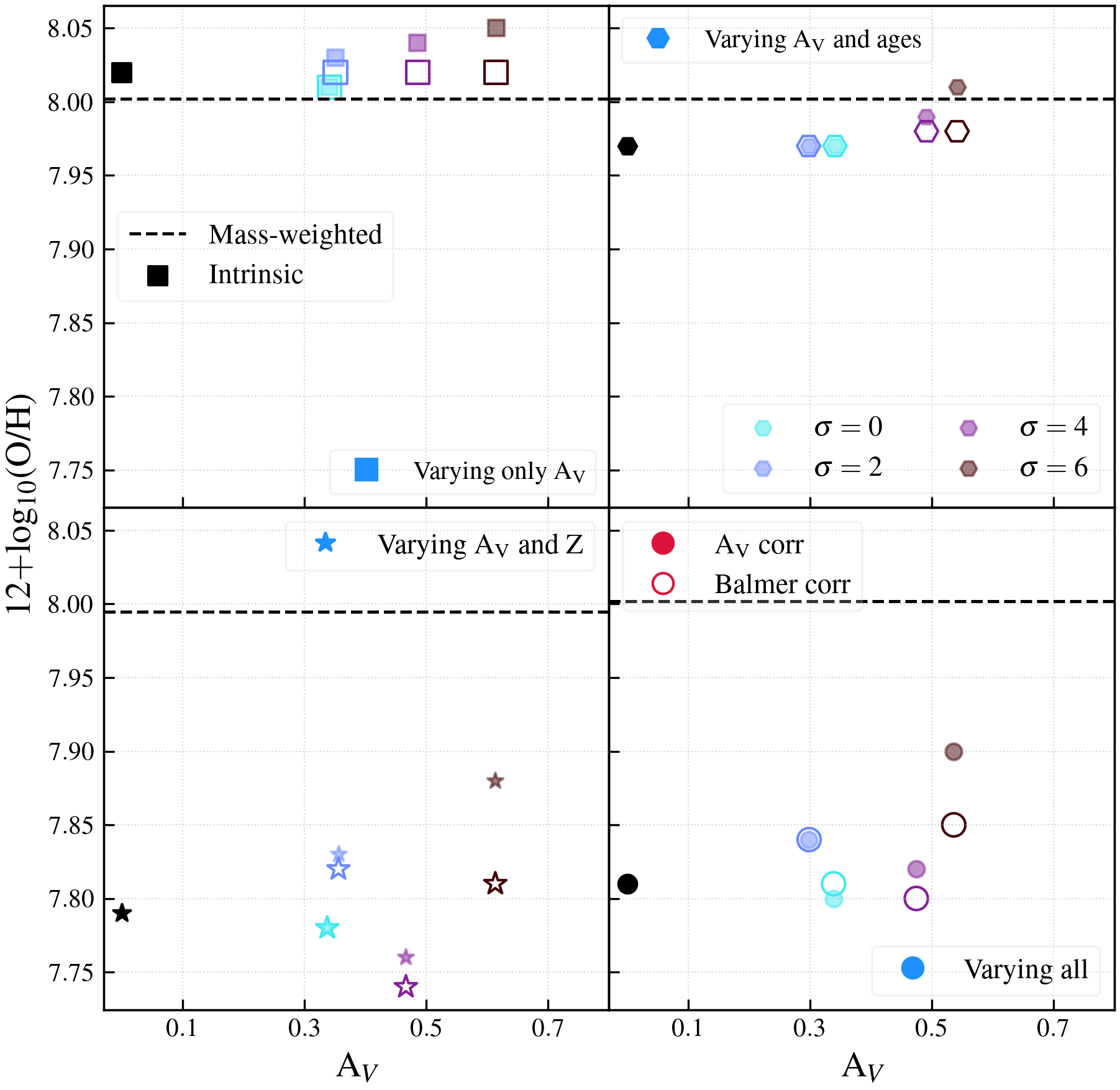}
    \caption{Recovered metallicity of the toy galaxy sample plotted against the attenuation in the V-band.  The different subplots show the 4 model galaxies: `Varying only A$_{\rm V}$', `Varying A$_{\rm V}$ and ages', `Varying A$_{\rm V}$ and Z' and `Varying all'. 
    The markers in black denote the metallicity derived from the intrinsic (dust-free) line luminosity ratios.
    The filled markers and open marker denote the metallicity recovered after performing dust correction using the `A$_{\rm V}$ method' and `Balmer decrement method' respectively on the individual lines used.
    We also plot the mass-weighted metallicity of the galaxy as the black dashed line.}
    \label{fig:toy_met_Av}
\end{figure*}

Figure~\ref{fig:toy_met_Av} shows the metallicity estimates of the toy models as a function of A$_{\rm V}$, obtained using the combination of R2, R3, O32, Ne3O2 and $\hat{R}$ as described before. We correct the constituent lines in the ratio using the Balmer decrement (open markers) and A$_{\rm V}$ method (filled markers). 
We also plot the metallicities obtained using the intrinsic (dust-free) line ratios (black symbols at A$_{\rm V}=0$). The intrinsic value could be considered as the `true metallicity' for a particular calibration, as that is the value obtained without dust. The mass-weighted metallicity of the star-forming region is $12+{\rm log}_{10}({\rm O/H})\sim8.0$, which is denoted by the dashed horizontal line.

The toy models consistently yield metallicities that are within 0.15 dex of the intrinsic value. Similar to the previous sections, when the dust attenuation is negligible, the estimates are closer to the intrinsic value. Here, the Balmer decrement values show less scatter (within 0.1 dex) from the intrinsic value, compared to the A$_{\rm V}$ method ($>0.1$ dex). The scatter increases when there is variation in both metallicity and increased spread in the star-dust geometry (`Varying A$_{\rm V}$ and Z' and `Varying all' models).
However, this scatter is within the uncertainty quoted for the metallicity derived from strong-line calibration against the `Direct-T$_{\rm e}$' method in \cite{Laseter2024}, and thus can be taken to be within the error margins of the different calibrations.

The mass-weighted metallicity of all the toy models, except the `Varying only A$_{\rm V}$' (whose inferred metallicities show negligible difference from the mass-weighted value), is higher compared to those obtained from the strong-line calibrations (light-weighted metallicity). 
The difference between the intrinsic value and the mass-weighted value is more pronounced in the case of the `Varying A$_{\rm V}$ and Z' and `Varying all' toy models. 
It is important to note that all the toy galaxies have similar
mass-weighted ages and metallicities, reinforcing the claim that the observed scatter is mainly driven by variation in
the underlying stellar population and star-dust geometry.

In Appendix~\ref{sec:app.mzr}, we will also explore how the metallicity estimates differ when using the strong-line calibrations from \cite{Sanders2024_calibration}.

\subsubsection{\flares}
\begin{figure}
    \centering
    \includegraphics[width=\columnwidth]{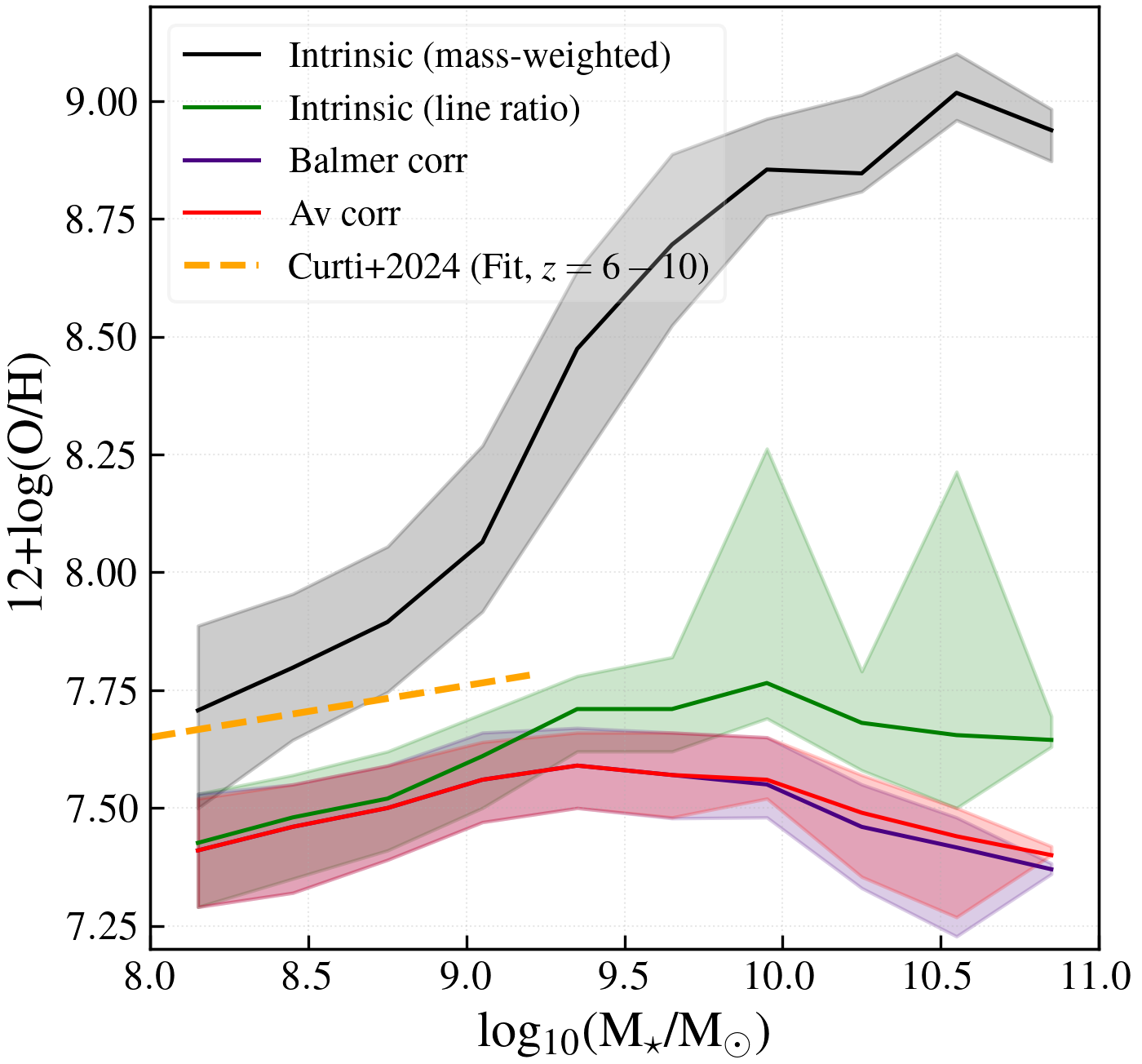}
    \caption{The mass-metallicity relation for the \flares\ galaxies at $z=6$. Observed and intrinsic are the value computed using the observed and intrinsic (dust-free) line luminosity ratios. In the black is plotted the \flares\ metallicity calculated directly from the simulation. We also plot alongside the fit to the  mass-metallicity from \cite{Curti2024_metallicity} for comparison, since we use the same method to derive the metallicities. 
    }
    \label{fig:flares_mzr_z6}
    % \vspace{0.5cm}
\end{figure}
\begin{figure*}
    \centering
    \includegraphics[width=0.95\textwidth]{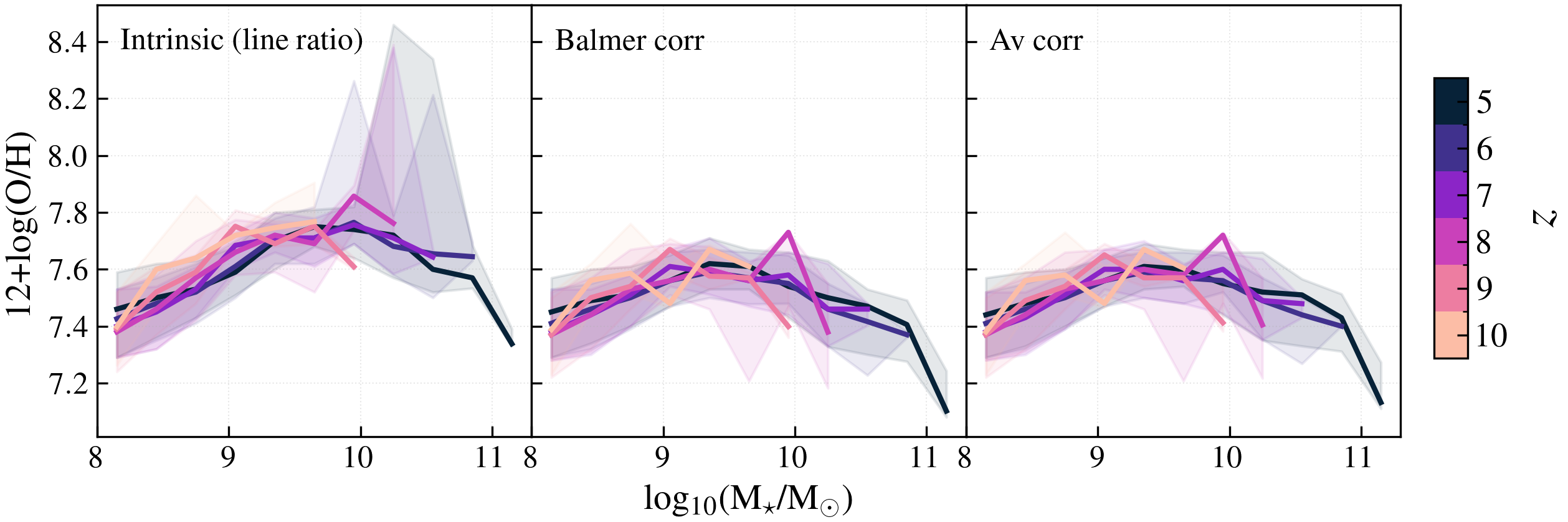}
    \caption{The redshift evolution of the mass-metallicity relation for \flares\ galaxies in the redshift range, $z\in[5,10]$. We show the mass-metallicity relation obtained using the strong-line calibrations in the different panel, with the Intrinsic (line ratio), Balmer corrected and A$_{\rm V}$ corrected values plotted in the left, middle and right panel respectively.
    }
    \label{fig:flares_mzr_z_evo}
    % \vspace{0.5cm}
\end{figure*}
\flares\ galaxies have been shown to reproduce the observed \Ha\ (and hence \Hb) and [O\textsc{iii}]5007\AA\ luminosity functions \cite[]{CoveloPaz2025_halpha,Meyer2024Oiii}, indicating that the simulation broadly reproduces the correct star formation and metal enrichment history, and the dust obscuration of the observed high-redshift galaxies. We explore some of the differences in the forward modelled observables of \flares\ galaxies in Appendix~\ref{sec:app.stronglines}, and the derived metallicities using strong-line calibrations from \cite{Sanders2024_calibration} in Appendix~\ref{sec:app.mzr}.

We now apply the empirical calibration to the \flares\ galaxies at $z=6$, and present the mass-metallicity relation in Figure~\ref{fig:flares_mzr_z6}. 
We also apply a [O\textsc{iii}]5007\AA\ flux cut of $3\times 10^{-18}$ erg/s/cm$^{-2}$, corresponding to typical flux limits of high-redshift observational surveys.
We show
the `Intrinsic (mass-weighted)' value that is obtained from the simulation by using the mass-weighted metallicity (converting the metal mass fraction of young stars ($\le 10$ Myr) to a $12+{\rm log}_{10}{\rm (O/H)}$ value assuming solar metallicity of 0.014) and the `Intrinsic (line ratio)' value obtained from applying the strong-line calibration to the intrinsic line ratio (dust-free). 
We plot the values obtained after dust correcting the constituent lines using the Balmer decrement (`Balmer corr') and the A$_{\rm V}$ (`A$_{\rm V}$ corr'). We show the fit to the mass-metallicity relation from \cite{Curti2024_metallicity}, which are mainly based on galaxies with stellar mass less than $10^{9.5}$ M$_{\odot}$.

It can be seen that the metallicity measurements are exactly similar for the `Balmer corr' and `A$_{\rm V}$ corr' across the full stellar mass range in \flares, implying that the strong-line ratios occupy a narrow range. 
We will discuss this in Appendix~\ref{sec:app.stronglines}, showing the distribution of the strong line measures in \flares\ in comparison to the observational data from \jwst. 
In contrast, the mass-weighted metallicity returns a relation with a positive slope throughout the full stellar mass range.

The intrinsic line ratio values deviate from the dust-corrected values at the high-mass end (M$_{\star}>10^9$ M$_{\odot}$). 
At the highest masses (M$_{\star} \ge 10^{10}$ M$_{\odot}$), all three of these relations show a downward trend. This is possibly due to the strong-line calibrations breaking down at the high-mass, high-metallicity end. 

In order to compare the shape of the mass-metallicity relation obtained in \cite{Curti2024_metallicity}, we fit the \flares\ data points with the same fitting function, with the form
\begin{equation}
    12+{\rm log}_{10}({\rm O/H}) = \beta\, {\rm log}_{10}({\rm M}_{\star}/10^{8} {\rm M}_{\odot}) + {\rm Z}_{\rm m8},
\end{equation}
where $\beta$ is the slope and Z$_{\rm m8}$ is the normalisation at ${\rm M}_{\star}/ {\rm M}_{\odot}=10^{8}$. The \cite{Curti2024_metallicity} sample for $6 \le z \le 10$ showed a slope of $0.11\pm0.05$. We fit the relation to the \flares\ galaxies with stellar mass $<10^{9.5}$ M$_{\odot}$ to match the range probed in \cite{Curti2024_metallicity}. We obtain slopes of $0.17\pm0.03$, $0.16\pm0.03$, $0.22\pm0.03$ and $0.48\pm0.03$ for the Balmer corrected, A$_{\rm V}$ corrected, Intrinsic (line ratio) and Intrinsic (mass-weighted) metallicities respectively. The slope of the relation matches better with the metallicities obtained after dust correcting the strong lines, instead of the intrinsic values, from the lines as well as the mass-weighted value. 

In Figure~\ref{fig:flares_mzr_z_evo} we show the evolution of the mass-metallicity relation for the \flares\ galaxies in $z\in [5,10]$ for the metallicities obtained using the strong-line calibrations. Across the different methods, we see that there is negligible evolution in the mass-metallicity relation in \flares\ for these redshifts ($\sim 600$ Myr separates $z=5$ from $z=10$).

These results highlight the sensitivity of metallicity estimates to both dust attenuation and the properties of the underlying stellar population. 
A key driver of the discrepancy between the mass-weighted metallicity and those inferred from strong-line calibrations is that the latter trace the ionising photon number weighted metallicity, which is further modulated by dust.
As stated in Section~\ref{sec:results.sfrd}, the ionising photon production rate can differ by an order of magnitude for young stars at the low ($10^{-4}$) and high end (0.02) of the metallicity spectrum.
This discrepancy is especially pronounced in FLARES galaxies, which exhibit strong radial gradients in both dust and metallicity \cite[]{FLARES-XII}. 
In contrast, simulations with flat gradients in dust or metals will produce minimal effects.

\section{Caveats}\label{sec:caveats}
We highlight below some important caveats that should be considered when interpreting our results and understanding the limitations of this work.
\begin{enumerate}
    \item The toy galaxy models explored in this study are based on a single random realisation and specific choices of star formation and metal enrichment history (age, metallicity ranges). 
    Alternative parametrisations will lead to different trends.
    However, the main takeaway from this exercise is not the precise spread in the derived quantities or the observables, but rather the general insight that the line ratio calibrations can become unreliable in the presence of varying star-dust geometries and a spread in the properties (age, metallicity) of the underlying stellar populations.
    
    \item Our models assume all the ionising photons are processed within the H\textsc{ii} regions, \ie the escape fraction, f$_{\rm esc} = 0$. 
    This is a common assumption in many observational studies, where f$_{\rm esc}$ is often set to zero. 
    However, in reality, high-redshift galaxies can have non-zero escape fraction. 
    A non-zero escape fraction implies that some of the ionising photons leave the nebula without contributing to any nebular continuum or line emission, potentially leading to an underestimation of the intrinsic ionising photon production rate.
    Some SED fitting codes \cite[\eg \texttt{prospector},][]{Prospector2021} allow for f$_{\rm esc}$ to be treated as a free parameter. 
    And recent machine learning approaches \cite[\eg][]{Choustikov2025_ILI} have demonstrated the ability to jointly constrain f$_{\rm esc}$ and the ionising photon production efficiency. 
    
    \item We assume that H\textsc{ii} regions are ionisation bounded rather than density bounded. 
    In a density bounded nebulae, the hydrogen recombination lines such as \Ha\ or \Hb\ deviate from the expected case-B recombination values. 
    This results in Balmer decrements that do not match the theoretical expectations, leading to  overestimation of dust attenuation.
    While this assumption is prevalent in observational analysis, distinguishing between ionisation and density bounded nebulae is extremely challenging. 
    One diagnostic is the detection of observed Balmer ratios below the theoretical case-B limit in dust-free systems, which is rare. However, since the motivation is to perform dust-correction, this is quite counter-intuitive.
    
    \item The calibration adopted in \cite{Curti2024_metallicity} is based on the direct method, which uses the auroral [O\textsc{iii}]$4363$ line.
    This approach becomes increasingly unreliable at high metallicities due to factors such as temperature fluctuations within the H\textsc{ii} regions, elevated O$^{+}$/O$^{++}$ abundance ratios, etc \cite[\eg][]{Stasinska1978,Yates2020,Cameron2023}.
    These effects can introduce systemic biases in the metallicity estimates derived from the direct method, particularly at the high-metallicity end.
\end{enumerate}
Despite these caveats, the assumptions adopted in this study are consistent with standard practices in the field and serve as a useful baseline for interpreting observed trends.

\section{Conclusions}\label{sec:conclusions}
In this work we investigated how the variations in star-dust geometry, coupled with diversity in the physical properties of the underlying stellar population, impact the reliable recovery of physical properties in galaxies. 
We demonstrated these effects using a combination of simplified toy models and forward modelled galaxies from the \flares\ suite of simulations.
We focused on three key physical quantities typically inferred from emission line diagnostics: the star formation rate (SFR), ionising photon production efficiency ($\xi_{\rm ion}$), and the gas-phase metallicity.
These quantities were derived using empirical/theoretical calibrations typically employed in literature. 
We also incorporate dust corrections using the Balmer decrement and the attenuation in the V-band (A$_{\rm V}$) under the assumption of a Calzetti attenuation curve.
Our key findings are as follows:
\begin{itemize}
    \item We estimated the SFR of the simulated galaxies using \Ha\ line strength, correcting for dust using the Balmer decrement and A$_{\rm V}$.
    We find that the combined variation in the stellar population properties and spread in the star-dust geometry lead to dust corrections systematically underpredicting the intrinsic SFR. Specifically, the Balmer decrement can underestimate the SFR by up to $\sim 50\%$, especially at the high-SFR end ($> 30$ M$_{\odot}$/yr) for the \flares\ galaxies. A similar suppression is also seen in the recovered cosmic SFRD. The \Ha\ derived SFRD is substantially higher than that inferred directly from the simulation by summing the mass of stars formed in the last 10 Myr (underestimated by $\le 50 \%$ compared to the intrinsic SFR from \Ha).

    \item We also derive $\xi_{\rm ion}$ from \Ha\ and \Hb\ using the empirical relation, correcting for dust in a similar manner. The dust-corrected $\xi_{\rm ion}$ values can be underestimated by $\sim 0.2$ dex when using the Balmer decrement and up to $\sim0.5$ dex using A$_{\rm V}$.
    The effect of dust is negligible at the low mass end (M$_{\star}\leq 10^{9}$ M$_{\odot}$).
    
    \item We estimated the gas-phase metallicities using a combination of strong-line diagnostics (R2, R3, O32, Ne3O2) as presented in \cite{Curti2024_metallicity}. 
    We find that the \flares\ galaxies display a mass-metallicity relation with a slope similar to that obtained in \cite{Curti2024_metallicity} after correcting the line ratios for dust. The intrinsic (dust-free) line ratios gives a relation with a negligibly higher slope ($\sim0.22$ as compared to $\sim0.17$). However, this is very different from that obtained directly from the simulation using the mass-weighted metallicity ($\sim0.5$), which also exhibits a higher normalisation. We also find negligible evolution in the mass-metallicity relation for \flares\ at $z \in [5,10]$.
\end{itemize}

These findings emphasise the importance of accounting for biases
introduced by dust geometry and varying stellar populations when interpreting nebular emission lines, and highlight important implications for current and upcoming high-redshift studies.
Our results demonstrate that no single dust-correction approach yields unbiased estimates of all physical quantities in galaxies with complex star–dust geometries. 
The Balmer decrement preferentially traces dust attenuation in H\textsc{ii} regions and can underestimate the total star formation rate in systems where a significant fraction of star formation is heavily obscured, while correction using the A$_{\rm V}$ method can introduce larger biases in quantities such as $\xi{\rm ion}$; in both cases, the inferred quantities depend sensitively on assumptions about the effective attenuation curve.
These effects arise because different diagnostics weight distinct stellar populations and dust environments within galaxies. 
Consequently, inferred trends in SFR, $\xi_{\rm ion}$, and metallicity, particularly as functions of stellar mass, can reflect systematic biases rather than intrinsic physical correlations. 
For high-redshift observations, this implies that comparisons to simulations should be performed in observable space using forward-modelled quantities wherever possible (an `apples-to-apples' comparison), and that dust-corrected inferred quantities should be interpreted with caution, especially when assuming uniform attenuation laws. 
Rather than applying universal empirical corrections, robust interpretation requires accounting for internal galaxy structure and the coupling between stellar populations and dust.

In the future, far-infrared emission lines arising from the nebular regions, such as those of nitrogen ([O\textsc{iii}]52\um, [O\textsc{iii}]88\um) or oxygen ([N\textsc{iii}]57\um, [O\textsc{ii}]122\um) would be instrumental in deciphering the effect of dust on derived physical properties of the ISM \cite[]{Nagao2011,Pereira-Santaella2017,Fernandez2021}. 
Recent work by \cite{Harikane2025},
combining ALMA and \jwst\ observations of [O\textsc{ii}]$\lambda\lambda 3726,3729$, [O\textsc{iii}]$\lambda 4363$, [O\textsc{iii}]88\um, and [O\textsc{iii}]52\um,
has revealed metallicities up to $\sim 0.8$ dex higher than those 
inferred from `Direct-Te' metallicity, attributed to electron density.
Furthermore, `Direct-Te' metallicities themselves will be biased by spatial variations in electron temperature and dust distribution across galaxies \cite[]{Riffel2021,Cameron2023}.
Even spatially resolved observations (unlensed) in the high-redshift Universe will have resolution of $\gtrsim 1$ kpc, and thus bigger than giant molecular cloud scales. So there can still be dust inhomogeneities within these scales.  
Hence, one would require spatially resolved, multi-wavelength observations, combined with forward modelling approaches that couple realistic star formation histories, metal enrichment, and dust geometry to predicted observables to interpret these data.
Forward modelling tools like \texttt{powderday} \cite[]{Powderday2021}, \texttt{skirt} \cite[]{skirt2020}, \texttt{synthesizer} \cite[]{synth2_2025,synth1_2025} will be extremely useful in these efforts to decipher galaxy physics in the early Universe.
Ultimately, treating star–dust geometry as a fundamental component of high-redshift galaxy modelling will be key to extracting robust constraints.

% \newpage

\section*{Acknowledgements}
We thank the anonymous referee for a constructive report that has improved this manuscript.
We thank the \eagle\, team for their efforts in developing the \eagle\, simulation code.  This work used the DiRAC@Durham facility managed by the Institute for Computational Cosmology on behalf of the STFC DiRAC HPC Facility (www.dirac.ac.uk). The equipment was funded by BEIS capital funding via STFC capital grants ST/K00042X/1, ST/P002293/1, ST/R002371/1 and ST/S002502/1, Durham University and STFC operations grant ST/R000832/1. DiRAC is part of the National e-Infrastructure. 

APV, SMW and WJR acknowledge support from the Sussex Astronomy Centre STFC Consolidated Grant (ST/X001040/1).
SL acknowledges the supports from the National Natural Science Foundation of China (No. 12588202, 12473015).

\section*{Author Contributions}
We list here the roles and contributions of the authors according to the Contributor Roles Taxonomy (CRediT)\footnote{\url{https://credit.niso.org/}}.
% Me, Me and Me again 
\textbf{Aswin P. Vijayan}: Conceptualization, Data curation, Methodology, Investigation, Formal Analysis, Software, Visualization, Writing - original draft.
\textbf{Robert M. Yates}: Data curation, Writing - review \& editing.
\textbf{Christopher C. Lovell, William J. Roper, Stephen M. Wilkins}: Data curation, Software, Writing - review \& editing.
\textbf{Hiddo S. B. Algera , Shihong Liao , Paurush Punyasheel, Lucie E. Rowland, Louise T. C. Seeyave}: Data curation, Writing - review \& editing.

We also wish to acknowledge the following open source software packages used in the analysis: \textsc{Numpy} \citep{numpy}, \textsc{Scipy} \citep{scipy}, \textsc{Astropy} \citep{astropy:2013, astropy:2018, astropy:2022}, \textsc{Cmasher} \citep{cmasher}, and \textsc{Matplotlib} \citep{matplotlib}. 

\section*{Data Availability Statement}
All the scripts used to make the plots are available at \href{https://github.com/aswinpvijayan/line_ratio_reliability}{https://github.com/aswinpvijayan/line\_ratio\_reliability} as python notebooks.

\appendix
\section{Impact of increasing A$_{\rm V}$, with no scatter}\label{sec:app.vary_mu}

\begin{figure}
    \centering
    \includegraphics[width=0.95\linewidth]{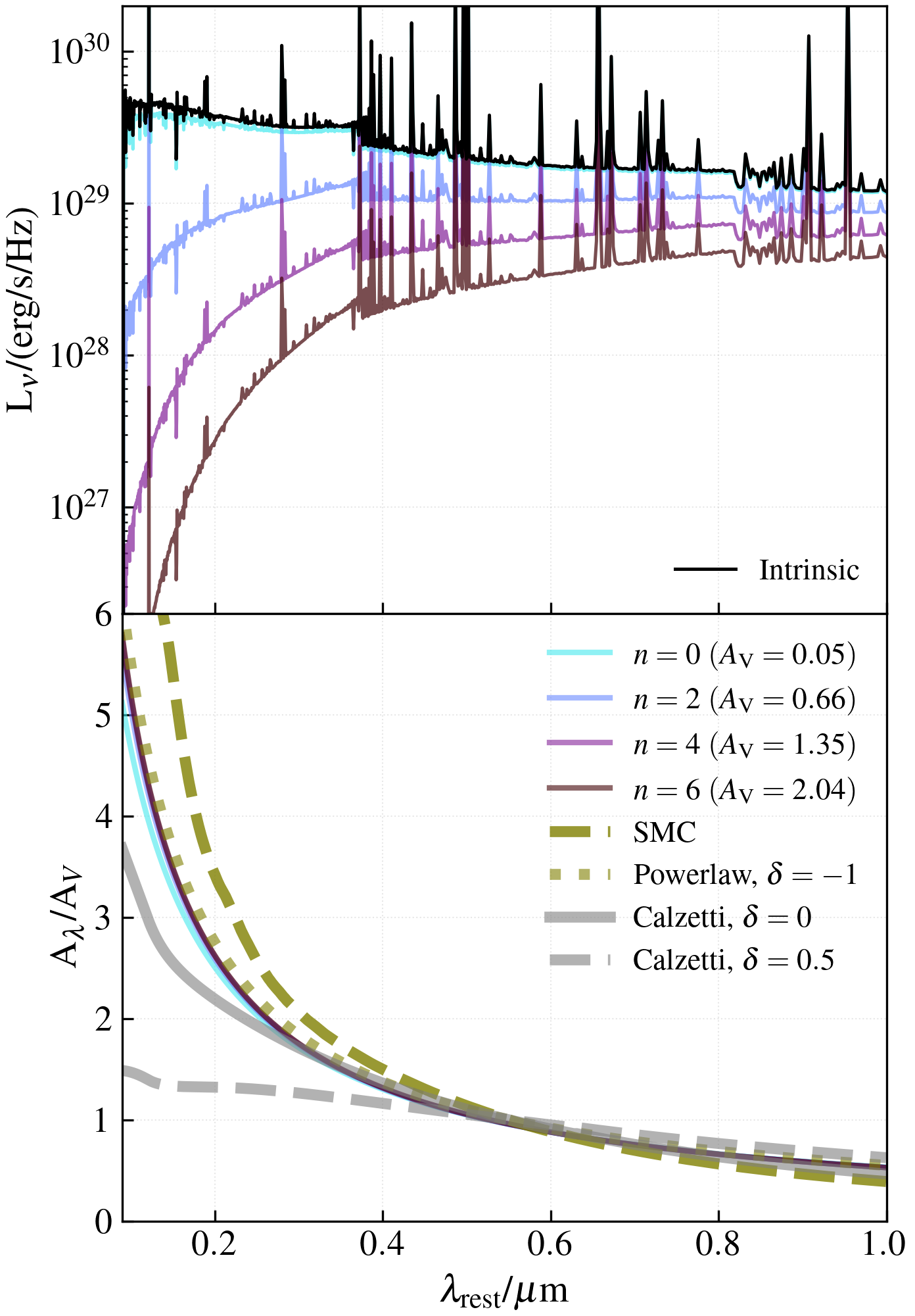}
    \caption{\textbf{Top}: The total intrinsic (black) and the dust-attenuated (for different value of the mean, in coloured) SEDs of the `Varying only A$_{\rm V}$' model galaxy, truncated to $0.1-1$ \um.  \textbf{Bottom}: Attenuation curves of the resultant galaxy when varying the mean value (with minimal spread) of the dust attenuation across the different star-forming clumps. We also show the power-law extinction curve (slope $=-1$) that was used for each stellar clumps, as well as the SMC and Calzetti (for slope$=0,0.5$) attenuation curve.}
    \label{fig:app_varymu_attcurve}
\end{figure}

\begin{figure}
    \centering
    \includegraphics[width=\linewidth]{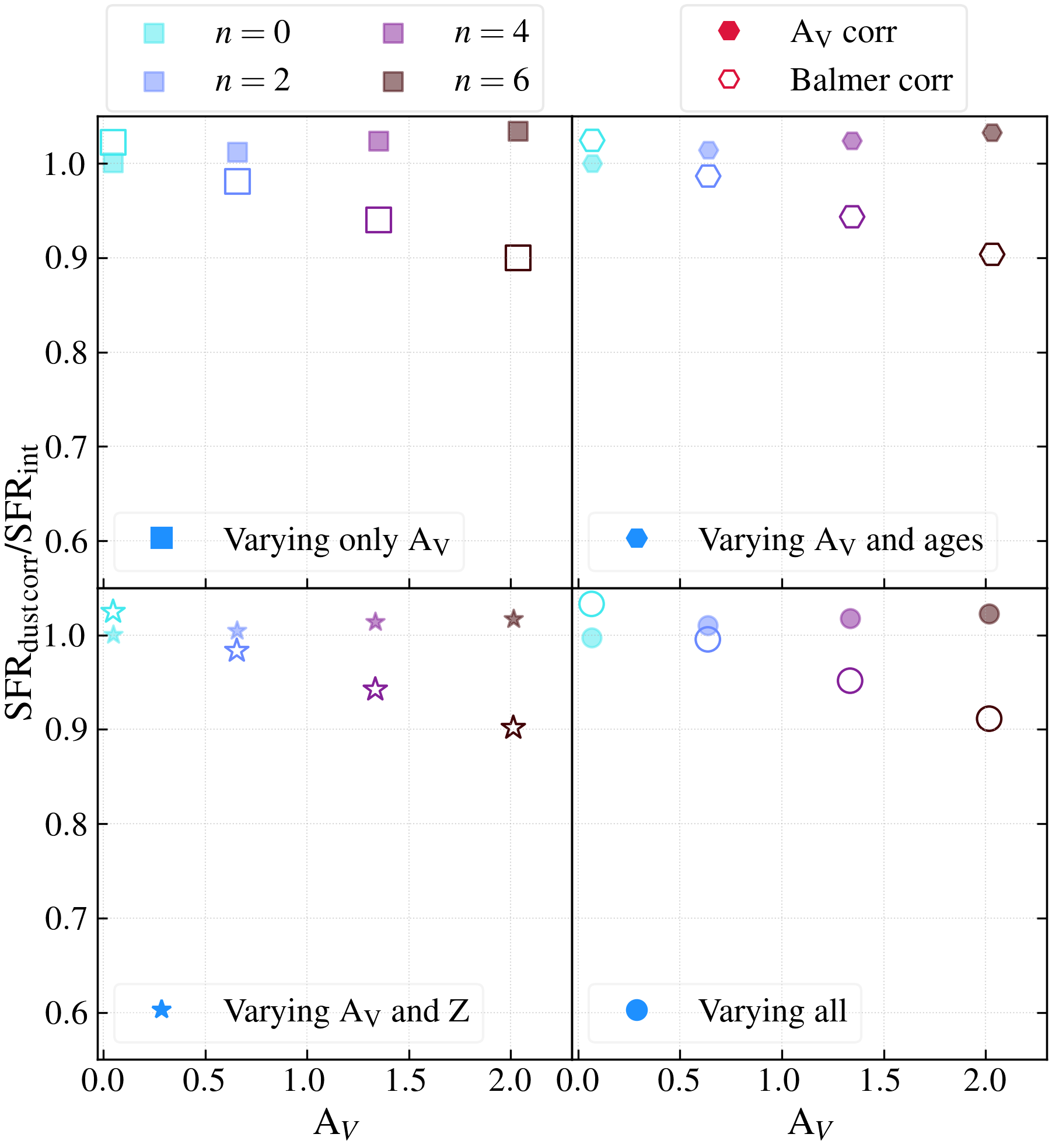}
    \caption{This is same plot as Figure~\ref{fig:toy_sfr_frac}, but now using toy galaxies with little spread in their dust optical depths, but increasing A$_{\rm V}$. Fraction of the recovered SFR after correcting for dust obscuration for the different toy galaxies in the model. The different markers denote the different models, with the different colours denoting the multiple of the mean obscuration along the line-of-sight to the different stellar clumps within the galaxy. The filled and open markers denote the SFR fraction recovered using the A$_{\rm V}$ method and the Balmer decrement method respectively.}
    \label{fig:app_varymu_sfr_frac}
\end{figure}
To drive home the point that it is the star-dust geometry that drives the unreliability in inferring the physical properties of galaxies, we adopt the same toy galaxy set-up, while increasing the dust attenuation progressively with little to no spread in the dust attenuation across the stellar populations.
We quantify the increasing dust attenuation for the toy galaxies by varying the mean from $0-6\mu_0$ as follows:
\begin{equation}
    \tau_{\rm V,i} = \mathcal{N}(0, n\mu_0) + \sigma_0,
\end{equation}
where $\tau_{\rm V,i}$ is the optical depth in the V-band of the i$^{\rm th}$ clump and $n\in[0,6]$. Here, we set the optical depth in the V -band for each clump by randomly drawing from a normal distribution with a mean ($mu_0$) of 0.3 and standard deviation ($\sigma_0$) of 0.1. 
We set any negative $\tau_{\rm V,i}$ to 0.01, if any. 
We do this for the four different galaxy models as in \S~\ref{sec:modelling.toygal}.

Figure~\ref{fig:app_varymu_attcurve} shows the SED (top panel) and the resultant attenuation curve (bottom panel) of the
‘Varying only A$_{\rm V}$’ model galaxy with the above set-up. 
The SED gets progressively more attenuated as `n' increases.
As expected, since there is little variation in the optical depths along different stellar populations, the resultant attenuation curve is the same as the input extinction curve, a powerlaw with slope=-1. 

Figure~\ref{fig:app_varymu_sfr_frac} is the same as Figure~\ref{fig:toy_sfr_frac}, but now showing the recovered fraction of SFR using the two dust correction methods for the above set-up.
As expected due to negligible variation in the resultant attenuation curve from the input one, the only variation is coming from the use of the Calzetti curve which is flatter than the input curve at shorter wavelength. At all times the recovered SFR is greater than $90\%$. This shows that the its the scatter in the optical depths that is the more important in driving the variations, compared to the overall A$_{\rm V}$.

\section{SFRF Schechter fit}\label{sec:app.schecter}
We fit the following form of the Schechter function to the SFRF:
% \begin{align}
%     {\rm log}_{10}{\phi} = {\rm log}_{10}\bigg(\frac{d{\rm n}}{d{\rm log_{10}L}}\bigg) \\= \phi^{\star}\, + \, ({\rm log_{10}(SFR) - log_{10}(SFR^{\star}}))(\alpha+1)\, + \, {\rm log}_{10}({\rm ln}(10)\, {\rm exp}(10^{\rm log_{10}(SFR) - log_{10}(SFR^{\star})}),
% \end{align}
\begin{align}
{\rm log}_{10}{\phi} &= {\rm log}_{10}\bigg(\frac{d{\rm n}}{d{\rm log_{10}L}}\bigg) \nonumber\\
  &= \phi^{\star}\, + \, ({\rm log_{10}(SFR) - log_{10}(SFR^{\star}}))(\alpha+1) \nonumber\\
  &+ {\rm log}_{10}({\rm ln}(10)\, {\rm exp}(10^{\rm log_{10}(SFR) - log_{10}(SFR^{\star})}),
\end{align}

where $\phi^{\star}$ is the characteristic number density, SFR$^{\star}$ is the characteristic star formation rate, and $\alpha$ is the faint-end slope of the function. To estimate the parameters we use the the \texttt{lmfit} \cite[]{lmfit_newville_2025} package in \texttt{Python}. Specifically we use the \texttt{emcee} model to perform a Bayesian inference using Markov Chain Monte Carlo (MCMC). For this purpose we provide uniform priors to the free parameters log$_{10}(\phi^{\star})$, log$_{10}$(SFR$^{\star}$) and $\alpha$ in the range $[-5,-1.5]$, $[0.5,3]$ and $[-3.5,-0.5]$, respectively. The exploration is performed with 4000 steps and 200 walkers, with a burn-in of 1000. We provide the fit parameters for the intrinsic, `A$_{\rm V}$ method' corrected, `Balmer decrement method' corrected and the observed SFRF respectively in Table~\ref{tab:schechter}.

\begin{table*}
\centering
\renewcommand\arraystretch{2}
\caption{SFRF Schechter function fits\label{tab:schechter}}
\resizebox{\textwidth}{!}{\begin{tabular}{|c|c|c|c|}
\hline
 {\large $z$} &  {\large log$_{10}$(SFR$^{\star}$)} & {\large log$_{10}$($\phi^{\star}$)} & {\large $\alpha$}   \\
 \hline
 5  & $2.11 \pm 0.08$, $1.89 \pm 0.04$, $1.70 \pm 0.03$, $1.61 \pm 0.03$ &  $-4.26 \pm 0.13$, $-4.1 \pm 0.07$, $-3.79 \pm 0.06$, $-3.83 \pm 0.07$  &  $-2.07 \pm 0.03$, $-2.11 \pm 0.03$, $-2.05 \pm 0.03$, $-2.14 \pm 0.03$ \\ \hline
 6  & $2.09 \pm 0.08$, $2.29 \pm 0.20$, $1.81 \pm 0.08$, $1.56 \pm 0.06$    & $-4.72 \pm 0.15$, $-5.17 \pm 0.34$, $-4.35 \pm 0.16$, $-4.09 \pm 0.13$     & $-2.26 \pm 0.03$, $-2.36 \pm 0.04$, $-2.26 \pm 0.04$, $-2.30 \pm 0.05$  \\ \hline
 7  & $2.39 \pm 0.24$, $2.69 \pm 0.23$, $1.69 \pm 0.06$, $1.70 \pm 0.12$    & $-5.66 \pm 0.44$, $-6.38 \pm 0.41$, $-4.5 \pm 0.15$, $-4.77 \pm 0.26$    & $-2.48 \pm 0.04$, $-2.59 \pm 0.04$, $-2.4 \pm 0.05$, $-2.57 \pm 0.06$ \\ \hline
 8  & $2.45 \pm 0.22$, $2.46 \pm 0.19$, $2.08 \pm 0.19$, $2.00 \pm 0.24$  & $-6.30 \pm 0.45$, $-6.48 \pm 0.39$, $-5.72 \pm 0.41$, $-5.84 \pm 0.54$  & $-2.68 \pm 0.05$, $-2.75 \pm 0.04$, $-2.71 \pm 0.06$, $-2.84 \pm 0.07$ \\ \hline
 9  & $2.31 \pm 0.14$, $2.30 \pm 0.14$, $2.25 \pm 0.17$, $1.87 \pm 0.22$  & $-6.59 \pm 0.33$, $-6.62 \pm 0.33$, $-6.52 \pm 0.39$, $-5.95 \pm 0.55$  & $-2.89 \pm 0.05$, $-2.90 \pm 0.05$, $-2.91 \pm 0.05$, $-2.99 \pm 0.09$  \\ \hline
 10 & $2.19 \pm 0.13$, $2.11 \pm 0.13$, $1.83 \pm 0.20$, $1.58 \pm 0.20$  &  $-6.67 \pm 0.30$, $-6.67 \pm 0.30$, $-6.07 \pm 0.51$, $-5.66 \pm 0.58$ &  $-2.95 \pm 0.06$, $-3.03 \pm 0.07$, $-3.03 \pm 0.09$, $-3.10 \pm 0.13$  \\ \hline
\end{tabular}}
\tablecomments{Schechter function fit parameters for the SFRF. The different fit parameters for the intrinsic (dust-free), `A$_{\rm V}$ method' corrected, `Balmer decrement method' corrected and the unobscured (dust attenuated) SFRF respectively are provided.}
\end{table*}

\section{Strong Line ratios in \flares}\label{sec:app.stronglines}
\begin{figure*}
    \centering
    \includegraphics[width=0.49\textwidth]{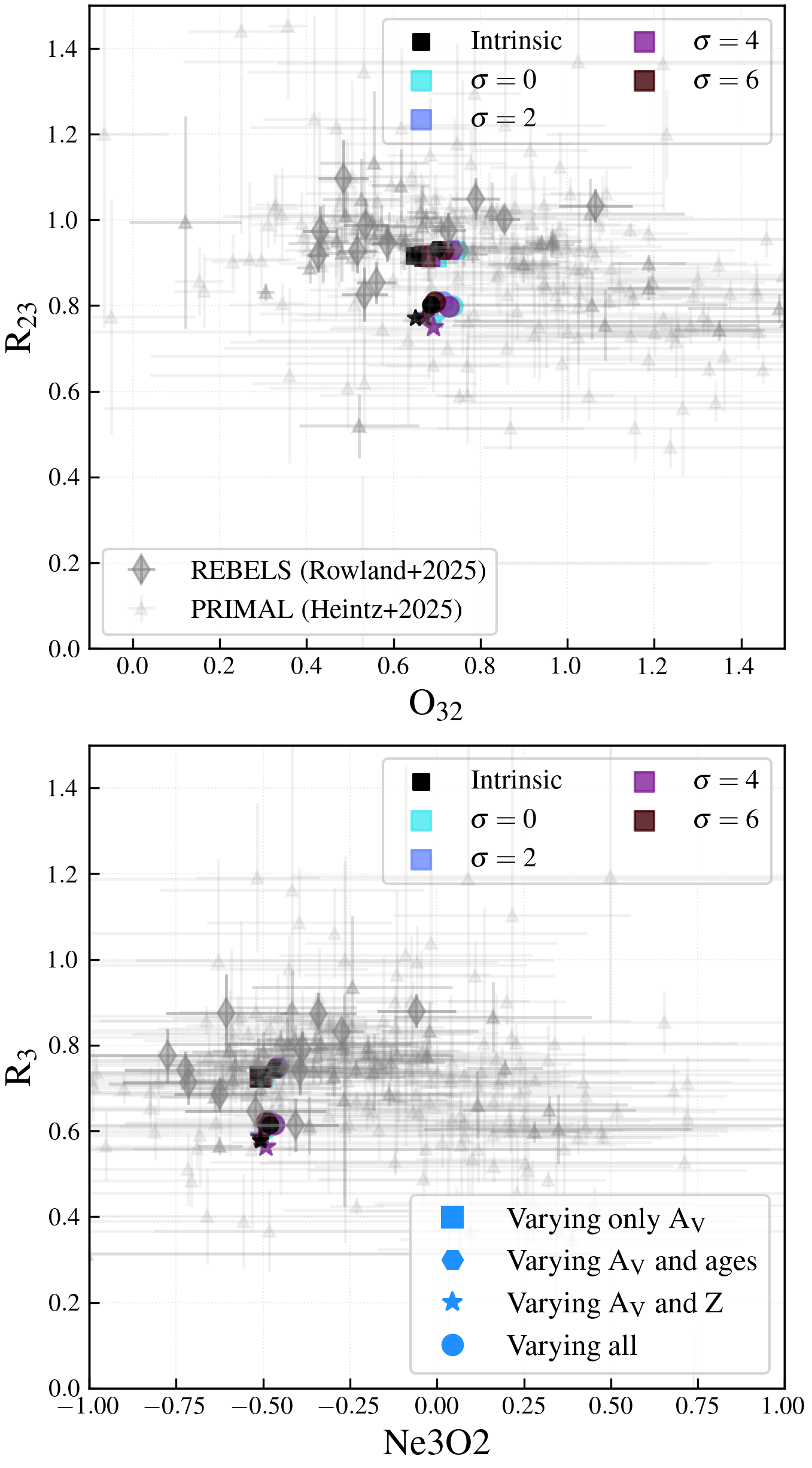}
    \includegraphics[width=0.49\textwidth]{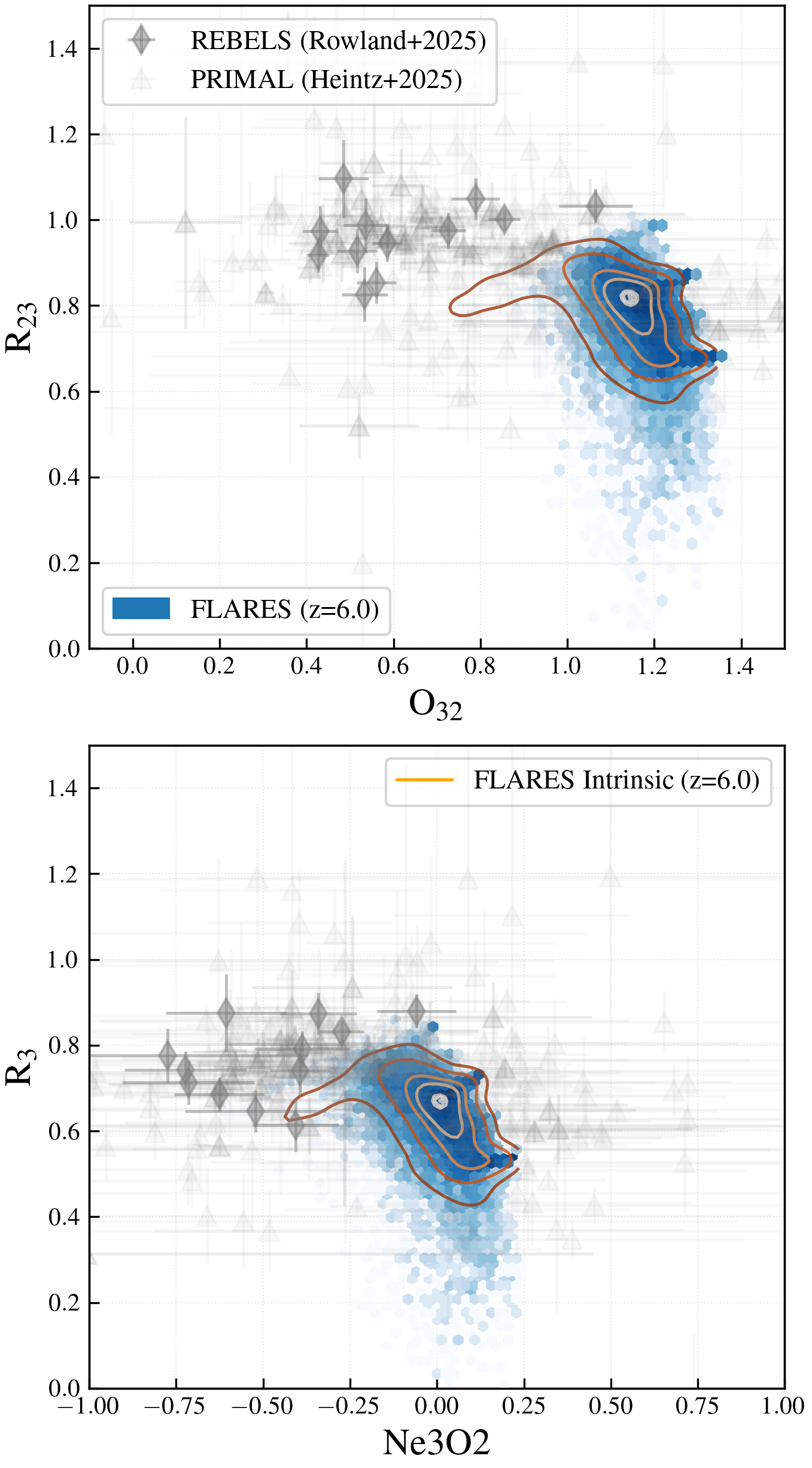}
    \caption{Figure shows the O32 ratio plotted against the R23 ratio (\textbf{top}), and the Ne3O2 ratio against the R3 ratio (\textbf{bottom}), for the toy galaxies (\textbf{left}) and \flares\ galaxies (\textbf{right}, at $z=6$). We make a [O\textsc{iii}]5007\AA\ flux cut of $3\times 10^{-18}$ erg/s/cm$^{-2}$. We also plot the observed line ratios for galaxies obtained using \jwst\ from \cite{Heintz2025,Rowland2025_rebels}.
    % The \cite{Heintz2025} sample contains more than 500 galaxies, we randomly select 50 galaxies that have low error estimates on their line emission.
    }
    \label{fig:R23_O32_R3_Ne3O2_comp}
\end{figure*}
To better understand the distribution of strong line ratios in \flares, we first explore the relationship between the different line ratios using the toy model galaxies introduced in Section~\ref{sec:modelling.toygal}.
We also use them to explore the impact of varying properties and star-dust geometry on emission line ratios. Figure~\ref{fig:R23_O32_R3_Ne3O2_comp} (left) shows the the O32 ratio as a function of R23 ratio (top), and the Ne3O2 ratio against the R3 (bottom) for the toy models, alongside observational data from \cite{Heintz2025,Rowland2025_rebels}. 
It is evident that the intrinsic values of R23 and R3 (y-axis) are more sensitive to the variations in the underlying stellar populations than O32 and Ne3O2 (x-axis), which remain similar.
Across the models, there is $\sim0.3$ dex in variation in R23 and R3. This can be explained as follows: the models which do not vary the stellar ages, have very similar \Hb\ fluxes (which R23 and R3  depend on), since they have similar SFRs. The \Hb\ flux is higher for these model galaxies compared to the ones that vary the ages, due to the rapid drop-off in the number of ionising photons after 10 Myr in SPS models (the x-axis of the figure is composed of only collisionally excited lines). 

Due to the variation in the metallicity, the `Varying A$_{\rm V}$ and ages' model has a higher [O\textsc{iii}]5007\AA\, (and [O\textsc{iii}]4959\AA) line strength compared to the `Varying A$_{\rm V}$ and Z' model for this realisation, making the R23 or R3 value of the former higher. There is negligible change in the O32 or Ne3O2 ratio, due to there being negligible variation in the shape of the ionising radiation.

The spread in the O32 ratio ($\sim 0.15$ dex) reflects the effect of dust attenuation. 
The Ne3O2 ratio remains mostly unaffected due to the constituent lines being closer in wavelength.
However, when both the age and metallicity are varied, a spread in the ratio appears, driven by changes in the star-dust geometry, as described in equation~\ref{eq:line_ratio}.
% , with the direction of the change depending on the star-dust geometry in the galaxy. 
The Ne3O2 ratio shows a variation of $\sim 0.1$ dex for the  `Varying A$_{\rm V}$ and Z' and `Varying all' models.
This is related to the fact that the former ratio is composed of two distinct metals, and thus their abundances and hence the dust attenuated line ratios can scale drastically based on equation~\ref{eq:line_ratio}.
It is important to note that all the toy galaxies have similar mass-weighted ages and metallicities, reinforcing the claim that the observed scatter is mainly driven by 
variation in the underlying stellar population and star-dust geometry.

\subsection{\flares}
The right panel of Figure~\ref{fig:R23_O32_R3_Ne3O2_comp} shows the O32 ratio as a function of R23 ratio (top), and the Ne3O2 ratio against R3 (bottom) for the \flares\ galaxies overlaid with observational data from \cite{Heintz2025,Rowland2025_rebels}. We also apply a [O\textsc{iii}]5007\AA\ flux cut of $3\times 10^{-18}$ erg/s/cm$^{-2}$, corresponding to typical flux limits of high-redshift observational surveys.

From the R23-O32 panel, it is apparent that the intrinsic and dust-attenuated values of O32 from \flares\ galaxies are concentrated at higher values than the bulk of the observations, which extend to lower O32 values by $\sim 0.4$ dex.
Since O32 is a strong tracer of the ionisation condition (compared to the metallicity), this indicates that in general the ionisation parameters of the simulated galaxies are higher than the bulk of the observations.
This higher on average O32 ratio can also be seen in other forward modelling works, for example in \cite{Hirschmann2023} (see their Figure 2).
However, both the R23 and R3 ratio are reproduced reasonably well, albeit the presence of a small fraction of galaxies in the observational data showing higher values (suggestive of higher ionisation parameter).

\flares\ galaxies have been shown to reproduce the observed \Ha\ (and hence \Hb) and [O\textsc{iii}]5007\AA\ luminosity functions \cite[]{CoveloPaz2025_halpha,Meyer2024Oiii}, indicating that the simulation broadly reproduces the correct star formation and metal enrichment history, and the dust obscuration of the observed high-redshift galaxies. 
This tension therefore highlights the fine balance between the required ionisation parameter, gas densities and star-dust geometry when forward modelling galaxies in simulations.
The effective ionisation parameter (equation~\ref{eq:Uref}) for the majority of the \flares\ galaxies deviates little from the fiducial U$_{\rm ref}$ (variation of $\sim 0.1$), as seen in Figure~8 of \cite{FLARES-XI}.
We show in Appendix~\ref{sec:app.varyphot} that a broader variation in the ionisation parameter is required to match the spread seen in the high-redshift observations.

Dust also plays a significant role in shaping the observed line ratios.
Even for lines that are close in wavelength, \ie R3 and Ne3O2, are affected due to the effect described as per equation~\ref{eq:line_ratio}. It is straightforward to explain the effect of dust on the O32 and Ne3O2 ratio, since both the ratios have in the numerator lines which are at a longer wavelength compared to the denominator, and thus experience less dust attenuation. 
This differential extinction naturally increases the observed ratios.
Additionally, a higher ionisation parameter shifts more oxygen into higher excited states, lowering the $[{\rm O\textsc{ii}}]\lambda 3727,29$ flux.

In contrast, the dust attenuation model (Section~\ref{sec:modelling.flares.sed}) employed in \flares\ generally suppresses the R23 and R3 ratio. 
For R23, the suppression arises primarily from the [O\textsc{ii}]$\lambda 3727,29$\AA\ doublet
being at shorter wavelength relative to [O\textsc{iii}]$\lambda 4959,5007$\AA\ and \Hb, thus suffering stronger dust attenuation.
The case for R3 is however counter-intuitive to what one would expect in case of a screen of dust or assuming a single attenuation curve. 
This is mainly due to the assumption of dust tracing metals made in \flares. 
This preferentially attenuates metal rich regions which then suffer higher-attenuation, driving the observed [O\textsc{iii}]$\lambda 5007$\AA\ luminosity down and decreasing the R3 ratio. This affects the Ne3O2 ratio as well. 
However, due to [O\textsc{iii}]$\lambda 5007$\AA\ being at slightly longer wavelength compared to \Hb, it suffers less attenuation and pushes the R3 ratio up. The competing effects result in a modest net decrease of the R3 ratio. 
\section{Varying default parameters in photoionisation modelling}\label{sec:app.varyphot}
We will describe here, the effect of varying the ionisation parameter and gas density on emission line strengths, followed by a demonstration of the effects using toy galaxy models.

The \Ha\ or \Hb\ lines trace the number of ionising photons and are largely insensitive to the variations in the ionisation parameter or gas density within typical conditions in the ISM. Their strength remain nearly constant in our photoionisation model variations (shown later in this section), and only changes under extreme temperatures or high-density regimes.

In general, increasing the ionisation parameter, increases the ionising photon flux available for the excitation of metal lines originating from higher excitation states. 
However, with increasing flux, a higher fraction of a metal can be excited to their higher excitation states, thus suppressing the emission from lower states (and thus the importance of using ionisation corrections). 
The emissivity of the metal lines, primarily due to collisional excitation, is approximately proportional to the gas density (and other factors such as temperature and abundances). 
Thus, there will be an increase in the emission line strengths with increasing density, only up to the critical density of the transition.
Above this threshold, collisional de-excitation becomes significant, leading to the suppression of the emission line.
For the [O\textsc{iii}]4959\AA, [O\textsc{iii}]5007\AA\ and [Ne\textsc{iii}]3869\AA\ lines in consideration here, the critical density is close to $10^7$ cm$^{-3}$, hence there is no suppression in typical star-forming regions. 
This is not the case for the [O\textsc{ii}]$\lambda 3727,29$\AA\ line whose critical density is $\sim10^{3}$ cm$^{-3}$, gets suppressed at densities higher.

\begin{figure*}
    \centering
    \includegraphics[width=0.4\textwidth]{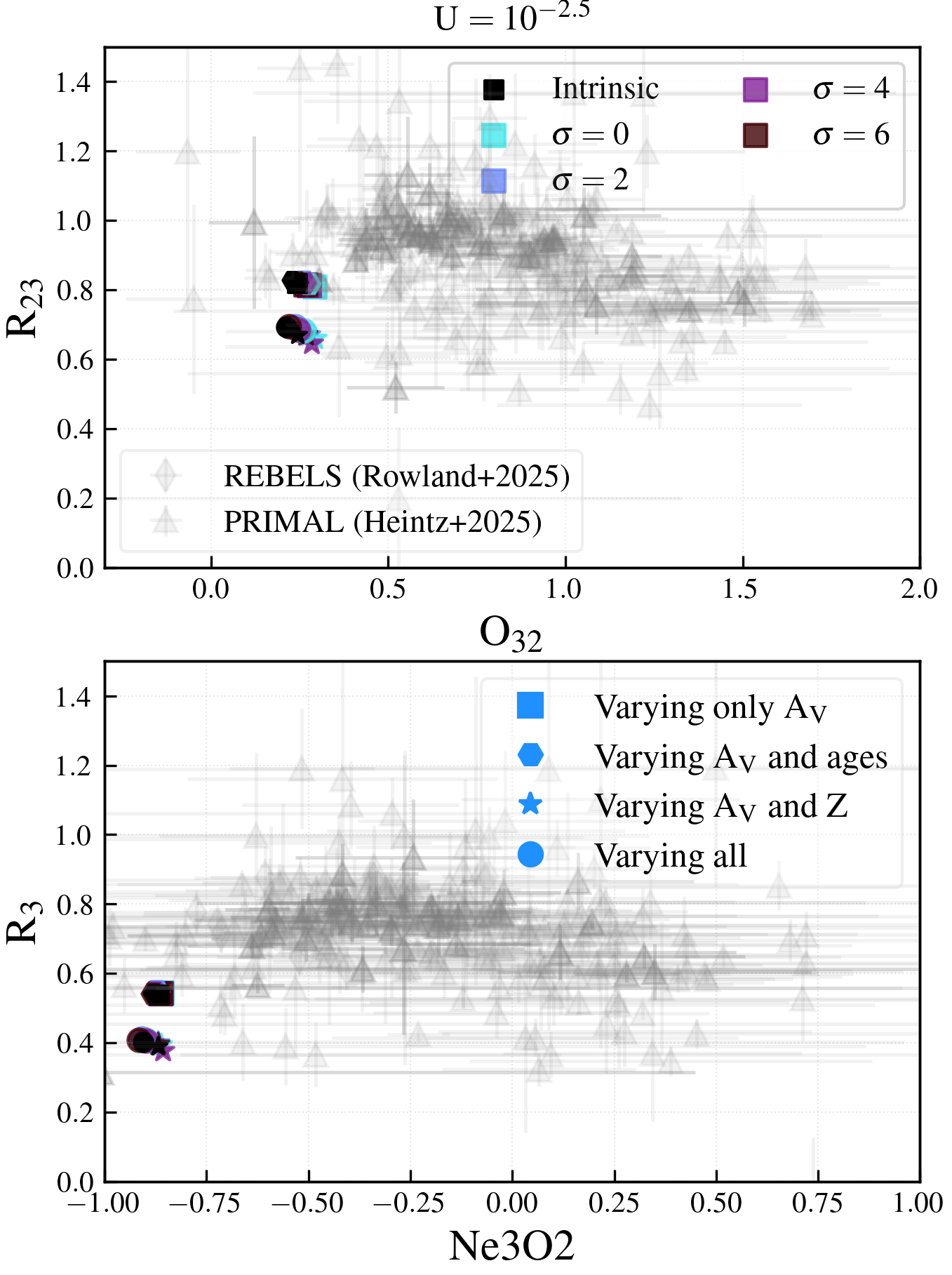}
    \includegraphics[width=0.4\textwidth]{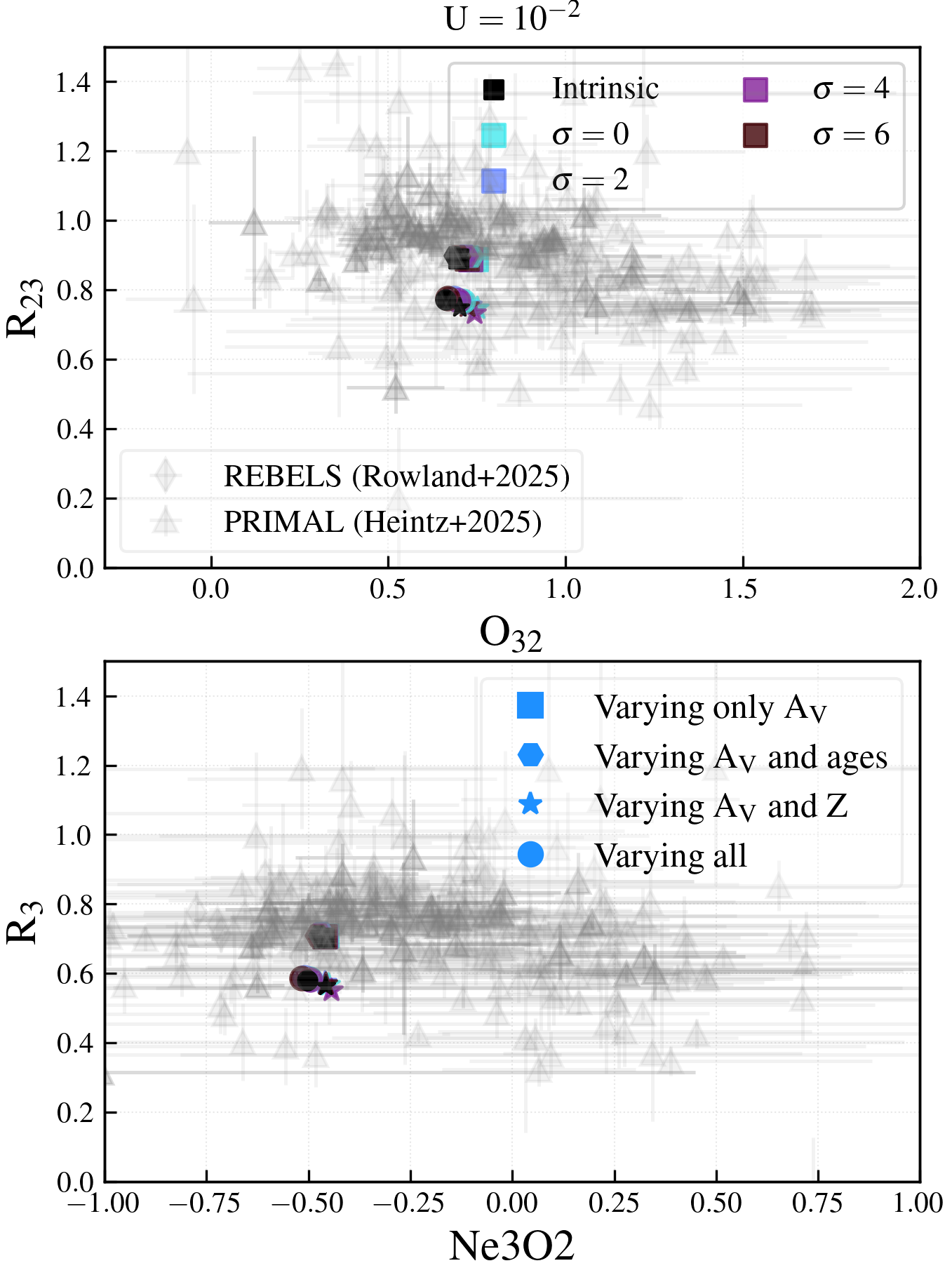}
    \includegraphics[width=0.4\textwidth]{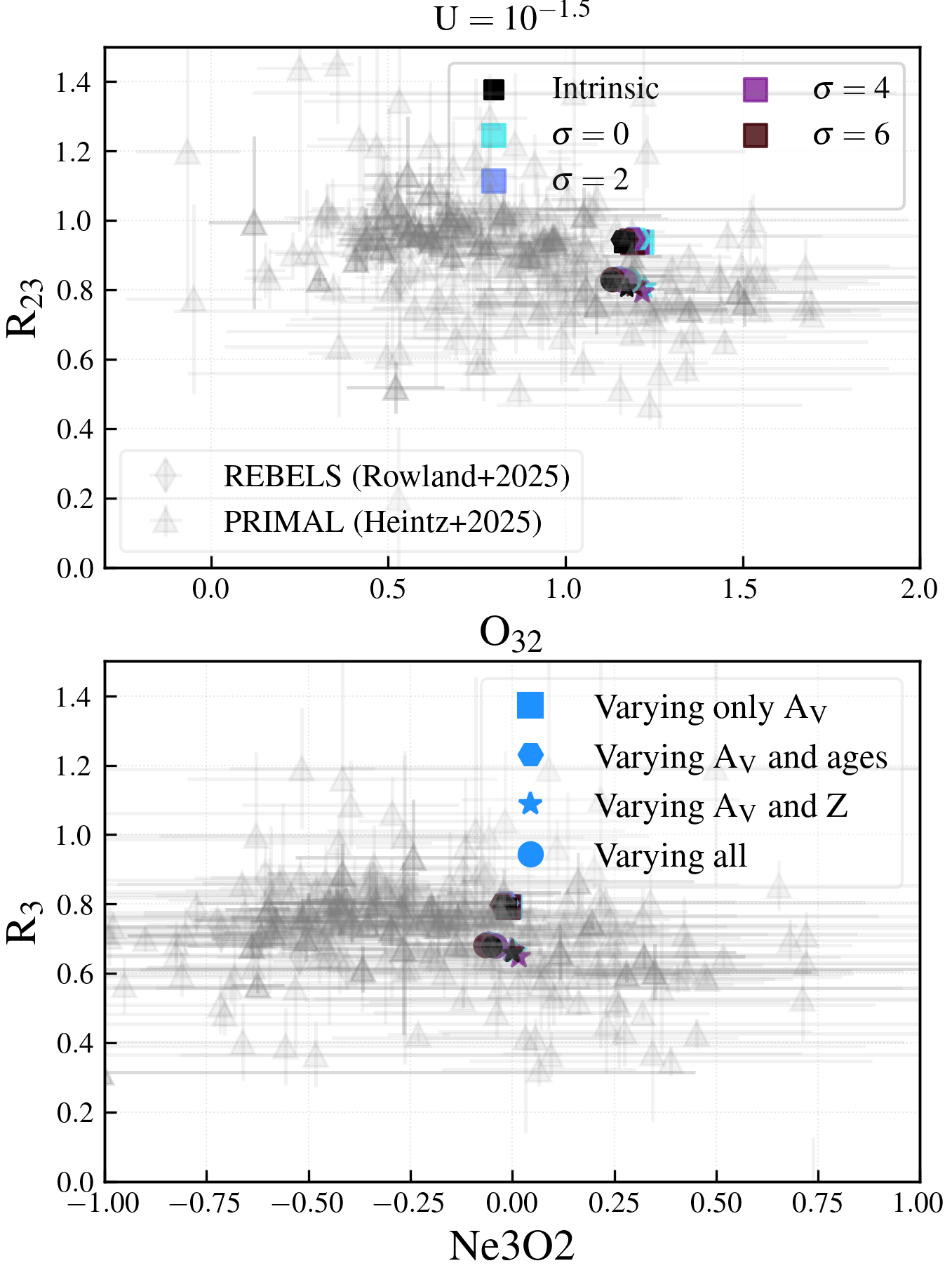}
    \includegraphics[width=0.4\textwidth]{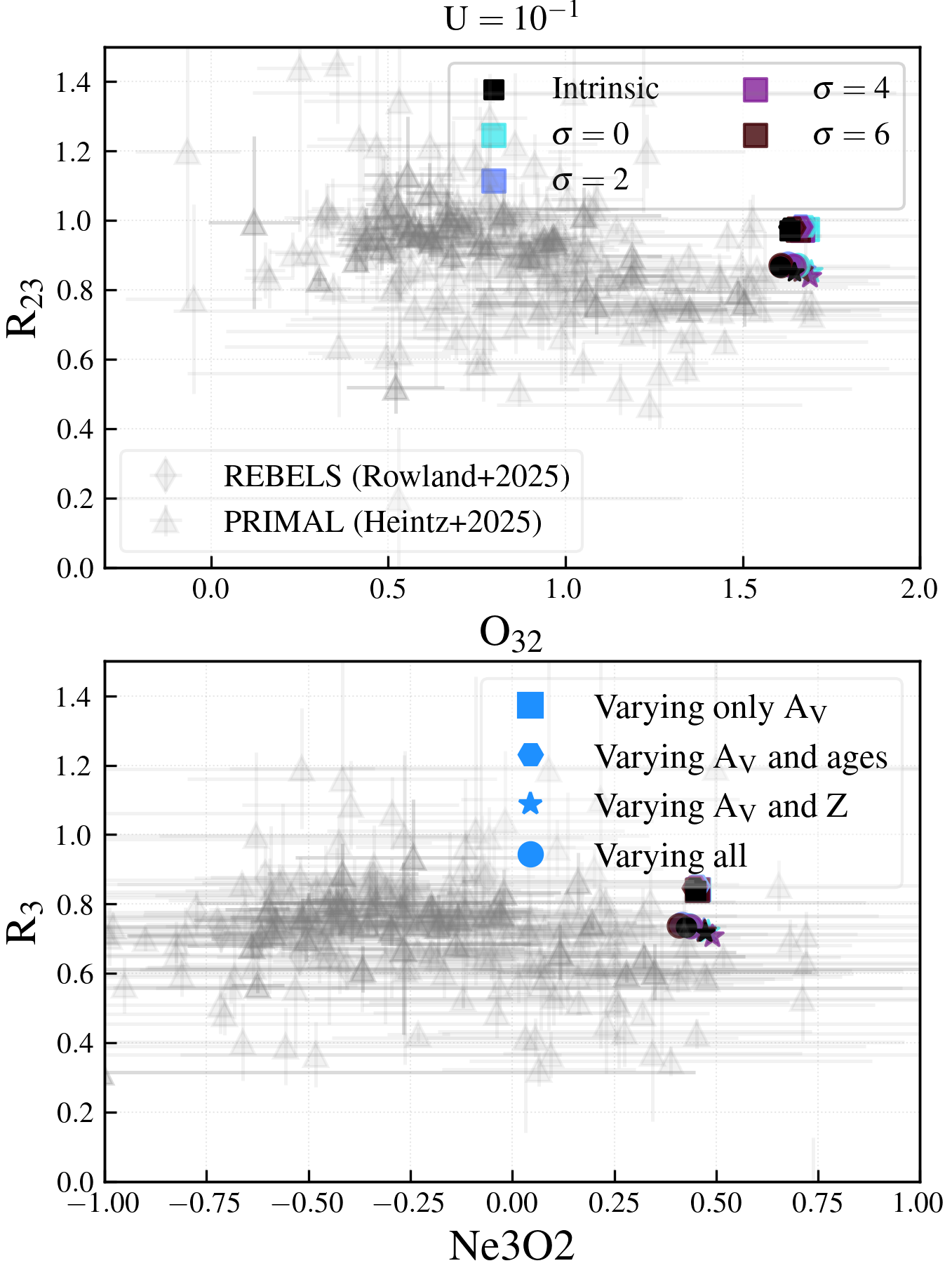}
    \caption{Same as Figure~\protect\ref{fig:R23_O32_R3_Ne3O2_comp}, showing the relationship between R23-O32 (left) and R3-Ne3O2 (right) for the toy model galaxies introduced in Section~\ref{sec:modelling.toygal}. Here we are changing the default photoionisation model used, and vary the ionisation parameter (U) from U$\in[10^{-2.5}, 10^{-2}, 10^{-1.5}, 10^{-1}]$. The quoted ionisation parameter is not a reference ionisation parameter as defined in Section~\ref{sec:modelling.forward}.}
    \label{fig:app_diff_U}
\end{figure*}

\begin{figure*}
    \centering
    \includegraphics[width=0.4\textwidth]{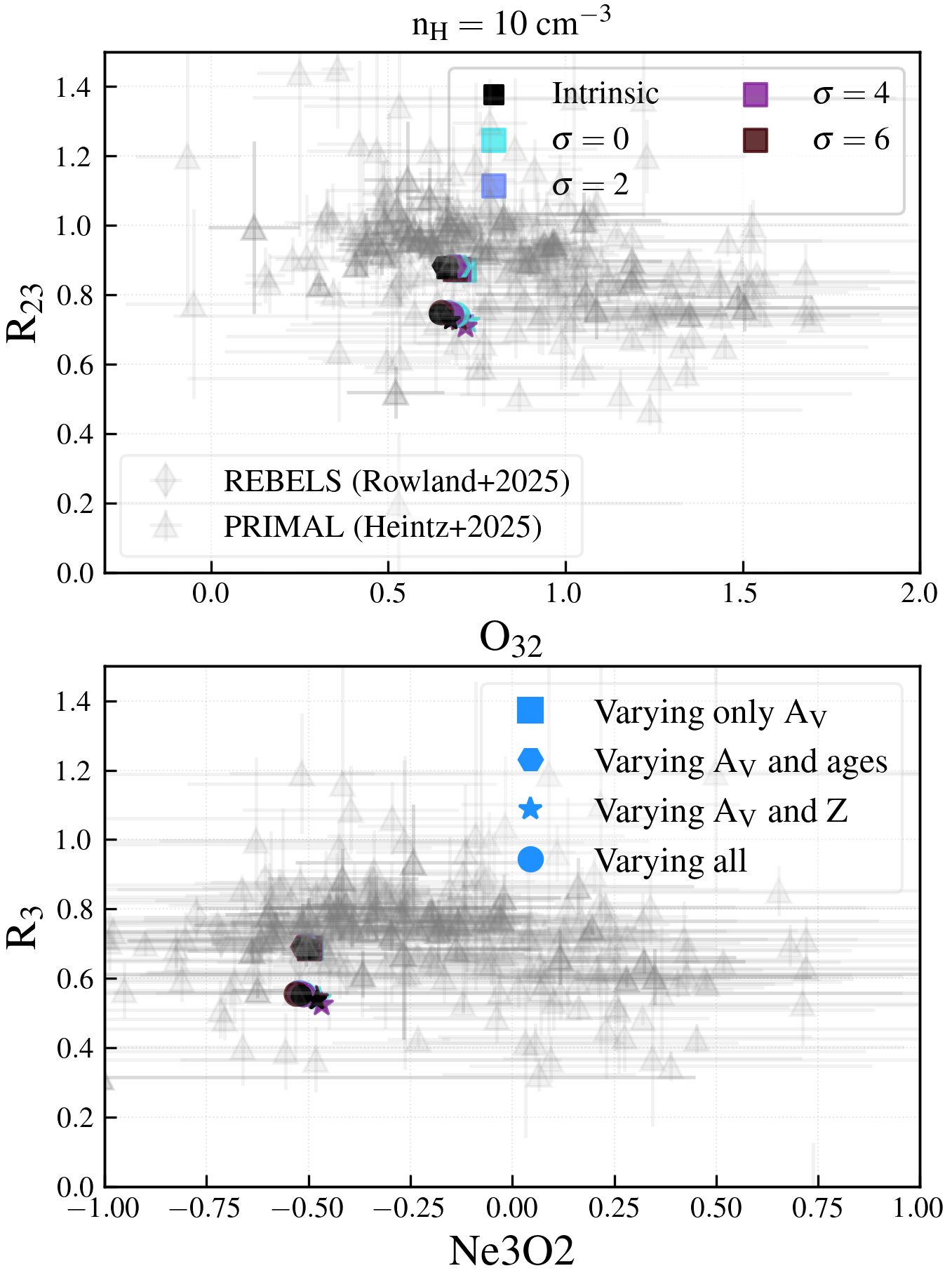}
    \includegraphics[width=0.4\textwidth]{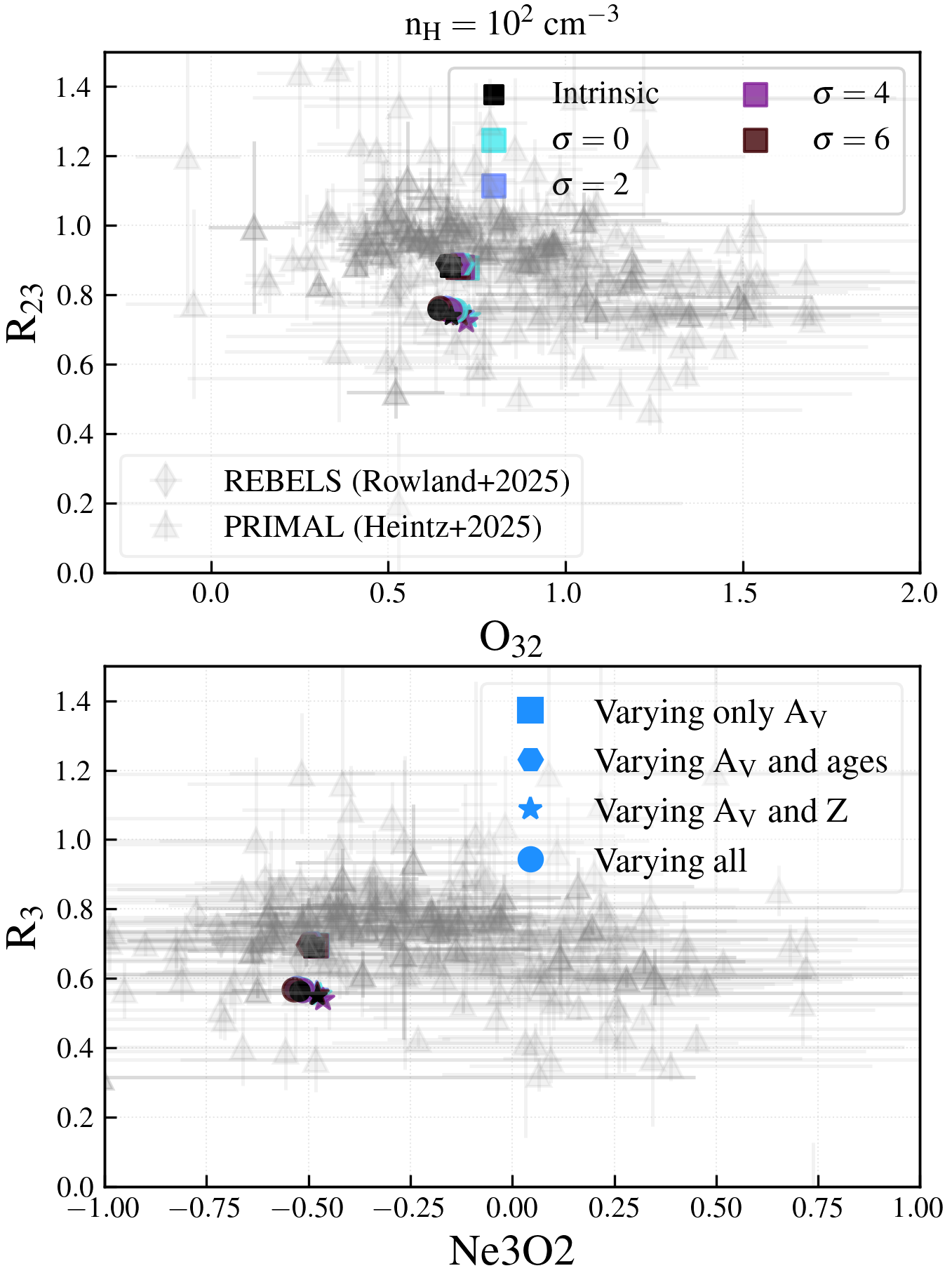}
    \includegraphics[width=0.4\textwidth]{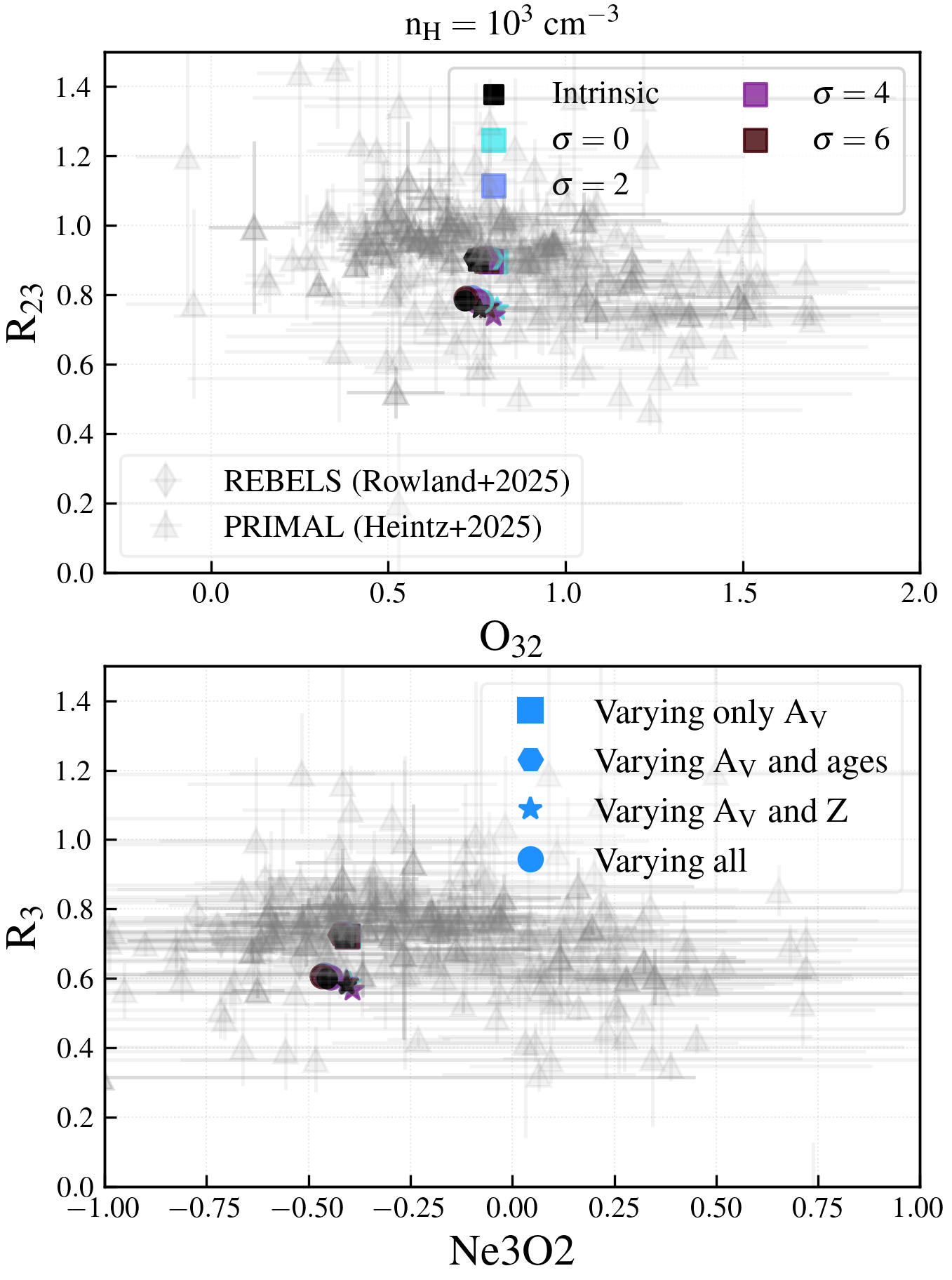}
    \includegraphics[width=0.4\textwidth]{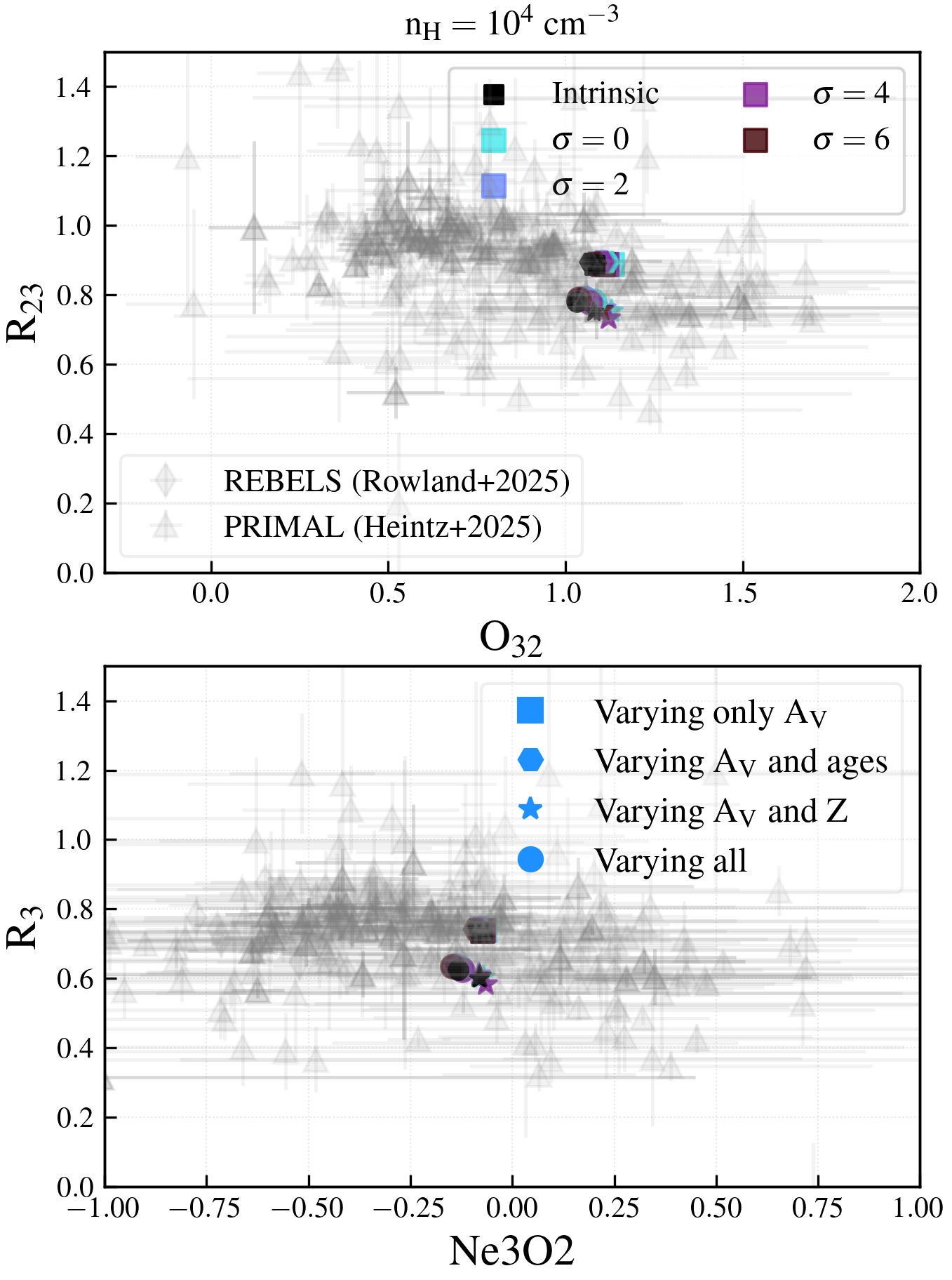}
    \caption{Same as Figure~\ref{fig:app_diff_U}, but now we vary the hydrogen density (n$_{\rm H}$) of the star-forming regions, with n$_{\rm H}\in[10,10^2,10^3,10^4]$ cm$^{-3}$.}
    \label{fig:app_diff_nh}
\end{figure*}

In Figure~\ref{fig:app_diff_U} and \ref{fig:app_diff_nh}, we modify the default assumptions in our photoionisation modelling of the toy galaxies and explain the changes to the R23-O32 and R3-Ne3O2 space based on the discussion above. In Figure~\ref{fig:app_diff_U}, we change the our assumption of a reference ionisation parameter for the model galaxies and fix the ionisation parameter (U). Thus all the star particles have the same U and hydrogen density (n$_{\rm H}=10^{2.5}$ cm$^{-3}$) in this figure. We vary U from $10^{-2.5} - 10^{-1}$. It can be clearly seen that O32 and Ne3O2 ratio increases for increasing U. This is due to [O\textsc{ii}]$\lambda 3727,29$\AA\ (which is in the denominator) line strength decreasing with increasing U, due to the higher excited states getting populated. R23 remains almost similar across different U, since the numerator has both [O\textsc{ii}]$\lambda 3727,29$\AA\ and [O\textsc{iii}]4959\AA, [O\textsc{iii}]5007\AA. R3 remains similar across different U, except at U$=10^{-2.5}$, where the lower excited state ([O\textsc{ii}]$\lambda 3727,29$\AA) is more populated compared to the higher state of [O\textsc{iii}]5007\AA. 

In Figure~\ref{fig:app_diff_nh}, we fix U to $10^{-2}$, while varying n$_{\rm H}$ from $10 - 10^{4}$ cm$^{-3}$. There is minimal changes to the plotted ratios, until n$_{\rm H}=10^{4}$ cm$^{-3}$, when the density is above the critical density of the [O\textsc{ii}]$\lambda 3727,29$\AA\ line, and gets suppressed. This increases the magnitude of the O32 and Ne3O2 ratios.

\section{MZR using other strong-line calibrations}\label{sec:app.mzr}
Here we use the strong line calibrations derived in \cite{Sanders2024_calibration}, which are based on direct method measurement of 46 galaxies in $1.4 \le z \le 8.7$, and apply the  method outlined below following \cite{Rowland2025_rebels} to estimate the metallicity \cite[works such as][use similar methods to derive metallicities]{Nakajima2023_metallicity,Chemerynska2024_metallicity} . 

We estimate the metallicity using a combination of R23 and O32 ($[{\rm O\textsc{iii}}]\lambda5007/[{\rm O\textsc{ii}}]\lambda3727,29$), or R3 ($[{\rm O\textsc{iii}}]\lambda 5007/ {\rm H\beta}$) and Ne3O2 ($[{\rm Ne\textsc{iii}}]\lambda3869/[{\rm O\textsc{ii}}]\lambda 3727,29$). 
To resolve the degeneracy inherent in R23 and R3 when estimating the metallicity, we compare the two solutions to the observed O32 or Ne3O2 ratio and select the metallicity closest to their empirical relation.
The combination of R23-O32 is usually used when one can correct the corresponding lines for dust using the Balmer decrement, which we perform here. We also perform dust corrections using the `A$_{\rm V}$ method' as described before. Usually, in observational studies, when the Balmer decrement is not available, one can choose R3 and Ne3O2, as the constituent lines are closer in wavelength, so dust attenuation is expected to be negligible. 
In cases where the observed line ratios fall outside the calibration range of R23 or R3 (\ie yielding complex solutions), we use the O32 or Ne3O2 calibration from \cite{Sanders2024_calibration}. 
This provides a single valued solution. 
Notably, a higher value of O32 or Ne3O2 typically indicates lower metallicity for the same ionisation parameter.
\begin{figure*}
    \centering
    \includegraphics[width=0.75\textwidth]{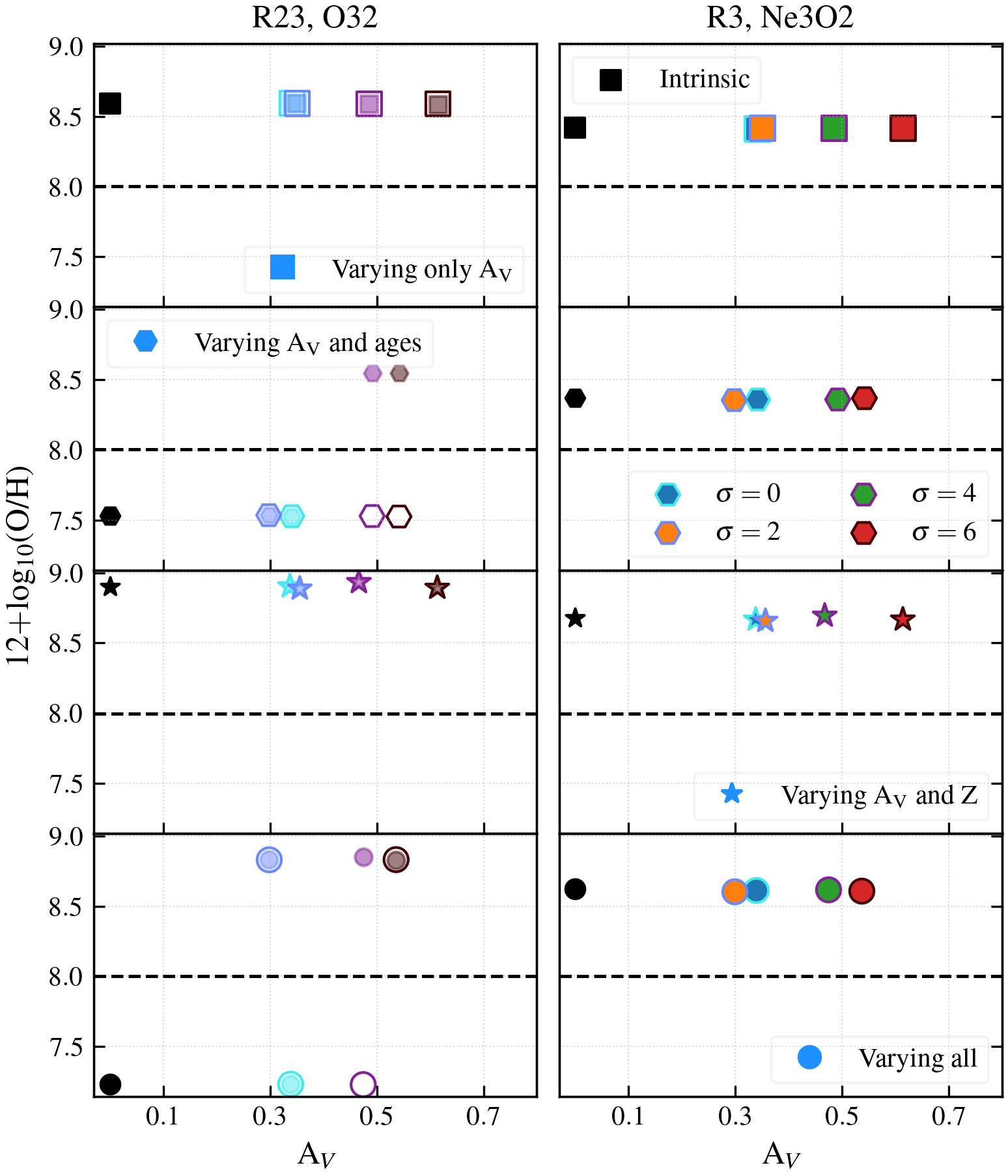}
    \caption{Figure shows the recovered metallicity of the toy galaxy sample plotted against the attenuation in the V-band.  The different rows show the 4 model galaxies: `Varying only A$_{\rm V}$', `Varying A$_{\rm V}$ and ages', `Varying A$_{\rm V}$ and Z' and `Varying all'. The two panels denote the value obtained using R23 and O32 (left), and R3 and Ne3O2 (right) respectively.
    The markers in black denote the metallicity derived from the intrinsic (dust-free) line luminosity ratios.
    In case of the `R23, O32' column, the filled markers and open marker denote the metallicity recovered after performing the `A$_{\rm V}$ method' and `Balmer decrement method' respectively on the individual lines used. There is no dust correction applied when using the `R3, Ne3O2' line ratios.
    We also plot the mass-weighted metallicity of the galaxy as the black dashed line.}
    \label{fig:toy_met_Av_sanders}
\end{figure*}

Figure~\ref{fig:toy_met_Av_sanders} shows the metallicity estimates of the toy models as a function of A$_{\rm V}$, using both R23-O32 and R3-Ne3O2. 
For the `R23-O32' case, we apply dust corrections to the constituent lines using the Balmer decrement and A$_{\rm V}$. 
We do not apply any dust-correction when using `R3-Ne3O2'.
We also plot the metallicities obtained using the intrinsic (dust-free) line ratios (black symbols at A$_{\rm V}=0$). The intrinsic value could be considered as the `true metallicity' for a particular calibration, as that is the value obtained without dust. The mass-weighted metallicity of the star-forming region is $12+{\rm log}_{10}({\rm O/H})\sim8.0$, which is denoted by the dashed horizontal line.

The toy models consistently yield R3-Ne3O2 metallicities (right panel) within 0.1 dex of the intrinsic value. It is also quite clear that due to the low value of the Ne3O2 ratio, the high-metallicity solution of the R3 ratio is preferred. 
This results in all the derived metallicities being $\gtrsim 0.4$ dex higher than the mass-weighted metallicity of the toy galaxies.
In case of the R23-O32 ratio, dust corrections introduce greater complexity to the derived metallicity.
Depending on the dust correction that is done for O32 ratio, the derived metallicity may swing to the lower or upper branch, thus care must be taken when interpreting metallicities derived using strong-line calibrations.
This does not affect the `Varying only A$_{\rm V}$' toy models, as all the star clusters have nearly identical line strength.
All the derived metallicity measurements are within $\geq 0.5$ dex or $\le 0.5$ dex of the mass-weighted value.

\subsection{\flares}
\begin{figure*}
    \centering
    \includegraphics[width=0.8\linewidth]{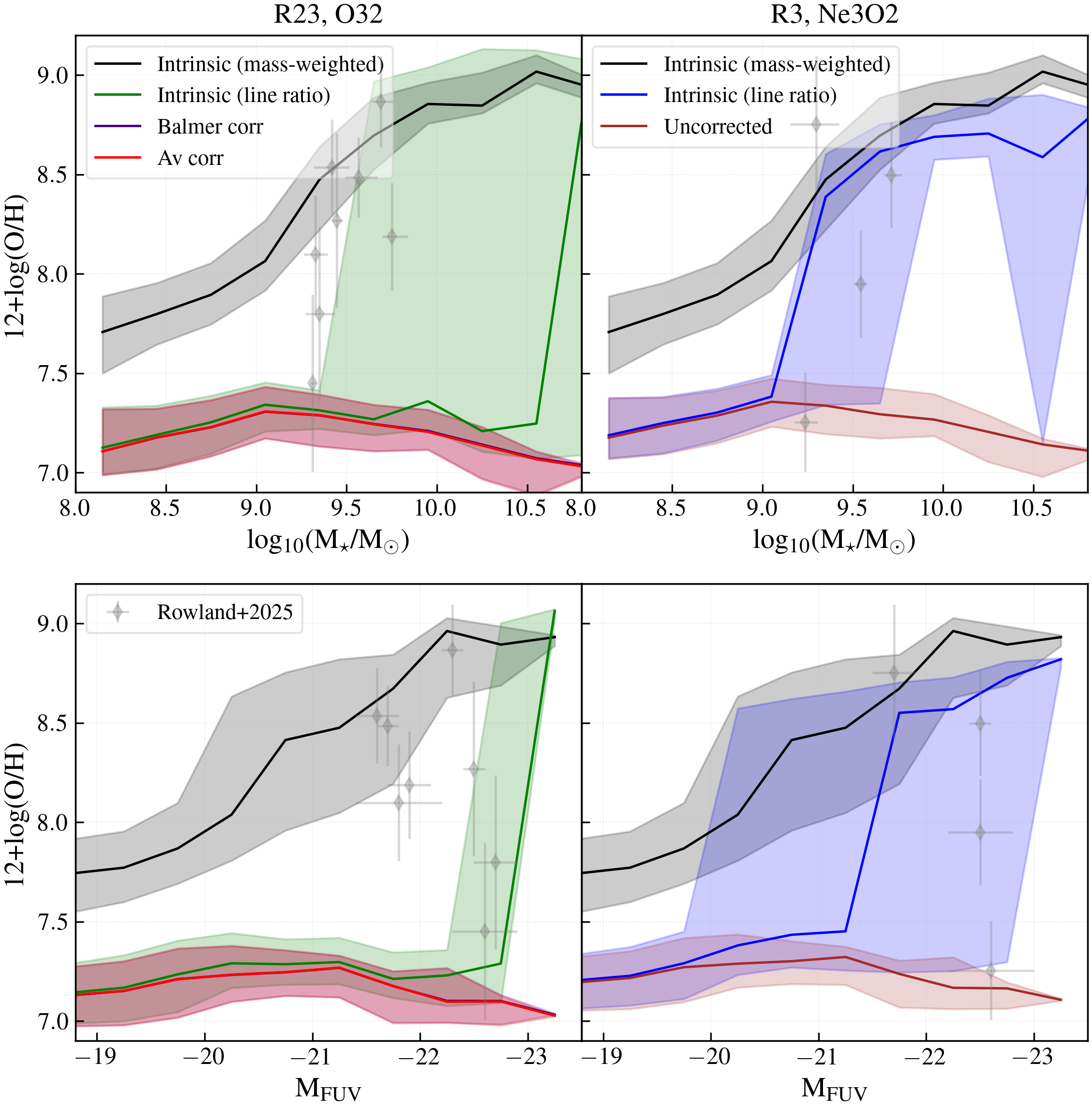}
    \caption{The different panels show the mass-metallicity and FUV luminosity-metallicity relation for the galaxies in \flares\ at $z=6$. The left panel shows the metallicity derived from using a combination of the R23 and O32 ratio, while the right panel is the metallicity derived using R3 and Ne3O2. Observed and intrinsic are the value computed using the observed and intrinsic (dust-free) line luminosity ratios. In the black is plotted the \flares\ metallicity calculated directly from the simulation. We also plot alongside the metallicity estimates from \cite{Rowland2025_rebels} for comparison, since we use the same method to derive the metallicities. 
    }
    \label{fig:flares_mzr_lratios}
    % \vspace{0.5cm}
\end{figure*}

We now apply the empirical calibration to the \flares\ galaxies at $z=6$, and present the mass-metallicity relation in Figure~\ref{fig:flares_mzr_lratios} (top panel). We also plot the metallicity against the observed UV luminosity in the bottom panel.
The left and right panels show the MZR based on the R23–O32 and R3–Ne3O2 calibrations, respectively.
For both methods we show
the `Intrinsic (mass-weighted)' value that is obtained from the simulation by using the mass-weighted metallicity (converting the metal mass fraction to a $12+{\rm log}_{10}{\rm (O/H)}$ value assuming solar metallicity of 0.014) and the `Intrinsic (line ratio)' value obtained from applying the strong-line calibration to the intrinsic line ratio (dust-free). 
For the `R23-O32' measured metallicities we plot the values obtained after dust correcting the constituent lines using the Balmer decrement (`Balmer corr') and the A$_{\rm V}$ (`A$_{\rm V}$ corr'). 
In case of the `R3-Ne3O2' derived value, we do not apply any dust correction to the observed line ratios from \flares\ and denote it as \revised{`Uncorrected'}. 
We also plot data from \cite{Rowland2025_rebels}, since they use the \cite{Sanders2024_calibration} strong-line calibrations for deriving the metallicities.

It can be seen that all metallicity measurements are similar across both combinations for the `Balmer corr', `A$_{\rm V}$ corr' and the `No dust corr' estimates. 
The derived mass-metallicity (or UV luminosity-metallicity) relation is flat in \flares, implying that the strong-line ratios occupy a narrow range. This was already discussed with respect to Figure~\ref{fig:R23_O32_R3_Ne3O2_comp}.
In contrast, the mass-weighted metallicity shows a mass (or luminosity) - metallicity relation, that has a positive slope, which flattens at the highest stellar (and brightest luminosity) bins.

The intrinsic line ratio values deviate from the dust-corrected (left panel of Figure~\ref{fig:flares_mzr_lratios}) and the non dust corrected (right panel of Figure~\ref{fig:flares_mzr_lratios}) at the high-mass/luminosity end.
For the `R23-O32' method, this deviation from the dust-corrected values occurs at stellar mass of $\sim10^{10.5}$ M$_{\odot}$, and  at the highest luminosity bin (M$_{\rm UV}\lesssim-22$). 
As shown in Figure~\ref{fig:R23_O32_R3_Ne3O2_comp} (top right panel), most \flares\ galaxies have high O32 values, favouring the low-metallicity branch of the R23 solutions.
However, galaxies at the upper end of mass or luminosity display lower O32, resulting in selection of the high-metallicity branch, explaining the flattening.

A similar pattern is seen for the `R3-Ne3O2' method. However, in this case the higher-metallicity branch is chosen at intermediate mass/luminosities (M$_{\star}\gtrsim10^9$ M$_{\odot}$ / M$_{\rm FUV}\lesssim -20$). 
The `Intrinsic' ratio in this case matches the mass-weighted relation at the high-mass (M$_{\star}\gtrsim10^{9.5}$ M$_{\odot}$) and the high-luminosity (M$_{\rm FUV}\lesssim -22$) end.

% This can be understood in the context of Figure~\ref{fig:R23_O32_R3_Ne3O2_comp} (top right panel), where most of the \flares\ galaxies occupy high values of the O32 ratio. This implies that in most cases the low-metallicity branch of the double-valued R23 solution is chosen, while bulk of the points at the high-mass/luminosity end have lower O32 values, picking the high-metallicity solution. A similar case happens for the `R3-Ne3O2' method, however, in this case the higher-metallicity branch is chosen at intermediate mass/luminosities (M$_{\star}\gtrsim10^9$ M$_{\odot}$ / M$_{\rm FUV}\lesssim -20$). 
% The `Intrinsic' ratio in this case matches the mass-weighted relation at the high-mass (M$_{\star}\gtrsim10^{9.5}$ M$_{\odot}$) and the high-luminosity (M$_{\rm FUV}\lesssim -22$) end. 
% This suggests that variations in the underlying stellar population and dust attenuation can lead to bias in metallicity estimates, when using these empirical methods.

These results highlight the sensitivity of metallicity estimates to both dust attenuation and the properties of the underlying stellar population. 
A key driver of the discrepancy between the mass-weighted metallicity and those inferred from strong-line calibrations is that the latter trace the ionising photon number weighted metallicity, which is further modulated by dust.
For example, a simple stellar population with metallicities of $10^{-4}$ ($12+{\rm log}_{10}({\rm O/H})\sim6.5$) and $0.02$ ($12+{\rm log}_{10}({\rm O/H})\sim8.8$) can differ approximately by an order of magnitude in the ionising photon production rate for stellar ages below 10 Myr \cite[see Figure A1 in][]{Wilkins2020_nebular}.
This discrepancy is especially pronounced in FLARES galaxies, which exhibit strong radial gradients in both dust and metallicity \cite[]{FLARES-XII}. 
In contrast, simulations with flat gradients in dust or metals produce minimal effects.

Moreover, the observed discrepancies also suggest that our assumptions about ISM conditions in the \textsc{cloudy} photoionisation modelling and metals tracing dust will need refinement. 
There is already evidence of increased hydrogen densities in the high-redshift Universe, which is important given that the [{O\textsc{ii}}]$\lambda3727,29$ doublet is very sensitive to the electron density at $10^2 - 10^3$ cm$^{-3}$, and will get suppressed at densities $>10^3$ cm$^{-3}$ due to approaching its critical density for excitation.
In Appendix~\ref{sec:app.varyphot}, we already explored how changes to our fiducial assumptions for the ionisation parameter (U) and hydrogen density (n$_{\rm H}$) affect the inferred metallicities using the toy models.

In conclusion, the analysis in this section shows that an incorrect estimate of the O32 or Ne3O2 ratio can cause a galaxy to be put on the wrong diagnostic branch when estimating metallicities using the strong-line methods assumed here. This can lead to a strongly underestimated (or overestimated) metallicity, even after applying a dust correction. While this  is a systematic issue for our FLARES galaxies, our toy models show that it can also be a problem for observations, if the physical properties of the line emitting gas are different from those in the diagnostic's original calibration sample.

\bibliographystyle{mnras}
\bibliography{line_ratios, flares} 

\begin{thebibliography}{}
\makeatletter
\relax
\def\mn@urlcharsother{\let\do\@makeother \do\$\do\&\do\#\do\^\do\_\do\%\do\~}
\def\mn@doi{\begingroup\mn@urlcharsother \@ifnextchar [ {\mn@doi@} {\mn@doi@[]}}
\def\mn@doi@[#1]#2{\def\@tempa{#1}\ifx\@tempa\@empty \href {http://dx.doi.org/#2} {doi:#2}\else \href {http://dx.doi.org/#2} {#1}\fi \endgroup}
\def\mn@eprint#1#2{\mn@eprint@#1:#2::\@nil}
\def\mn@eprint@arXiv#1{\href {http://arxiv.org/abs/#1} {{\tt arXiv:#1}}}
\def\mn@eprint@dblp#1{\href {http://dblp.uni-trier.de/rec/bibtex/#1.xml} {dblp:#1}}
\def\mn@eprint@#1:#2:#3:#4\@nil{\def\@tempa {#1}\def\@tempb {#2}\def\@tempc {#3}\ifx \@tempc \@empty \let \@tempc \@tempb \let \@tempb \@tempa \fi \ifx \@tempb \@empty \def\@tempb {arXiv}\fi \@ifundefined {mn@eprint@\@tempb}{\@tempb:\@tempc}{\expandafter \expandafter \csname mn@eprint@\@tempb\endcsname \expandafter{\@tempc}}}

\bibitem[\protect\citeauthoryear{{Arellano-C{\'o}rdova} et~al.,}{{Arellano-C{\'o}rdova} et~al.}{2022}]{Arellano2022}
{Arellano-C{\'o}rdova} K.~Z.,  et~al., 2022, \mn@doi [\apjl] {10.3847/2041-8213/ac9ab2}, \href {https://ui.adsabs.harvard.edu/abs/2022ApJ...940L..23A} {940, L23}

\bibitem[\protect\citeauthoryear{{Asplund}, {Grevesse}, {Sauval}  \& {Scott}}{{Asplund} et~al.}{2009}]{Asplund2009}
{Asplund} M.,  {Grevesse} N.,  {Sauval} A.~J.,   {Scott} P.,  2009, \mn@doi [\araa] {10.1146/annurev.astro.46.060407.145222}, \href {https://ui.adsabs.harvard.edu/abs/2009ARA&A..47..481A} {47, 481}

\bibitem[\protect\citeauthoryear{{Astropy Collaboration} et~al.,}{{Astropy Collaboration} et~al.}{2013}]{astropy:2013}
{Astropy Collaboration} et~al., 2013, \mn@doi [\aap] {10.1051/0004-6361/201322068}, \href {http://adsabs.harvard.edu/abs/2013A%26A...558A..33A} {558, A33}

\bibitem[\protect\citeauthoryear{{Astropy Collaboration} et~al.,}{{Astropy Collaboration} et~al.}{2018}]{astropy:2018}
{Astropy Collaboration} et~al., 2018, \mn@doi [\aj] {10.3847/1538-3881/aabc4f}, \href {https://ui.adsabs.harvard.edu/abs/2018AJ....156..123A} {156, 123}

\bibitem[\protect\citeauthoryear{{Astropy Collaboration} et~al.,}{{Astropy Collaboration} et~al.}{2022}]{astropy:2022}
{Astropy Collaboration} et~al., 2022, \mn@doi [\apj] {10.3847/1538-4357/ac7c74}, \href {https://ui.adsabs.harvard.edu/abs/2022ApJ...935..167A} {935, 167}

\bibitem[\protect\citeauthoryear{{Boquien} et~al.,}{{Boquien} et~al.}{2022}]{Boquien2022}
{Boquien} M.,  et~al., 2022, \mn@doi [\aap] {10.1051/0004-6361/202142537}, \href {https://ui.adsabs.harvard.edu/abs/2022A&A...663A..50B} {663, A50}

\bibitem[\protect\citeauthoryear{{Bouwens} et~al.,}{{Bouwens} et~al.}{2012}]{Bouwens2012}
{Bouwens} R.~J.,  et~al., 2012, \mn@doi [\apj] {10.1088/0004-637X/754/2/83}, \href {https://ui.adsabs.harvard.edu/abs/2012ApJ...754...83B} {754, 83}

\bibitem[\protect\citeauthoryear{{Bouwens} et~al.,}{{Bouwens} et~al.}{2014}]{Bouwens2014}
{Bouwens} R.~J.,  et~al., 2014, \mn@doi [\apj] {10.1088/0004-637X/793/2/115}, \href {https://ui.adsabs.harvard.edu/abs/2014ApJ...793..115B} {793, 115}

\bibitem[\protect\citeauthoryear{{Bouwens} et~al.,}{{Bouwens} et~al.}{2015}]{Bouwens2015}
{Bouwens} R.~J.,  et~al., 2015, \mn@doi [\apj] {10.1088/0004-637X/803/1/34}, \href {https://ui.adsabs.harvard.edu/abs/2015ApJ...803...34B} {803, 34}

\bibitem[\protect\citeauthoryear{{Burgarella} et~al.,}{{Burgarella} et~al.}{2025}]{Burgarella2025}
{Burgarella} D.,  et~al., 2025, \mn@doi [\aap] {10.1051/0004-6361/202554231}, \href {https://ui.adsabs.harvard.edu/abs/2025A&A...699A.336B} {699, A336}

\bibitem[\protect\citeauthoryear{{Byler}, {Dalcanton}, {Conroy}  \& {Johnson}}{{Byler} et~al.}{2017}]{Byler2017}
{Byler} N.,  {Dalcanton} J.~J.,  {Conroy} C.,   {Johnson} B.~D.,  2017, \mn@doi [\apj] {10.3847/1538-4357/aa6c66}, \href {https://ui.adsabs.harvard.edu/abs/2017ApJ...840...44B} {840, 44}

\bibitem[\protect\citeauthoryear{{Calzetti}}{{Calzetti}}{2013}]{Calzetti2013seg}
{Calzetti} D.,  2013, in {Falc{\'o}n-Barroso} J.,  {Knapen} J.~H.,  eds, , Secular Evolution of Galaxies.
p.~419, \mn@doi{10.48550/arXiv.1208.2997}

\bibitem[\protect\citeauthoryear{{Calzetti}, {Armus}, {Bohlin}, {Kinney}, {Koornneef}  \& {Storchi-Bergmann}}{{Calzetti} et~al.}{2000}]{Calzetti2000}
{Calzetti} D.,  {Armus} L.,  {Bohlin} R.~C.,  {Kinney} A.~L.,  {Koornneef} J.,   {Storchi-Bergmann} T.,  2000, \mn@doi [\apj] {10.1086/308692}, \href {https://ui.adsabs.harvard.edu/abs/2000ApJ...533..682C} {533, 682}

\bibitem[\protect\citeauthoryear{{Cameron}, {Katz}  \& {Rey}}{{Cameron} et~al.}{2023}]{Cameron2023}
{Cameron} A.~J.,  {Katz} H.,   {Rey} M.~P.,  2023, \mn@doi [\mnras] {10.1093/mnrasl/slad046}, \href {https://ui.adsabs.harvard.edu/abs/2023MNRAS.522L..89C} {522, L89}

\bibitem[\protect\citeauthoryear{{Camps} \& {Baes}}{{Camps} \& {Baes}}{2020}]{skirt2020}
{Camps} P.,  {Baes} M.,  2020, \mn@doi [Astronomy and Computing] {10.1016/j.ascom.2020.100381}, \href {https://ui.adsabs.harvard.edu/abs/2020A&C....3100381C} {31, 100381}

\bibitem[\protect\citeauthoryear{Chabrier}{Chabrier}{2003}]{ChabrierIMF}
Chabrier G.,  2003, \mn@doi [\pasp] {10.1086/376392}, 115, 763

\bibitem[\protect\citeauthoryear{{Chakraborty} et~al.,}{{Chakraborty} et~al.}{2025}]{Chakraborty2025}
{Chakraborty} P.,  et~al., 2025, \mn@doi [\apj] {10.3847/1538-4357/adc7b5}, \href {https://ui.adsabs.harvard.edu/abs/2025ApJ...985...24C} {985, 24}

\bibitem[\protect\citeauthoryear{{Charlot} \& {Fall}}{{Charlot} \& {Fall}}{2000}]{CF00}
{Charlot} S.,  {Fall} S.~M.,  2000, \mn@doi [\apj] {10.1086/309250}, \href {https://ui.adsabs.harvard.edu/abs/2000ApJ...539..718C} {539, 718}

\bibitem[\protect\citeauthoryear{{Chemerynska} et~al.,}{{Chemerynska} et~al.}{2024}]{Chemerynska2024_metallicity}
{Chemerynska} I.,  et~al., 2024, \mn@doi [\apjl] {10.3847/2041-8213/ad8dc9}, \href {https://ui.adsabs.harvard.edu/abs/2024ApJ...976L..15C} {976, L15}

\bibitem[\protect\citeauthoryear{{Chen} et~al.,}{{Chen} et~al.}{2020}]{Chen2020}
{Chen} C.-C.,  et~al., 2020, \mn@doi [\aap] {10.1051/0004-6361/201936286}, \href {https://ui.adsabs.harvard.edu/abs/2020A&A...635A.119C} {635, A119}

\bibitem[\protect\citeauthoryear{{Choustikov}, {Stiskalek}, {Saxena}, {Katz}, {Devriendt}  \& {Slyz}}{{Choustikov} et~al.}{2025}]{Choustikov2025_ILI}
{Choustikov} N.,  {Stiskalek} R.,  {Saxena} A.,  {Katz} H.,  {Devriendt} J.,   {Slyz} 2025, \mn@doi [\mnras] {10.1093/mnras/staf126}, \href {https://ui.adsabs.harvard.edu/abs/2025MNRAS.537.2273C} {537, 2273}

\bibitem[\protect\citeauthoryear{{Cornejo-C{\'a}rdenas}, {Sillero}, {Tissera}, {Boquien}, {Vilchez}, {Bruzual}  \& {Jofr{\'e}}}{{Cornejo-C{\'a}rdenas} et~al.}{2025}]{Cornejo2025}
{Cornejo-C{\'a}rdenas} A.,  {Sillero} E.,  {Tissera} P.~B.,  {Boquien} M.,  {Vilchez} J.,  {Bruzual} G.,   {Jofr{\'e}} P.,  2025, \mn@doi [\aap] {10.1051/0004-6361/202452226}, \href {https://ui.adsabs.harvard.edu/abs/2025A&A...699A.380C} {699, A380}

\bibitem[\protect\citeauthoryear{{Covelo-Paz} et~al.,}{{Covelo-Paz} et~al.}{2025}]{CoveloPaz2025_halpha}
{Covelo-Paz} A.,  et~al., 2025, \mn@doi [\aap] {10.1051/0004-6361/202452363}, \href {https://ui.adsabs.harvard.edu/abs/2025A&A...694A.178C} {694, A178}

\bibitem[\protect\citeauthoryear{Crain \& van~de Voort}{Crain \& van~de Voort}{2023}]{crain_voort_review}
Crain R.~A.,  van~de Voort F.,  2023, \mn@doi [\araa] {https://doi.org/10.1146/annurev-astro-041923-043618}, 61, 473

\bibitem[\protect\citeauthoryear{Crain et~al.,}{Crain et~al.}{2015}]{crain_eagle_2015}
Crain R.~A.,  et~al., 2015, \mn@doi [\mnras] {10.1093/mnras/stv725}, 450, 1937

\bibitem[\protect\citeauthoryear{{Cullen} et~al.,}{{Cullen} et~al.}{2024}]{Cullen2024uvslope}
{Cullen} F.,  et~al., 2024, \mn@doi [\mnras] {10.1093/mnras/stae1211}, \href {https://ui.adsabs.harvard.edu/abs/2024MNRAS.531..997C} {531, 997}

\bibitem[\protect\citeauthoryear{{Curti} et~al.,}{{Curti} et~al.}{2023}]{Curti2023}
{Curti} M.,  et~al., 2023, \mn@doi [\mnras] {10.1093/mnras/stac2737}, \href {https://ui.adsabs.harvard.edu/abs/2023MNRAS.518..425C} {518, 425}

\bibitem[\protect\citeauthoryear{{Curti} et~al.,}{{Curti} et~al.}{2024}]{Curti2024_metallicity}
{Curti} M.,  et~al., 2024, \mn@doi [\aap] {10.1051/0004-6361/202346698}, \href {https://ui.adsabs.harvard.edu/abs/2024A&A...684A..75C} {684, A75}

\bibitem[\protect\citeauthoryear{{Davis}, {Efstathiou}, {Frenk}  \& {White}}{{Davis} et~al.}{1985}]{Davis1985FOF}
{Davis} M.,  {Efstathiou} G.,  {Frenk} C.~S.,   {White} S.~D.~M.,  1985, \mn@doi [\apj] {10.1086/163168}, \href {https://ui.adsabs.harvard.edu/abs/1985ApJ...292..371D} {292, 371}

\bibitem[\protect\citeauthoryear{{De Barros}, {Oesch}, {Labb{\'e}}, {Stefanon}, {Gonz{\'a}lez}, {Smit}, {Bouwens}  \& {Illingworth}}{{De Barros} et~al.}{2019}]{deBarros19_OIIIHbeta}
{De Barros} S.,  {Oesch} P.~A.,  {Labb{\'e}} I.,  {Stefanon} M.,  {Gonz{\'a}lez} V.,  {Smit} R.,  {Bouwens} R.~J.,   {Illingworth} G.~D.,  2019, \mn@doi [\mnras] {10.1093/mnras/stz940}, \href {https://ui.adsabs.harvard.edu/abs/2019MNRAS.489.2355D} {489, 2355}

\bibitem[\protect\citeauthoryear{{Dolag}, {Borgani}, {Murante}  \& {Springel}}{{Dolag} et~al.}{2009}]{Dolag2009subfind}
{Dolag} K.,  {Borgani} S.,  {Murante} G.,   {Springel} V.,  2009, \mn@doi [\mnras] {10.1111/j.1365-2966.2009.15034.x}, \href {https://ui.adsabs.harvard.edu/abs/2009MNRAS.399..497D} {399, 497}

\bibitem[\protect\citeauthoryear{{Feltre}, {Charlot}  \& {Gutkin}}{{Feltre} et~al.}{2016}]{Feltre2016}
{Feltre} A.,  {Charlot} S.,   {Gutkin} J.,  2016, \mn@doi [\mnras] {10.1093/mnras/stv2794}, \href {https://ui.adsabs.harvard.edu/abs/2016MNRAS.456.3354F} {456, 3354}

\bibitem[\protect\citeauthoryear{{Ferland} et~al.,}{{Ferland} et~al.}{2017}]{Cloudy17.02}
{Ferland} G.~J.,  et~al., 2017, \rmxaa, \href {https://ui.adsabs.harvard.edu/abs/2017RMxAA..53..385F} {53, 385}

\bibitem[\protect\citeauthoryear{{Fern{\'a}ndez-Ontiveros}, {P{\'e}rez-Montero}, {V{\'\i}lchez}, {Amor{\'\i}n}  \& {Spinoglio}}{{Fern{\'a}ndez-Ontiveros} et~al.}{2021}]{Fernandez2021}
{Fern{\'a}ndez-Ontiveros} J.~A.,  {P{\'e}rez-Montero} E.,  {V{\'\i}lchez} J.~M.,  {Amor{\'\i}n} R.,   {Spinoglio} L.,  2021, \mn@doi [\aap] {10.1051/0004-6361/202039716}, \href {https://ui.adsabs.harvard.edu/abs/2021A&A...652A..23F} {652, A23}

\bibitem[\protect\citeauthoryear{{Gunasekera}, {van Hoof}, {Chatzikos}  \& {Ferland}}{{Gunasekera} et~al.}{2023}]{cloudy2023_23.01}
{Gunasekera} C.~M.,  {van Hoof} P. A.~M.,  {Chatzikos} M.,   {Ferland} G.~J.,  2023, \mn@doi [Research Notes of the American Astronomical Society] {10.3847/2515-5172/ad0e75}, \href {https://ui.adsabs.harvard.edu/abs/2023RNAAS...7..246G} {7, 246}

\bibitem[\protect\citeauthoryear{{Gutkin}, {Charlot}  \& {Bruzual}}{{Gutkin} et~al.}{2016}]{Gutkin2016}
{Gutkin} J.,  {Charlot} S.,   {Bruzual} G.,  2016, \mn@doi [\mnras] {10.1093/mnras/stw1716}, \href {https://ui.adsabs.harvard.edu/abs/2016MNRAS.462.1757G} {462, 1757}

\bibitem[\protect\citeauthoryear{{Harikane} et~al.,}{{Harikane} et~al.}{2025}]{Harikane2025}
{Harikane} Y.,  et~al., 2025, \mn@doi [\apj] {10.3847/1538-4357/ae0e53}, \href {https://ui.adsabs.harvard.edu/abs/2025ApJ...993..204H} {993, 204}

\bibitem[\protect\citeauthoryear{Harris et~al.,}{Harris et~al.}{2020}]{numpy}
Harris C.~R.,  et~al., 2020, \mn@doi [Nature] {10.1038/s41586-020-2649-2}, 585, 357

\bibitem[\protect\citeauthoryear{{Harvey} et~al.,}{{Harvey} et~al.}{2025}]{Harvey2025}
{Harvey} T.,  et~al., 2025, \mn@doi [\mnras] {10.1093/mnras/staf1396}, \href {https://ui.adsabs.harvard.edu/abs/2025MNRAS.542.2998H} {542, 2998}

\bibitem[\protect\citeauthoryear{{Heintz} et~al.,}{{Heintz} et~al.}{2023}]{Heintz2023}
{Heintz} K.~E.,  et~al., 2023, \mn@doi [Nature Astronomy] {10.1038/s41550-023-02078-7}, \href {https://ui.adsabs.harvard.edu/abs/2023NatAs...7.1517H} {7, 1517}

\bibitem[\protect\citeauthoryear{{Heintz} et~al.,}{{Heintz} et~al.}{2025}]{Heintz2025}
{Heintz} K.~E.,  et~al., 2025, \mn@doi [\aap] {10.1051/0004-6361/202450243}, \href {https://ui.adsabs.harvard.edu/abs/2025A&A...693A..60H} {693, A60}

\bibitem[\protect\citeauthoryear{{Hirschmann}, {Charlot}  \& {Somerville}}{{Hirschmann} et~al.}{2023}]{Hirschmann2023}
{Hirschmann} M.,  {Charlot} S.,   {Somerville} R.~S.,  2023, \mn@doi [\mnras] {10.1093/mnras/stad2745}, \href {https://ui.adsabs.harvard.edu/abs/2023MNRAS.526.3504H} {526, 3504}

\bibitem[\protect\citeauthoryear{Hunter}{Hunter}{2007}]{matplotlib}
Hunter J.~D.,  2007, \mn@doi [Computing in Science \& Engineering] {10.1109/MCSE.2007.55}, 9, 90

\bibitem[\protect\citeauthoryear{{Inoue}}{{Inoue}}{2005}]{Inoue2005}
{Inoue} A.~K.,  2005, \mn@doi [\mnras] {10.1111/j.1365-2966.2005.08890.x}, \href {https://ui.adsabs.harvard.edu/abs/2005MNRAS.359..171I} {359, 171}

\bibitem[\protect\citeauthoryear{{Isobe}, {Ouchi}, {Nakajima}, {Harikane}, {Ono}, {Xu}, {Zhang}  \& {Umeda}}{{Isobe} et~al.}{2023}]{Isobe2023ne}
{Isobe} Y.,  {Ouchi} M.,  {Nakajima} K.,  {Harikane} Y.,  {Ono} Y.,  {Xu} Y.,  {Zhang} Y.,   {Umeda} H.,  2023, \mn@doi [\apj] {10.3847/1538-4357/acf376}, \href {https://ui.adsabs.harvard.edu/abs/2023ApJ...956..139I} {956, 139}

\bibitem[\protect\citeauthoryear{{Ji} et~al.,}{{Ji} et~al.}{2023}]{XihanJi2023dust}
{Ji} X.,  et~al., 2023, \mn@doi [\aap] {10.1051/0004-6361/202245072}, \href {https://ui.adsabs.harvard.edu/abs/2023A&A...670A.125J} {670, A125}

\bibitem[\protect\citeauthoryear{{Johnson}, {Leja}, {Conroy}  \& {Speagle}}{{Johnson} et~al.}{2021}]{Prospector2021}
{Johnson} B.~D.,  {Leja} J.,  {Conroy} C.,   {Speagle} J.~S.,  2021, \mn@doi [\apjs] {10.3847/1538-4365/abef67}, \href {https://ui.adsabs.harvard.edu/abs/2021ApJS..254...22J} {254, 22}

\bibitem[\protect\citeauthoryear{{Kennicutt}}{{Kennicutt}}{1998}]{Kennicut1998}
{Kennicutt} Jr. R.~C.,  1998, \mn@doi [\apj] {10.1086/305588}, \href {https://ui.adsabs.harvard.edu/abs/1998ApJ...498..541K} {498, 541}

\bibitem[\protect\citeauthoryear{{Kewley}, {Nicholls}  \& {Sutherland}}{{Kewley} et~al.}{2019}]{Kewley2019}
{Kewley} L.~J.,  {Nicholls} D.~C.,   {Sutherland} R.~S.,  2019, \mn@doi [\araa] {10.1146/annurev-astro-081817-051832}, \href {https://ui.adsabs.harvard.edu/abs/2019ARA&A..57..511K} {57, 511}

\bibitem[\protect\citeauthoryear{{Kobayashi}, {Karakas}  \& {Lugaro}}{{Kobayashi} et~al.}{2020}]{Kobayashi2020}
{Kobayashi} C.,  {Karakas} A.~I.,   {Lugaro} M.,  2020, \mn@doi [\apj] {10.3847/1538-4357/abae65}, \href {https://ui.adsabs.harvard.edu/abs/2020ApJ...900..179K} {900, 179}

\bibitem[\protect\citeauthoryear{{Laseter} et~al.,}{{Laseter} et~al.}{2024}]{Laseter2024}
{Laseter} I.~H.,  et~al., 2024, \mn@doi [\aap] {10.1051/0004-6361/202347133}, \href {https://ui.adsabs.harvard.edu/abs/2024A&A...681A..70L} {681, A70}

\bibitem[\protect\citeauthoryear{{Leitherer} \& {Heckman}}{{Leitherer} \& {Heckman}}{1995}]{Leitherer1995}
{Leitherer} C.,  {Heckman} T.~M.,  1995, \mn@doi [\apjs] {10.1086/192112}, \href {https://ui.adsabs.harvard.edu/abs/1995ApJS...96....9L} {96, 9}

\bibitem[\protect\citeauthoryear{{Li} et~al.,}{{Li} et~al.}{2025}]{Li2025_ne}
{Li} S.,  et~al., 2025, \mn@doi [\apjl] {10.3847/2041-8213/ad9eac}, \href {https://ui.adsabs.harvard.edu/abs/2025ApJ...979L..13L} {979, L13}

\bibitem[\protect\citeauthoryear{{Llerena} et~al.,}{{Llerena} et~al.}{2025}]{Llerena2024}
{Llerena} M.,  et~al., 2025, \mn@doi [\aap] {10.1051/0004-6361/202453251}, \href {https://ui.adsabs.harvard.edu/abs/2025A&A...698A.302L} {698, A302}

\bibitem[\protect\citeauthoryear{{Lovell}, {Vijayan}, {Thomas}, {Wilkins}, {Barnes}, {Irodotou}  \& {Roper}}{{Lovell} et~al.}{2021}]{FLARESI}
{Lovell} C.~C.,  {Vijayan} A.~P.,  {Thomas} P.~A.,  {Wilkins} S.~M.,  {Barnes} D.~J.,  {Irodotou} D.,   {Roper} W.,  2021, \mn@doi [\mnras] {10.1093/mnras/staa3360}, \href {https://ui.adsabs.harvard.edu/abs/2021MNRAS.500.2127L} {500, 2127}

\bibitem[\protect\citeauthoryear{{Lovell}, {Roper}, {Vijayan}, {Wilkins}, {Newman}  \& {Seeyave}}{{Lovell} et~al.}{2025}]{synth2_2025}
{Lovell} C.~C.,  {Roper} W.~J.,  {Vijayan} A.~P.,  {Wilkins} S.~M.,  {Newman} S.,   {Seeyave} L.,  2025, \mn@doi [The Open Journal of Astrophysics] {10.33232/001c.145766}, \href {https://ui.adsabs.harvard.edu/abs/2025OJAp....8E.152L} {8, 152}

\bibitem[\protect\citeauthoryear{{Maiolino} \& {Mannucci}}{{Maiolino} \& {Mannucci}}{2019}]{MM2019review}
{Maiolino} R.,  {Mannucci} F.,  2019, \mn@doi [\aapr] {10.1007/s00159-018-0112-2}, \href {https://ui.adsabs.harvard.edu/abs/2019A&ARv..27....3M} {27, 3}

\bibitem[\protect\citeauthoryear{{Meyer} et~al.,}{{Meyer} et~al.}{2024}]{Meyer2024Oiii}
{Meyer} R.~A.,  et~al., 2024, \mn@doi [\mnras] {10.1093/mnras/stae2353}, \href {https://ui.adsabs.harvard.edu/abs/2024MNRAS.535.1067M} {535, 1067}

\bibitem[\protect\citeauthoryear{{Nagao}, {Maiolino}, {Marconi}  \& {Matsuhara}}{{Nagao} et~al.}{2011}]{Nagao2011}
{Nagao} T.,  {Maiolino} R.,  {Marconi} A.,   {Matsuhara} H.,  2011, \mn@doi [\aap] {10.1051/0004-6361/201015471}, \href {https://ui.adsabs.harvard.edu/abs/2011A&A...526A.149N} {526, A149}

\bibitem[\protect\citeauthoryear{{Nakajima}, {Ouchi}, {Isobe}, {Harikane}, {Zhang}, {Ono}, {Umeda}  \& {Oguri}}{{Nakajima} et~al.}{2023}]{Nakajima2023_metallicity}
{Nakajima} K.,  {Ouchi} M.,  {Isobe} Y.,  {Harikane} Y.,  {Zhang} Y.,  {Ono} Y.,  {Umeda} H.,   {Oguri} M.,  2023, \mn@doi [\apjs] {10.3847/1538-4365/acd556}, \href {https://ui.adsabs.harvard.edu/abs/2023ApJS..269...33N} {269, 33}

\bibitem[\protect\citeauthoryear{{Narayanan} et~al.,}{{Narayanan} et~al.}{2021}]{Powderday2021}
{Narayanan} D.,  et~al., 2021, \mn@doi [\apjs] {10.3847/1538-4365/abc487}, \href {https://ui.adsabs.harvard.edu/abs/2021ApJS..252...12N} {252, 12}

\bibitem[\protect\citeauthoryear{{Narayanan} et~al.,}{{Narayanan} et~al.}{2024}]{Narayanan2024}
{Narayanan} D.,  et~al., 2024, \mn@doi [\apj] {10.3847/1538-4357/ad0966}, \href {https://ui.adsabs.harvard.edu/abs/2024ApJ...961...73N} {961, 73}

\bibitem[\protect\citeauthoryear{{Newman}, {Lovell}, {Maraston}, {Roper}, {Vijayan}, {Wilkins}, {Giavalisco}  \& {Saxena}}{{Newman} et~al.}{2026}]{Newman2025}
{Newman} S.~L.,  {Lovell} C.~C.,  {Maraston} C.,  {Roper} W.~J.,  {Vijayan} A.~P.,  {Wilkins} S.~M.,  {Giavalisco} M.,   {Saxena} A.,  2026, \mn@doi [\mnras] {10.1093/mnras/staf1866}, \href {https://ui.adsabs.harvard.edu/abs/2026MNRAS.545f1866N} {545, staf1866}

\bibitem[\protect\citeauthoryear{Newville et~al.,}{Newville et~al.}{2025}]{lmfit_newville_2025}
Newville M.,  et~al., 2025, LMFIT: Non-Linear Least-Squares Minimization and Curve-Fitting for Python, \mn@doi{10.5281/zenodo.15014437}, \url {https://doi.org/10.5281/zenodo.15014437}

\bibitem[\protect\citeauthoryear{{Nicholls}, {Sutherland}, {Dopita}, {Kewley}  \& {Groves}}{{Nicholls} et~al.}{2017}]{GalacticConcordance2017}
{Nicholls} D.~C.,  {Sutherland} R.~S.,  {Dopita} M.~A.,  {Kewley} L.~J.,   {Groves} B.~A.,  2017, \mn@doi [\mnras] {10.1093/mnras/stw3235}, \href {https://ui.adsabs.harvard.edu/abs/2017MNRAS.466.4403N} {466, 4403}

\bibitem[\protect\citeauthoryear{{Pei}}{{Pei}}{1992}]{SMC1992}
{Pei} Y.~C.,  1992, \mn@doi [\apj] {10.1086/171637}, \href {https://ui.adsabs.harvard.edu/abs/1992ApJ...395..130P} {395, 130}

\bibitem[\protect\citeauthoryear{{Pereira-Santaella}, {Rigopoulou}, {Farrah}, {Lebouteiller}  \& {Li}}{{Pereira-Santaella} et~al.}{2017}]{Pereira-Santaella2017}
{Pereira-Santaella} M.,  {Rigopoulou} D.,  {Farrah} D.,  {Lebouteiller} V.,   {Li} J.,  2017, \mn@doi [\mnras] {10.1093/mnras/stx1284}, \href {https://ui.adsabs.harvard.edu/abs/2017MNRAS.470.1218P} {470, 1218}

\bibitem[\protect\citeauthoryear{{Pillepich} et~al.,}{{Pillepich} et~al.}{2018}]{Pillepich2018a}
{Pillepich} A.,  et~al., 2018, \mn@doi [\mnras] {10.1093/mnras/stx2656}, \href {https://ui.adsabs.harvard.edu/abs/2018MNRAS.473.4077P} {473, 4077}

\bibitem[\protect\citeauthoryear{{Planck Collaboration} et~al.,}{{Planck Collaboration} et~al.}{2014}]{planck_collaboration_2014}
{Planck Collaboration} et~al., 2014, \mn@doi [A\&A] {10.1051/0004-6361/201321529}, 571, A1

\bibitem[\protect\citeauthoryear{{Reddy}, {Topping}, {Sanders}, {Shapley}  \& {Brammer}}{{Reddy} et~al.}{2023a}]{Reddy2023_decrement}
{Reddy} N.~A.,  {Topping} M.~W.,  {Sanders} R.~L.,  {Shapley} A.~E.,   {Brammer} G.,  2023a, \mn@doi [\apj] {10.3847/1538-4357/acc869}, \href {https://ui.adsabs.harvard.edu/abs/2023ApJ...948...83R} {948, 83}

\bibitem[\protect\citeauthoryear{{Reddy}, {Topping}, {Sanders}, {Shapley}  \& {Brammer}}{{Reddy} et~al.}{2023b}]{Reddy2023}
{Reddy} N.~A.,  {Topping} M.~W.,  {Sanders} R.~L.,  {Shapley} A.~E.,   {Brammer} G.,  2023b, \mn@doi [\apj] {10.3847/1538-4357/acd754}, \href {https://ui.adsabs.harvard.edu/abs/2023ApJ...952..167R} {952, 167}

\bibitem[\protect\citeauthoryear{{Riffel}, {Dors}, {Krabbe}  \& {Esteban}}{{Riffel} et~al.}{2021}]{Riffel2021}
{Riffel} R.~A.,  {Dors} O.~L.,  {Krabbe} A.~C.,   {Esteban} C.,  2021, \mn@doi [\mnras] {10.1093/mnrasl/slab064}, \href {https://ui.adsabs.harvard.edu/abs/2021MNRAS.506L..11R} {506, L11}

\bibitem[\protect\citeauthoryear{Robertson}{Robertson}{2022}]{robertson2022_review}
Robertson B.~E.,  2022, \mn@doi [\araa] {https://doi.org/10.1146/annurev-astro-120221-044656}, 60, 121

\bibitem[\protect\citeauthoryear{{Roper} et~al.,}{{Roper} et~al.}{2025}]{synth1_2025}
{Roper} W.~J.,  et~al., 2025, \mn@doi [arXiv e-prints] {10.48550/arXiv.2506.15811}, \href {https://ui.adsabs.harvard.edu/abs/2025arXiv250615811R} {p. arXiv:2506.15811}

\bibitem[\protect\citeauthoryear{{Rowland} et~al.,}{{Rowland} et~al.}{2026}]{Rowland2025_rebels}
{Rowland} L.~E.,  et~al., 2026, \mn@doi [\mnras] {10.1093/mnras/staf2023}, \href {https://ui.adsabs.harvard.edu/abs/2026MNRAS.546f2023R} {546, staf2023}

\bibitem[\protect\citeauthoryear{{Salim} \& {Narayanan}}{{Salim} \& {Narayanan}}{2020}]{salim_narayanan_2020}
{Salim} S.,  {Narayanan} D.,  2020, \mn@doi [\araa] {10.1146/annurev-astro-032620-021933}, \href {https://ui.adsabs.harvard.edu/abs/2020ARA&A..58..529S} {58, 529}

\bibitem[\protect\citeauthoryear{{Sanders}, {Shapley}, {Topping}, {Reddy}  \& {Brammer}}{{Sanders} et~al.}{2023}]{Sanders2023}
{Sanders} R.~L.,  {Shapley} A.~E.,  {Topping} M.~W.,  {Reddy} N.~A.,   {Brammer} G.~B.,  2023, \mn@doi [\apj] {10.3847/1538-4357/acedad}, \href {https://ui.adsabs.harvard.edu/abs/2023ApJ...955...54S} {955, 54}

\bibitem[\protect\citeauthoryear{{Sanders}, {Shapley}, {Topping}, {Reddy}  \& {Brammer}}{{Sanders} et~al.}{2024}]{Sanders2024_calibration}
{Sanders} R.~L.,  {Shapley} A.~E.,  {Topping} M.~W.,  {Reddy} N.~A.,   {Brammer} G.~B.,  2024, \mn@doi [\apj] {10.3847/1538-4357/ad15fc}, \href {https://ui.adsabs.harvard.edu/abs/2024ApJ...962...24S} {962, 24}

\bibitem[\protect\citeauthoryear{{Sandles} et~al.,}{{Sandles} et~al.}{2024}]{Sandles2024}
{Sandles} L.,  et~al., 2024, \mn@doi [\aap] {10.1051/0004-6361/202347119}, \href {https://ui.adsabs.harvard.edu/abs/2024A&A...691A.305S} {691, A305}

\bibitem[\protect\citeauthoryear{{Schady} et~al.,}{{Schady} et~al.}{2024}]{Schady2024}
{Schady} P.,  et~al., 2024, \mn@doi [\mnras] {10.1093/mnras/stae677}, \href {https://ui.adsabs.harvard.edu/abs/2024MNRAS.529.2807S} {529, 2807}

\bibitem[\protect\citeauthoryear{Schaye et~al.,}{Schaye et~al.}{2015}]{schaye_eagle_2015}
Schaye J.,  et~al., 2015, \mn@doi [\mnras] {10.1093/mnras/stu2058}, 446, 521

\bibitem[\protect\citeauthoryear{{Schechter}}{{Schechter}}{1976}]{Schechter1976}
{Schechter} P.,  1976, \mn@doi [\apj] {10.1086/154079}, \href {https://ui.adsabs.harvard.edu/abs/1976ApJ...203..297S} {203, 297}

\bibitem[\protect\citeauthoryear{{Seeyave} et~al.,}{{Seeyave} et~al.}{2023}]{FLARES_XIII}
{Seeyave} L. T.~C.,  et~al., 2023, \mn@doi [\mnras] {10.1093/mnras/stad2487}, \href {https://ui.adsabs.harvard.edu/abs/2023MNRAS.525.2422S} {525, 2422}

\bibitem[\protect\citeauthoryear{{Somerville} \& {Dav{\'e}}}{{Somerville} \& {Dav{\'e}}}{2015}]{Dave_somerville_2015review}
{Somerville} R.~S.,  {Dav{\'e}} R.,  2015, \mn@doi [\araa] {10.1146/annurev-astro-082812-140951}, \href {https://ui.adsabs.harvard.edu/abs/2015ARA&A..53...51S} {53, 51}

\bibitem[\protect\citeauthoryear{{Speagle}, {Steinhardt}, {Capak}  \& {Silverman}}{{Speagle} et~al.}{2014}]{Speagle2014}
{Speagle} J.~S.,  {Steinhardt} C.~L.,  {Capak} P.~L.,   {Silverman} J.~D.,  2014, \mn@doi [\apjs] {10.1088/0067-0049/214/2/15}, \href {https://ui.adsabs.harvard.edu/abs/2014ApJS..214...15S} {214, 15}

\bibitem[\protect\citeauthoryear{{Springel}, {White}, {Tormen}  \& {Kauffmann}}{{Springel} et~al.}{2001}]{Springel2001subfind}
{Springel} V.,  {White} S. D.~M.,  {Tormen} G.,   {Kauffmann} G.,  2001, \mn@doi [\mnras] {10.1046/j.1365-8711.2001.04912.x}, \href {https://ui.adsabs.harvard.edu/abs/2001MNRAS.328..726S} {328, 726}

\bibitem[\protect\citeauthoryear{{Stanway} \& {Eldridge}}{{Stanway} \& {Eldridge}}{2018}]{BPASS2.2.1}
{Stanway} E.~R.,  {Eldridge} J.~J.,  2018, \mn@doi [\mnras] {10.1093/mnras/sty1353}, \href {https://ui.adsabs.harvard.edu/abs/2018MNRAS.479...75S} {479, 75}

\bibitem[\protect\citeauthoryear{Stark}{Stark}{2016}]{stark_2016_review}
Stark D.~P.,  2016, \mn@doi [\araa] {https://doi.org/10.1146/annurev-astro-081915-023417}, 54, 761

\bibitem[\protect\citeauthoryear{{Stasi{\'n}ska}}{{Stasi{\'n}ska}}{1978}]{Stasinska1978}
{Stasi{\'n}ska} G.,  1978, \aap, \href {https://ui.adsabs.harvard.edu/abs/1978A&A....66..257S} {66, 257}

\bibitem[\protect\citeauthoryear{{Tacchella} et~al.,}{{Tacchella} et~al.}{2022}]{Tacchella2022}
{Tacchella} S.,  et~al., 2022, \mn@doi [\apj] {10.3847/1538-4357/ac4cad}, \href {https://ui.adsabs.harvard.edu/abs/2022ApJ...927..170T} {927, 170}

\bibitem[\protect\citeauthoryear{{Torrey} et~al.,}{{Torrey} et~al.}{2019}]{Torrey2019}
{Torrey} P.,  et~al., 2019, \mn@doi [\mnras] {10.1093/mnras/stz243}, \href {https://ui.adsabs.harvard.edu/abs/2019MNRAS.484.5587T} {484, 5587}

\bibitem[\protect\citeauthoryear{{Vijayan}, {Clay}, {Thomas}, {Yates}, {Wilkins}  \& {Henriques}}{{Vijayan} et~al.}{2019}]{Vijayan2019}
{Vijayan} A.~P.,  {Clay} S.~J.,  {Thomas} P.~A.,  {Yates} R.~M.,  {Wilkins} S.~M.,   {Henriques} B.~M.,  2019, \mn@doi [\mnras] {10.1093/mnras/stz1948}, \href {https://ui.adsabs.harvard.edu/abs/2019MNRAS.489.4072V} {489, 4072}

\bibitem[\protect\citeauthoryear{{Vijayan}, {Lovell}, {Wilkins}, {Thomas}, {Barnes}, {Irodotou}, {Kuusisto}  \& {Roper}}{{Vijayan} et~al.}{2021}]{FLARESII}
{Vijayan} A.~P.,  {Lovell} C.~C.,  {Wilkins} S.~M.,  {Thomas} P.~A.,  {Barnes} D.~J.,  {Irodotou} D.,  {Kuusisto} J.,   {Roper} W.~J.,  2021, \mn@doi [\mnras] {10.1093/mnras/staa3715}, \href {https://ui.adsabs.harvard.edu/abs/2021MNRAS.501.3289V} {501, 3289}

\bibitem[\protect\citeauthoryear{{Vijayan}, {Thomas}, {Lovell}, {Wilkins}, {Greve}, {Irodotou}, {Roper}  \& {Seeyave}}{{Vijayan} et~al.}{2024}]{FLARES-XII}
{Vijayan} A.~P.,  {Thomas} P.~A.,  {Lovell} C.~C.,  {Wilkins} S.~M.,  {Greve} T.~R.,  {Irodotou} D.,  {Roper} W.~J.,   {Seeyave} L. T.~C.,  2024, \mn@doi [\mnras] {10.1093/mnras/stad3594}, \href {https://ui.adsabs.harvard.edu/abs/2024MNRAS.527.7337V} {527, 7337}

\bibitem[\protect\citeauthoryear{{Virtanen} et~al.,}{{Virtanen} et~al.}{2020}]{scipy}
{Virtanen} P.,  et~al., 2020, \mn@doi [Nature Methods] {https://doi.org/10.1038/s41592-019-0686-2}, \href {https://rdcu.be/b08Wh} {17, 261}

\bibitem[\protect\citeauthoryear{{Vogelsberger}, {Marinacci}, {Torrey}  \& {Puchwein}}{{Vogelsberger} et~al.}{2020}]{Vogelsberger2020}
{Vogelsberger} M.,  {Marinacci} F.,  {Torrey} P.,   {Puchwein} E.,  2020, \mn@doi [Nature Reviews Physics] {10.1038/s42254-019-0127-2}, \href {https://ui.adsabs.harvard.edu/abs/2020NatRP...2...42V} {2, 42}

\bibitem[\protect\citeauthoryear{{Wiersma}, {Schaye}, {Theuns}, {Dalla Vecchia}  \& {Tornatore}}{{Wiersma} et~al.}{2009}]{Wiersma2009}
{Wiersma} R. P.~C.,  {Schaye} J.,  {Theuns} T.,  {Dalla Vecchia} C.,   {Tornatore} L.,  2009, \mn@doi [\mnras] {10.1111/j.1365-2966.2009.15331.x}, \href {https://ui.adsabs.harvard.edu/abs/2009MNRAS.399..574W} {399, 574}

\bibitem[\protect\citeauthoryear{{Wilkins}, {Lovell}  \& {Stanway}}{{Wilkins} et~al.}{2019}]{Wilkins2019}
{Wilkins} S.~M.,  {Lovell} C.~C.,   {Stanway} E.~R.,  2019, \mn@doi [\mnras] {10.1093/mnras/stz2894}, \href {https://ui.adsabs.harvard.edu/abs/2019MNRAS.490.5359W} {490, 5359}

\bibitem[\protect\citeauthoryear{{Wilkins} et~al.,}{{Wilkins} et~al.}{2020}]{Wilkins2020_nebular}
{Wilkins} S.~M.,  et~al., 2020, \mn@doi [\mnras] {10.1093/mnras/staa649}, \href {https://ui.adsabs.harvard.edu/abs/2020MNRAS.493.6079W} {493, 6079}

\bibitem[\protect\citeauthoryear{{Wilkins} et~al.,}{{Wilkins} et~al.}{2022}]{FLARES-VI}
{Wilkins} S.~M.,  et~al., 2022, \mn@doi [\mnras] {10.1093/mnras/stac2548}, \href {https://ui.adsabs.harvard.edu/abs/2022MNRAS.517.3227W} {517, 3227}

\bibitem[\protect\citeauthoryear{{Wilkins} et~al.,}{{Wilkins} et~al.}{2023}]{FLARES-XI}
{Wilkins} S.~M.,  et~al., 2023, \mn@doi [\mnras] {10.1093/mnras/stad1126}, \href {https://ui.adsabs.harvard.edu/abs/2023MNRAS.522.4014W} {522, 4014}

\bibitem[\protect\citeauthoryear{{Witt} \& {Gordon}}{{Witt} \& {Gordon}}{2000}]{Witt2000}
{Witt} A.~N.,  {Gordon} K.~D.,  2000, \mn@doi [\apj] {10.1086/308197}, \href {https://ui.adsabs.harvard.edu/abs/2000ApJ...528..799W} {528, 799}

\bibitem[\protect\citeauthoryear{{Yates}, {Schady}, {Chen}, {Schweyer}  \& {Wiseman}}{{Yates} et~al.}{2020}]{Yates2020}
{Yates} R.~M.,  {Schady} P.,  {Chen} T.~W.,  {Schweyer} T.,   {Wiseman} P.,  2020, \mn@doi [\aap] {10.1051/0004-6361/201936506}, \href {https://ui.adsabs.harvard.edu/abs/2020A&A...634A.107Y} {634, A107}

\bibitem[\protect\citeauthoryear{{Yates}, {Henriques}, {Fu}, {Kauffmann}, {Thomas}, {Guo}, {White}  \& {Schady}}{{Yates} et~al.}{2021}]{Yates2021a}
{Yates} R.~M.,  {Henriques} B. M.~B.,  {Fu} J.,  {Kauffmann} G.,  {Thomas} P.~A.,  {Guo} Q.,  {White} S. D.~M.,   {Schady} P.,  2021, \mn@doi [\mnras] {10.1093/mnras/stab741}, \href {https://ui.adsabs.harvard.edu/abs/2021MNRAS.503.4474Y} {503, 4474}

\bibitem[\protect\citeauthoryear{{van der Velden}}{{van der Velden}}{2020}]{cmasher}
{van der Velden} E.,  2020, \mn@doi [The Journal of Open Source Software] {10.21105/joss.02004}, \href {https://ui.adsabs.harvard.edu/abs/2020JOSS....5.2004V} {5, 2004}

\makeatother
\end{thebibliography}
%%%%%%%%%%%%%%%%%%%%%%%%%%%%%%%%%%%%%%%%%%%%%%%%%%

%%%%%%%%%%%%%%%%% APPENDICES %%%%%%%%%%%%%%%%%%%%%

% \appendix

% \input sections/appendixA

\end{document}